\documentclass[12pt]{aastex61}
\begin{document}
\title{Kinematics of Parsec-Scale Jets of Gamma-Ray Blazars at 43~GHz within the VLBA-BU-BLAZAR Program}
\author{Svetlana G. Jorstad}
\affiliation{Institute for Astrophysical Research, Boston University, 725 Commonwealth Avenue, Boston, MA 02215}
\affiliation{Astronomical Institute, St. Petersburg University, Universitetskij Pr. 28, Petrodvorets, 
198504 St. Petersburg, Russia}

\author{Alan P. Marscher}
\affiliation{Institute for Astrophysical Research, Boston University, 725 Commonwealth Avenue, Boston, MA 02215}
 
\author{Daria A. Morozova}
\affiliation{Astronomical Institute, St. Petersburg University, Universitetskij Pr. 28, Petrodvorets, 
198504 St. Petersburg, Russia}
 
\author{Ivan S. Troitsky}
\affiliation{Astronomical Institute, St. Petersburg University, Universitetskij Pr. 28, Petrodvorets, 
198504 St. Petersburg, Russia}
 
\author{Iv\'an Agudo} 
\affiliation{Instituto de Astrof\'{\i}sica de Andaluc\'{\i}a (IAA), CSIC, Apartado 3004, 18080, Granada, Spain}

\author{Carolina Casadio}
\affiliation{Max-Planck-Institut f\"{u}r Radioastronomie, Auf dem Hügel 69, 53121 Bonn, Germany}
\affiliation{Instituto de Astrof\'{\i}sica de Andaluc\'{\i}a (IAA), CSIC, Apartado 3004, 18080, Granada, Spain}

\author{Adi Foord}
\affiliation{Astronomy Department, University of Michigan, 830, Dennison, 500 Church St., Ann Arbor, MI, 84109-1042}
\affiliation{Institute for Astrophysical Research, Boston University, 725 Commonwealth Avenue, Boston, MA 02215}

\author{Jos\'e L. G\'omez}
\affiliation{Instituto de Astrof\'{\i}sica de Andaluc\'{\i}a (IAA), CSIC, Apartado 3004, 18080, Granada, Spain} 

\author{Nicholas R. MacDonald}
\affiliation{Institute for Astrophysical Research, Boston University, 725 Commonwealth Avenue, Boston, MA 02215}

\author{Sol N. Molina}
\affiliation{Instituto de Astrof\'{\i}sica de Andaluc\'{\i}a (IAA), CSIC, Apartado 3004, 18080, Granada, Spain}  

\author{Anne L\"{a}hteenm\"{a}ki\altaffiliation}
\affiliation{Aalto University Mets\"ahovi Radio Observatory Mets\"ahovintie 114, FIN-02540 Kylm\"al\"a, Finland} 
\affiliation{Aalto University Dept of Electronics and Nanoengineering, P.O. BOX 15500, FI-00076, AALTO, Finland}
\affiliation{Tartu Observatory, Observatooriumi 1, 61602 T\~{o}ravere, Estonia}

\author{Joni Tammi}
\affiliation{Aalto University Mets\"ahovi Radio Observatory Mets\"ahovintie 114, FIN-02540 Kylm\"al\"a, Finland}

\author{Merja Tornikoski}
\affiliation{Aalto University Mets\"ahovi Radio Observatory Mets\"ahovintie 114, FIN-02540 Kylm\"al\"a, Finland}

\correspondingauthor{Svetlana G. Jorstad}
\email{jorstad@bu.edu}

\shorttitle{Kinematics of Gamma-Ray Blazar Jets}
\shortauthors{Jorstad et al.}
\begin{abstract}
We analyze the parsec-scale jet kinematics from 2007 June to 2013 January of a sample of $\gamma$-ray bright blazars monitored roughly monthly with the Very Long Baseline Array at 43~GHz. In a total of 1929 images, we measure apparent speeds of 252 emission knots in 21 quasars, 12 BL~Lacertae objects (BLLacs), and 3 radio galaxies, ranging from 0.02$c$ to 78$c$; 21\% of the knots are quasi-stationary. Approximately 1/3 of the moving knots execute non-ballistic motions, with the quasars exhibiting  acceleration along the jet within 5~pc (projected) of the core, and knots in the BLLacs tending to decelerate near the core. Using apparent speeds of components and timescales of variability from their light curves, we derive physical parameters of 120 superluminal knots, including variability Doppler factors, Lorentz factors, and viewing angles. We estimate the half-opening angle of each jet based on the projected opening angle and scatter of intrinsic viewing angles of knots. We determine characteristic values of physical parameters for each jet and  AGN class based on the range of values obtained for individual features. We calculate intrinsic brightness temperatures of the cores, $T_{\rm b,int}^{\rm core}$, at all epochs, finding that the radio galaxies usually maintain equipartition conditions in the cores, while $\sim$30\% of $T_{\rm b,int}^{\rm core}$ measurements in the quasars and BLLacs deviate from equipartition values by a factor $>$10. This probably occurs during transient events connected with active states. In the Appendix we briefly describe the behavior of each blazar during the period analyzed.   
\end{abstract}
\keywords{galaxies: active --- galaxies: BL Lacertae objects: individual  --- galaxies: individual --- galaxies: jets --- 
galaxies: quasars: individual --- techniques: interferometric}
\section{Introduction \label{Intro}}
Blazars \citep{AS80} are active galactic nuclei (AGN) that possess extreme characteristics 
across the electromagnetic spectrum, such as ultra-luminous (up to $\sim10^{50}$erg~s$^{-1}$) emission \citep[e.g.,][]{1FLG,ABDO11,GPL12,SEMM13}, high amplitudes of variability on various timescales as short as several minutes \citep[e.g.,][]{HESS10,J13}, and high degrees of optical linear polarization, which can exceed 40\%  \citep[e.g.,][]{PAUL16}. These observed phenomena are thought to be generated by relativistic jets of high-energy plasma flowing out of the nucleus at nearly the speed of light in a direction close to the line of sight \citep[e.g.,][]{LIST16}.

Observational evidence for a connection between high energy emission and radio jets of AGN emerged during the 1990s when the $\gamma$-ray telescope EGRET on board the {\it Compton Gamma Ray Observatory} found blazars to be the most numerous class of identified $\gamma$-ray sources \citep{Hartman99}, although theoretical predictions for such a connection arose a decade earlier \citep[e.g.,][]{KA81,MAR87}.
Blazars consist of two subclasses: flat-spectrum radio quasars (FSRQs) and objects of BL Lacertae type (BLLacs or BLs), which have different optical properties \citep{UP95} and different radio morphology \citep{WMA84}. Spectral energy distributions (SEDs) of blazars possess two humps, with the first (synchrotron radiation) located 
from 10$^{13}$ to 10$^{17}$~Hz and the second (probably inverse Compton scattering) between 1~MeV and 100~GeV.
According to locations of these peaks, FSRQs and BLLacs form a ``blazar sequence,'' with FSRQs and more luminous
BLLacs (LBLs) peaking at the lowest frequencies and less luminous BLLacs at higher frequencies within the above intervals \citep[HBLs;][]{GG98,EM11}.
During the EGRET era, analysis of radio light curves of radio-loud AGN revealed that $\gamma$-rays are detected mostly in quasars with high optical polarization when they exhibit a high-radio-frequency outburst \citep{VT95}, while BLLacs
are weaker $\gamma$-ray emitters, with no clear correlation between radio and $\gamma$-ray variations \citep{LV03}. Using multi-epoch images obtained with the Very Long Baseline 
Array (VLBA),
\cite{J01} found a statistically significant correlation between the flux density 
of the compact ``core'' and the $\gamma$-ray flux in a sample of FSRQs and BLLacs detected by EGRET, while \cite{J01b} reported a connection between epochs of high $\gamma$-ray states and ejections of superluminal components from the core. These findings placed the origin of $\gamma$-ray flaring emission in relativistic radio jets with high Doppler factors, several parsecs downstream of the central engine associated with a supermassive black hole (BH).

The launch of the {\it Fermi Gamma-Ray Space Telescope} in 2008, with the Large Area Telescope (LAT) surveying almost the entire sky every 3~hr, has opened a new era in the investigation of blazars. Multi-wavelength (MW) campaigns of simultaneous monitoring from $\gamma$-ray to radio bands have become routine  \citep[e.g.,][]{ABDO10a,RAI11,HAY12,VLAR13,Aleksic14, Aleksic15,ANN16}. In addition, regular imaging observations of AGN jets at 15~GHz (the MOJAVE program\footnote{http://www.physics.purdue.edu/MOJAVE/}) and monitoring of $\gamma$-ray blazars at 43~GHz (the VLBA-BU-BLAZAR program\footnote{http://www.bu.edu/blazars/VLBAproject.html}) with the VLBA have been established to probe the jets of blazars.   

During the Fermi era, the number of $\gamma$-ray sources detected by LAT has been growing 
significantly with every new catalog release \citep{1FLG,2FLG,3FLG}, although blazars remain the largest class of identified $\gamma$-ray sources \citep{3FLG}.
Differences between LAT-detected BLLacs and FSRQs have become more apparent as the size of the detected population has increased. Slower apparent speeds are measured in BLLacs despite their higher overall LAT detection rate with respect to quasars \citep{3FLG}. This suggests that LAT-detected BLLacs have higher intrinsic (unbeamed) $\gamma$-ray to radio luminosity ratios \citep{LIST09a}.
In addition, the $\gamma$-ray spectra of BLLacs are harder than those of FSRQs, and therefore dominate the sources detected at very high $\gamma$-ray energies, E$>$100~GeV
\citep[VHE; e.g.,][]{BW16}.

Early analyses comparing contemporaneous LAT and radio observations indicated a tight connection between the parsec-scale jet behavior and $\gamma$-ray properties. For example, the jets of $\gamma$-ray bright quasars in the MOJAVE survey have faster apparent motions on average and brighter parsec-scale cores than do quasars that have not yet been detected by the LAT. This implies that the jets of brighter $\gamma$-ray quasars have preferentially higher Doppler-boosting factors \citep{LIST09a,YY09, LIST16}.
A statistically significant correlation between both the flux densities and luminosities in $\gamma$-ray and radio bands has been found by \cite{N37} using radio light curves at 37~GHz from the Mets\"{a}hovi quasar monitoring  program (Mets\"{a}hovi Radio Obs., Aalto Univ., Finland). A comparison of $\gamma$-ray and mm-wave flares has revealed that the brightest $\gamma$-ray events coincide with the initial stages of a millimeter-wave outburst \citep{L37}. On the other hand, using radio light curves at 15~GHz of 41 $\gamma$-ray sources from the Owens Valley Radio Observatory 40~m monitoring program, \citet{MM14} have found a significant correlation in only 4 blazars, with the radio lagging the $\gamma$-ray variations. However, a cross-correlation analysis between radio (2.7 to 350 GHz, F-GAMMA program) and $\gamma$-ray light curves of 54 $\gamma$-ray bright blazars has revealed a statistically significant correlation when averaging over the entire sample, with radio variations lagging $\gamma$-ray variations; the delays systematically decrease from cm to mm/sub-mm, with a minimum delay of 7$\pm$9 days \citep{LARS14}. 

It is now clear that, despite short timescales of variability, the typical duration of a high $\gamma$-ray state in a blazar is several months \citep{KAREN14}.
Studies of individual sources during such major $\gamma$-ray events have revealed striking complexity in the behavior of blazars. Although
the majority of $\gamma$-ray outbursts are connected with the propagation of a disturbance in parsec-scale radio jets
seen on VLBA images as a region of enhanced brightness moving with an apparently superluminal speed \citep[e.g.,][]{MAR10,IVAN11a,ANN12,J13,DASHA14,TANAKA15,Carolina15a}, 
the physics behind the observed behavior of such jets has been extremely challenging to unravel. In the majority of these cases, 
$\gamma$-ray outbursts coincide with the passage of superluminal knots through the
millimeter-wave ``core,'' or even features downstream 
of the core \citep{JM16}. This locates the origin of high energy emission parsecs from the BH. The latter is difficult to 
reconcile with the short timescales of variability during  $\gamma$-ray flares \citep{TAV10,NK14}. In addition, according to the current picture of AGN, this location 
lacks the intense external photon field needed for the most popular mechanism 
for $\gamma$-ray production. 
However, an increase of emission line fluxes in blazars coinciding with the ejection of superluminal knots during major 
$\gamma$-ray events has been found in the quasar 3C454.3 \citep{LT13}, which challenges our current understanding of 
the location of the BLR. Emission-line flares in close temporal proximity to prominent $\gamma$-ray flares have been found in several other $\gamma$-ray blazars, which suggests that the jet might provide additional photo-ionization of the BLR \citep{Isler15,LT15}. Moreover, the detection of VHE $\gamma$-rays up to 600~GeV from several quasars \citep[e.g.,][]{Aleksic11} questions the origin of $\gamma$-rays within the BLR located $< 1$~pc from the BH, since such high energy photons would not then be able to escape without pair producing off the intense photon field.

In this paper we present detailed analyses of the parsec-scale jet behavior of a sample of $\gamma$-ray bright AGN based on 
observations with the VLBA contemporaneous with the first 4.5~yr of monitoring with the {\it Fermi} LAT. We determine kinematic 
properties of emission features seen in the VLBA images with a resolution of $\sim$0.1~mas, and derive their physical parameters. 
Throughout the paper we use cosmological parameters corresponding to a $\Lambda$CDM model of the Universe, 
with $H_\circ = 70$~km~s$^{-1}$Mpc$^{-1}$, $\Omega_{\rm m}=0.3$, and $\Omega_\Lambda=0.7$.

\section{Observations and Data Analysis \label{Obs}}

The VLBA-BU-BLAZAR monitoring program consists of approximately monthly observations with the VLBA at 43 GHz of a sample of AGN detected as $\gamma$-ray sources. In this paper we 
present results of observations from 2007 June to 2013 January. The sample consists of 21~FSRQs, 12~BLLacs, and 3 radio galaxies (RGs). It includes the blazars and radio 
galaxies detected at $\gamma$-ray energies by EGRET with average flux density at 43~GHz exceeding 0.5~Jy, declination north of $-30^\circ$, and optical magnitude in R band brighter than 18.5$^m$. The limit on the radio flux density allows us to 
perform an efficient set of observations of up to 33 sources within a time span of 24~hr, with an average total on-source time of 45~min per object 
(6 to 10 scans) over 7-8~hr of optimal visibility of the source at the VLBA antennas. Such observations during favorable weather conditions result in radio images with dynamic ranges of 400:1 or higher for uniform weighting of interferometric visibilities, which optimizes the angular resolution for a given coverage of the ({\it u,v}) spatial frequencies. The limit on optical magnitude is determined by the size of the mirror of the 1.83~m Perkins telescope (Lowell Obs., Flagstaff, AZ), which we use for optical monitoring, and by the desired accuracy of $\pm0.5\%$ for measurements of the degree of optical linear polarization with integration times $\le1$~hr. The declination limit of $\sim-30^\circ$ corresponds to the most southerly declination accessible by the VLBA and Perkins telescope.

A list of sources is given in Table~\ref{Sample}, along with main characteristics of each source: 
1 - name based on B1950 coordinates; 2 - commonly used name; 3 - subclass of object; 
4 - redshift according to the NASA Extragalactic Database; 5 - average flux density at 43 GHz 
and its standard deviation based on VLBA observations considered here; 6 - average $\gamma$-ray 
photon flux at 0.1-200 GeV and its standard deviation based on $\gamma$-ray light curves, which we 
calculate using photon and spacecraft data provided by the Fermi Space Science Center
(for details see \citealt{KAREN14}); and 7 - average magnitude in R band and its standard deviation 
according to our observations at the Perkins telescope. Parameters given in Table~\ref{Sample} 
correspond to the time period mentioned above. Six sources marked in Table~\ref{Sample} 
by symbols ``$\star$'',``$\star\star$'', and ``+''  were added to the program after their 
detection by the Fermi LAT during the first year of operation \citep{1FLG}. In addition to the 
regular monitoring, we also performed 1-2 campaigns per year of 15-16 sources from our sample 
(those whose placement on the sky allowed simultaneous optical observations) involving three observations 
with the VLBA at 43~GHz (16~hr per epoch) and nightly optical observations with the Perkins telescope over a 2-week period.

\begin{deluxetable*}{lccrrcc}
\singlespace
\tablecolumns{7}
\tablecaption{\small\bf The VLBA-BU-BLAZAR Sample \label{Sample}}
\tabletypesize{\footnotesize}
\tablehead{
\colhead{Source}&\colhead{Name}&\colhead{Type}&\colhead{z}&\colhead{$\langle S_{\rm 43}\rangle$}&\colhead{$\langle S_\gamma\rangle$}&\colhead{$\langle R\rangle$}\\
\colhead{}&\colhead{}&\colhead{}&\colhead{}&\colhead{Jy}&\colhead{10$^{-8}ph~cm^{-2}s^{-1}$}&\colhead{mag}\\
\colhead{(1)}&\colhead{(2)}&\colhead{(3)}&\colhead{(4)}&\colhead{(5)}&\colhead{(6)}&\colhead{(7)}
}
\startdata
0219+428$^*$&3C66A& BL &0.444?&0.43$\pm$0.12&12.5$\pm$7.0&14.01$\pm$0.49\\
0235+164&AO0235+16&BL&0.940?&1.77$\pm$1.38&18.8$\pm$27.4&17.57$\pm$1.34\\
0316+413$^{**}$&3C84&RG&0.0176&15.51$\pm$3.65&21.3$\pm$9.23&12.81$\pm$0.14\\
0336$-$019&CTA26&FSRQ&0.852&1.68$\pm$0.60&12.3$\pm$8.2&17.08$\pm$0.40\\
0415+379&3C111&RG&0.0485&3.26$\pm$2.13&6.35$\pm$3.22&17.31$\pm$0.21\\
0420$-$014&OA129&FSRQ&0.916&4.74$\pm$1.34&13.3$\pm$6.8&17.12$\pm$0.62\\
0430+05$^{+}$&3C120&RG&0.033 &1.67$\pm$0.60 &6.42$\pm$2.54&14.24$\pm$0.27\\
0528+134&PKS0528+134&FSRQ&2.060&2.02$\pm$1.15&11.0$\pm$8.3&19.27$\pm$0.25\\
0716+714&S4 0716+71&BL&0.3?&2.12$\pm$1.10&21.7$\pm$13.8&13.21$\pm$0.50\\
0735+178&PKS0735+17&BL&0.424&0.40$\pm$0.12&7.08$\pm$3.02&15.91$\pm$0.33\\
0827+243&OJ248&FSRQ&0.939&1.28$\pm$0.68&15.1$\pm$13.0&16.77$\pm$0.53\\
0829+046&OJ049&BL&0.182&0.57$\pm$0.22&6.41$\pm$3.78&15.61$\pm$0.31\\
0836+710&4C+71.07&FSRQ&2.17&1.73$\pm$0.69&15.8$\pm$14.2&16.61$\pm$0.08\\
0851+202&OJ287&BL&0.306&4.68$\pm$1.99&10.5$\pm$8.11&14.45$\pm$0.48\\
0954+658&S4 0954+65&BL&0.368?&1.05$\pm$0.34&6.86$\pm$4.71&16.24$\pm$0.58\\
1055+018$^*$&4C+01.28&BL&0.890&4.02$\pm$1.13&11.7$\pm$6.70&16.58$\pm$0.46\\
1101+384$^{**}$&Mkn421&BL&0.030&0.28$\pm$0.07&17.9$\pm$7.29&12.36$\pm$0.34\\
1127$-$145&PKS1127$-$14&FSRQ&1.184&1.76$\pm$0.95&10.3$\pm$6.01&16.63$\pm$0.06\\
1156+295&4C+29.45&FSRQ&0.729&1.40$\pm$0.68&14.3$\pm$10.4&16.02$\pm$1.05\\
1219+285&WCom&BL&0.102&0.25$\pm$0.08&6.03$\pm$2.62&14.94$\pm$0.44\\
1222+216&4C+21.35&FSRQ&0.435&1.19$\pm$0.67&20.2$\pm$33.7&15.48$\pm$0.29\\
1226+023&3C273&FSRQ&0.158&11.88$\pm$5.70&31.4$\pm$33.4&12.53$\pm$0.05\\
1253$-$055&3C279&FSRQ&0.538&18.05$\pm$8.25&34.3$\pm$27.1&15.62$\pm$0.81\\
1308+326&B2 1308+32&FSRQ&0.998&2.14$\pm$0.42&7.38$\pm$3.64&17.57$\pm$0.56\\
1406$-$076&PKS1406$-$07&FSRQ&1.494&0.59$\pm$0.14&8.50$\pm$4.02&18.35$\pm$0.21\\
1510$-$089&PKS1510-08&FSRQ&0.361&2.44$\pm$1.07&61.1$\pm$62.9&15.87$\pm$0.57\\
1611+343&DA406&FSRQ&1.40&1.53$\pm$0.63&2.02$\pm$1.92&17.20$\pm$0.12\\
1622$-$297&PKS1622$-$29&FSRQ&0.815& 1.35$\pm$0.62&9.98$\pm$5.61&18.15$\pm$0.28\\
1633+382&4C+38.41&FSRQ&1.814&2.93$\pm$0.76&24.4$\pm$18.7&16.83$\pm$0.58\\
1641+399&3C345&FSRQ&0.595&4.47$\pm$1.33&12.6$\pm$6.56&16.96$\pm$0.40\\
1730$-$130&NRAO530&FSRQ&0.902&3.31$\pm$0.85&18.2$\pm$12.5&17.52$\pm$0.56\\
1749+096$^*$&OT081&BL&0.322&3.60$\pm$1.30&8.39$\pm$5.17&16.42$\pm$0.62\\
2200+420&BLLAC&BL&0.069&4.21$\pm$1.62&24.0$\pm$16.3&14.23$\pm$0.55\\
2223$-$052&3C446&FSRQ&1.404&3.82$\pm$1.66&7.87$\pm$4.33&17.77$\pm$0.40\\
2230+114&CTA102&FSRQ&1.037&2.71$\pm$0.99&20.5$\pm$21.6&16.29$\pm$0.60\\
2251+158&3C454.3&FSRQ&0.859&14.44$\pm$9.85&77.9$\pm$158.0&15.13$\pm$0.90\\
\enddata
\vspace{2mm}
$^*$ - Source added to the sample in 2009; $^{**}$ - Source added to the sample in 2010; 
$^+$ - Source added to the sample in 2012; ? - Redshift has not been confirmed. 
\end{deluxetable*}
The observations were carried out in continuum mode, recording both left (L) and right (R) circular polarization signals, at a central frequency of 43.10349~GHz using four intermediate frequency bands (IFs), each of 16~MHz width over 8 channels. The data were recorded in Mark~VIA format with a 512~Mb per second rate, correlated at the National Radio Astronomy Observatory (NRAO; Soccoro, NM) using the VLBA DiFX software
correlator, and downloaded from the NRAO archive to the Boston University (BU) blazar group's computers within 2-4 weeks after each observation. To reduce the data, we used the {\it Astronomical Image Processing System} software (AIPS) \citep{AIPS96} provided by NRAO, with the latest version corresponding to the epoch of observation, and {\it Difmap} \citep{Difmap}. The data reduction was performed in a manner similar to that described in \cite{J05} (J05 hereafter) using utilities within AIPS. The reduction of each epoch's observations includes the following:
1) loading visibilities into AIPS (FITLD); 2) running VLBAMCAL to remove redundant calibration records; 3) analyzing quality of the data through different AIPS procedures (LISTR, UVPLT, SNPLT) and flagging unreliable data (UVFLG); 4) determining the reference 
antenna based on quality of the data and system temperature values (usually, VLBA\_KP or VLBA\_LA); 5) performing a priori amplitude corrections
(VLBACALA) including sky opacity correction; 6) correcting phases for parallactic angle effects (VLBAPANG); 7) removing the frequency dependence of the phase caused by a residual instrumental delay (FRING), using a scan on a bright source (usually 3C~279 or 3C~454.3), and applying  derived solutions to the data (CLCAL); 8) performing a global fringe fit on the full data set to determine the residual instrumental delays and rates based on the time dependence of the phase (FRING); 9) smoothing the delays and rates with a 2~hr timescale (SNSMO), and applying derived solutions to the data (CLCAL); 10) performing a bandpass calibration (BPASS) using a bright source as a calibrator (usually 3C~279 or 3C~454.3); 11) performing a cross-hand fringe fit on a scan
of a bright source and applying the resulting R-L phase and rate delay correction to the full data set (VLBACPOL); 12) splitting
the data into single-source files with averaging over channels and IFs and transferring into {\it Difmap}; 13) preliminary imaging using iterations of the CLEAN algorithm in {\it Difmap} plus self-calibration of the visibility phases and amplitudes based on the CLEAN model; 14) loading preliminary images into AIPS and performing both R and L phase calibration of single-source visibility data using the corresponding preliminary image (CALIB); 
15) transferring calibrated single-source visibility files into {\it Difmap} and performing final total-intensity imaging and self-calibration with uniform weighting. The single-source visibility files obtained after the final imaging are transferred into AIPS one more time to perform polarization calibration, which includes corrections for the instrumental polarization leakage (``D-terms'') and absolute value of the electric  vector position angle (EVPA). Since this paper deals with total intensity images, the polarization calibration mentioned above will be described in more detail in a future paper devoted to analysis of linear polarization in the parsec-scale jets of the sources in our sample. The fully calibrated visibility files and final images are posted at website {\bf http://www.bu.edu/blazars/VLBAproject.html} within 3-6 months after each observation. 
%In this paper we discuss jet kinematics of the 33 blazars and 3 radio galaxies based on 1929 total intensity images obtained from 2007 June to 2013 January.

\subsection{Modeling of Total Intensity Images}
We use a traditional approach to model the total intensity visibility files, (e.g.,   
\citealt{J01,Homan01}; J05; \citealt{LIST09,LIST13}), which is less sensitive to the morphology of a jet than recently developed algorithms \citep[e.g.,][]{COHEN14,LOB16}. The latter require well-developed extended emission regions beyond the ``core,'' and hence are more suitable for analysis of jets at longer wavelengths. As shown in many papers devoted to very long baseline interferometry (VLBI) of blazars \citep[e.g.,][]{KEN04,LIST13,LIST16}, the parsec-scale jet morphology of blazars, especially at high frequencies \citep{J01,J05}, is dominated by the VLBI ``core,'' which is a presumably stationary feature located at one end of the jet seen in VLBI images. The core is usually, but not always \citep[e.g.,][]{J13}, the brightest feature in the jet. We represent the total intensity structure of each source at each epoch by a model consisting of circular Gaussian components that best fits the visibility data as determined by iterations with the MODELFIT task in {\it Difmap}. We use the term ``knot'' to refer to a Gaussian component, which corresponds to a usually compact feature of enhanced brightness in the jet. The modeling starts with a Gaussian component approximating the brightness distribution of the core, and continues with the addition of knots at the approximate locations where a bright feature appears in the image of the jet. The addition of each knot is accompanied by hundreds of iterations to determine the parameters of the knot that give the best agreement between the model and the $uv$ data according to the $\chi^2$ test. The knot parameters are: $S$ - flux density; $R_{\rm obs}$ - distance with respect to the core, $\Theta_{\rm obs}$ - relative position angle (PA) with respect to the core (measured north through east), and $a$ - angular size, corresponding to the FWHM of the circular Gaussian component. The process is terminated when the addition of a new knot does not improve the $\chi^2$ value determined by MODELFIT. Since we perform roughly monthly monitoring of the sample, a model of the previous epoch is often used as the starting model for the next epoch for a given source. Table~\ref{Model} gives total intensity jet models as follows: 1 - source name according to its B1950 coordinates;
2 - epoch of the start of the observation in decimal years; 3 - epoch of the start of the observation in MJD; 4 - flux density, $S$, and its 1$\sigma$ uncertainty in Jy; 5 - distance with respect to the core, $R$, and its 1$\sigma$ uncertainty, in mas; 6 - PA with respect to the core, $\Theta$, and its 1$\sigma$ uncertainty in degrees; 7 - angular size, $a$, and its 1$\sigma$ uncertainty in mas; and 8 - observed brightness temperature, $T_{\rm b,obs}= 7.5\times10^8S/a^2$~K~(J05) (see \S~\ref{STB} for details
of $T_{\rm b,obs}$ calculations). Note that the core is located at position $(X,Y)$ = (0,0), where $X$ is relative right ascension
and $Y$ is relative declination).
\clearpage
\begin{deluxetable*}{lrrrrrrl}
\singlespace
\tablecolumns{8}
\tablecaption{\small\bf Modeling of Jets by Gaussian Components \label{Model}}
\tabletypesize{\footnotesize}
\tablehead{
\colhead{Source}&\colhead{Epoch}&\colhead{$MJD$}&\colhead{$S\pm\sigma_S$}&\colhead{$R\pm\sigma_R$}&\colhead{$\Theta\pm\sigma_\Theta$}&\colhead{$a\pm\sigma_a$}&\colhead{$T_{\rm b,obs}$}\\
\colhead{}&\colhead{}&\colhead{}&\colhead{Jy}&\colhead{mas}&\colhead{deg}&\colhead{mas}&\colhead{K} \\
\colhead{(1)}&\colhead{(2)}&\colhead{(3)}&\colhead{(4)}&\colhead{(5)}&\colhead{(6)}&\colhead{(7)}&\colhead{(8)}
}
\startdata
0219+428& 2008.8060&  54761&   0.238$\pm$  0.013&   0.000       &     0.0       &  0.003$\pm$ 0.008&   0.446E+12L \\
        &  2008.8060&  54761&   0.139$\pm$  0.011&   0.111$\pm$  0.015&  -156.9$\pm$    3.8&  0.082$\pm$ 0.018&   0.156E+11\\
        &  2008.8060&  54761&   0.052$\pm$  0.011&   0.304$\pm$  0.043&  -166.0$\pm$    4.1&  0.143$\pm$ 0.031&   0.191E+10\\
        &  2008.8060&  54761&   0.035$\pm$  0.011&   0.530$\pm$  0.053&  -167.8$\pm$    2.9&  0.142$\pm$ 0.034&   0.130E+10\\
        &  2008.8060&  54761&   0.021$\pm$  0.012&   0.930$\pm$  0.089&  -176.2$\pm$    2.7&  0.187$\pm$ 0.045&   0.451E+09\\
         & 2008.8060&  54761&   0.019$\pm$  0.015&   1.599$\pm$  0.348&  -173.1$\pm$    6.2&  0.532$\pm$ 0.077&   0.499E+08\\
         & 2008.8060&  54761&   0.040$\pm$  0.014&   2.324$\pm$  0.180&  -177.4$\pm$    2.2&  0.471$\pm$ 0.060&   0.135E+09\\
0219+428& 2008.8142&  54764&   0.269$\pm$  0.015&   0.000       &     0.0       &  0.031$\pm$ 0.010&   0.206E+12\\
         & 2008.8142&  54764&   0.090$\pm$  0.009&   0.141$\pm$  0.009&  -153.6$\pm$    1.9&  0.036$\pm$ 0.014&   0.533E+11\\
         & 2008.8142&  54764&   0.053$\pm$  0.011&   0.326$\pm$  0.047&  -167.7$\pm$    4.1&  0.158$\pm$ 0.033&   0.160E+10\\
         & 2008.8142&  54764&   0.025$\pm$  0.011&   0.560$\pm$  0.054&  -169.9$\pm$    2.8&  0.125$\pm$ 0.035&   0.120E+10\\
         & 2008.8142&  54764&   0.024$\pm$  0.013&   0.941$\pm$  0.111&  -175.2$\pm$    3.4&  0.237$\pm$ 0.049&   0.321E+09\\
         & 2008.8142&  54764&   0.015$\pm$  0.015&   1.494$\pm$  0.324&  -172.4$\pm$    6.2&  0.446$\pm$ 0.075&   0.569E+08\\
         & 2008.8142&  54764&   0.046$\pm$  0.014&   2.354$\pm$  0.160&  -176.4$\pm$    1.9&  0.452$\pm$ 0.057&   0.168E+09\\
\enddata
\vspace{3mm}
Values of $T_{\rm b,obs}$ denoted by letter ``L'' represent lower limits of the brightness temperature, for details see  \S~\ref{STB}.
The table is available entirely in a machine-readable format in the online journal (Jorstad et al. 2017, ApJ,
846, 98).
\end{deluxetable*}
 
All model parameters are subject to errors, which, as shown in  e.g., \cite{Homan02}, J05, \cite{LIST09}, depend
on the angular resolution of the observation, as well as the flux density and compactness of the knot. According to \cite{Homan02},
typical errors are $\sim$5\% for the total flux density and, for the position, $\sim$1/5 of the synthesized beam corresponding to the {\it uv} coverage. We use empirical relations between the brightness temperature of the knot and the total intensity and position errors derived in \cite{Carolina15b} that include the analysis of errors performed in \cite{J01,J05}. The uncertainties can be approximated as follows:  
$\sigma_{\rm X,Y}\approx 1.3\times10^4 T_{\rm b,obs}^{-0.6}$ and $\sigma_{\rm S}\approx 0.09T_{\rm b,obs}^{-0.1}$, where $\sigma_{\rm X,Y}$ is the 1$\sigma$ uncertainty in right ascension or declination in mas and  $\sigma_{\rm S}$ is the 1$\sigma$ uncertainty in the flux density in Jy. We add (in quadrature) a minimum positional error of 0.005~mas related to the resolution of the observations and a typical amplitude calibration error of 5\% to the values of the uncertainties calculated via these relations. We have used task MODELFIT in {\it Difmap} for a dozen knots with different brightness temperatures, as well as results obtained in J05, to derive a relation between the 1$\sigma$ uncertainty of the angular size, $\sigma_{\rm a}$, and $T_{\rm b,obs}$ of a knot. To do this, we fixed the parameters of the best-fit model for a given knot in MODELFIT, and changed the size of the knot until the value of $\chi^2$ increased by 1. As a result, we have obtained the following relation: $\sigma_{\rm a}\approx 6.5T_{\rm b,obs}^{-0.25}$ mas. Table~\ref{Model} gives brightness temperatures of all knots identified in the jets that were employed to calculate uncertainties of kinematic parameters.

\subsection{Identification of Components and Calculation of Apparent Speeds}
We define the VLBI core as the bright, compact (either unresolved or partially resolved) emission feature at the upstream end of the jet. We define the positions of other components relative to the centroid of the core as found in the model fitting procedure described above. We use all 4 model parameters for each knot, $S$, $R$, $\Theta$, and $a$, to aid in the identification of components across epochs by assuming that none of these parameters should change abruptly with time given the high temporal density of our monitoring observations. We search for continuity in motion (or no motion), relative stability in values of $\Theta$, and consistency in evolution of flux density \citep{Homan02} and size (J05) for each knot. If a knot is identified at $\ge$4 epochs, a kinematic classification (see \S{\ref{structure}}) and an identification number (ID) is assigned to it. 

We use the same formalism and software package DATAN \citep{DATAN} as described in J05 to calculate kinematic parameters of all knots with IDs. The procedure consists of the following steps:
\begin{enumerate}
\item The values of the $X$ and $Y$ coordinates of a knot detected at $N$ epochs ($N\ge$4) are fit by different polynomials of order $l$ as described in J05:
\begin{equation}
X(t_i)=a_o+a_1(t_i-t_{mid})+a_2(t_i-t_{mid})^2+ ... a_l(t_i-t_{mid})^l,\label{e1}
\end{equation}
\begin{equation}
Y(t_i)=b_o+b_1(t_i-t_{mid})+b_2(t_i-t_{mid})^2+ ... b_l(t_i-t_{mid})^l,\label{e2}
\end{equation}
where $t_i$ is the epoch of observation, $i$=1,...,N, and $t_{mid} \equiv (t_1+t_N)/2$, with $l$ varying from 0 to 4. We use the program LSQPOL of the DATAN package, which returns the best-fit polynomial of each order $l$ according to the least-square method, along with the value of $M_l$ characterizing the goodness of fit. The program takes into account the uncertainties of the data to estimate the contribution of each measurement. Values of $M_l$ are compared
with values of $M_{\chi^2}$, where $M_{\chi^2}$ is the $\chi^2$ value corresponding to a significance level of $\zeta$=0.05 for $f=N-l-1$ degrees of freedom \citep{BL72}. A polynomial of the lowest order for which $M_l<M_{\chi^2}$ is employed to fit the data (see Tables 3 \& 4 in J05). However, $l>1$ is used only if $N>$6, given the number of degrees of freedom. 
Polynomials with $l>4$ are not employed, since these do not generally improve the $\chi^2$ value.
In cases for which the $\chi^2$ criterion is not reached for $l=4$, we set $l=4$ and note this in the table of jet kinematics. There are only several such cases.
\item We use subroutine LSQASN in the DATAN package to calculate uncertainties of the parameters of 
the best-fit polynomial, which gives the uncertainty for each polynomial coefficient corresponding to a confidence level of $\zeta$=0.05. 
\item Based on the polynomials that best fit the $X$ and $Y$ data, we calculate
the proper motion $\mu$, direction of motion $\Phi$, and apparent speed $\beta_{\rm app}$ for each knot. In the case of $l>1$, values of the acceleration along, $\mu_\parallel$, and perpendicular to, $\mu_\perp$, the jet direction are computed as well. The uncertainties of the polynomial coefficients are propagated to obtain the uncertainties of the kinematic parameters.
\item We calculate the epoch of ejection, $T_\circ$, for each knot with a statistically significant proper motion. The epoch of ejection is defined as the time when the centroid of a knot coincides with the centroid of the core, with motion of the knot extrapolated back to $(X,Y)=(0,0)$. We search for roots, $t_{\rm xo}$ and $t_{\rm yo}$, of the polynomials that represent the motion of the knot along the $X$ and $Y$ axes, respectively. This is straightforward for $l\le$2. Polynomials with $l>$2 are usually employed when a knot is detected at $N\ge$15 epochs. While for the analysis of kinematics all measurements are important, for determining $T_\circ$ the most important are measurements closest to the core. For such knots we limit the data to 10 epochs when the knot was located closest to the core, and find the best-fit polynomials of order $l\le$2 that fit these $X$ and $Y$ measurements. When $t_{\rm xo}$ and $t_{\rm yo}$ are calculated, 
$T_\circ = [t_{\rm xo}/\sigma_{t_{\rm xo}}^2 + t_{\rm yo}/\sigma_{t_{\rm yo}}^2]/[1/\sigma_{t_{\rm xo}}^2 + 1/\sigma_{t_{\rm yo}}^2].$ The uncertainty of $T_\circ$ is computed as 
$\sigma T_\circ =\sqrt{[(T_\circ-t_{\rm xo})^2/\sigma_{t_{\rm xo}}^2 + (T_\circ-t_{\rm yo})^2/\sigma_{t_{\rm yo}}^2]/[1/\sigma_{t_{\rm xo}}^2 + 1/\sigma_{t_{\rm yo}}^2]}.$
In a few cases $t_{\rm xo}$ and $t_{\rm yo}$ are significantly different from each other, and weighting by uncertainties does not provide a robust solution. These correspond to a few knots whose ejections occurred before the start of our monitoring. In these cases we construct the best-fit polynomial to the $R$ coordinate and find $T_\circ$ as its root, and mark such values of $T_\circ$ in the table of jet kinematics.
\end{enumerate}

\section {Structure of the Parsec-Scale Jets} \label{structure}

Determination of the kinematic parameters of the jet components allows us to classify them into different types according to their properties. Images of all sources at all epochs contain a core, as described above, which we designate as $A0$. Knots detected at $\ge$10 epochs and with proper motion $\mu < 2\sigma_\mu$ are classified as quasi-stationary (which we often shorten to ``stationary'') features, type $St$. These are labeled by the letter $A$ plus an integer starting with 1 for the stationary feature closest to the core and increasing sequentially with distance from the core. Knots with proper motion $\mu\ge 2\sigma_\mu$ are classified as moving features, type $M$. Among moving knots we separate features with $T_\circ$ falling within
the VLBA monitoring period considered here. They are designated by the letter $B$ followed by a number, with
higher numbers corresponding to later epochs of ejection. The remaining moving knots are labeled by letters
$C$ or $D$, with a number according to distance from the core, where $D$ knots are the most diffuse 
features in our 43 GHz images, with uncertain epochs of ejection. There are also moving knots 
that we classify as trailing components \citep{IVAN01}, type $T$. These appear at some distance from the core, behind faster moving features. Such knots are not detected closer to the core at earlier epochs given the resolution of our images, and hence appear to have split off from the related faster moving components. A trailing knot has the same designation as the faster component with which it is associated, except that the latter is lower case. Finally, there are knots in a few jets that are detected upstream of the core at $\ge$6 epochs. We classify these as counter-jet features, type $CJ$, and designate them as either stationary or moving knots, depending on their kinematic properties. We have two exceptions to
the scheme described above: For the radio galaxy 3C~111 and the quasar 3C~279, we continue the designation of moving components based on the long history of observations of their jet kinematics at 43~GHz (see Appendix~\ref{Notes}).

For all classified features in the sample, we calculate average parameters, which we list in Table~\ref{Parm}: 
1 - name of the source according to its B1950 coordinates; 2 - designation of the knot; 3 - number of 
epochs at which the knot is detected; 4 - mean flux density, $\langle S\rangle$, of the knot over the epochs when it
was detected, and its standard deviation, in Jy; 5 - average distance, $\langle R\rangle$, over the epochs when the knot was detected \footnote{$\langle R\rangle$ defined in this manner represents the typical distance at which a knot was observed, although it is not necessarily the mid-point of its trajectory} and its standard deviation, in mas; 6 - 
average position angle, $\langle\Theta\rangle$, over the epochs and its standard deviation, in degrees; for the core, $\langle\Theta\rangle$ represents the average projected direction of the inner jet, which is the weighted average of $\Theta$ of all knots in the jet over all epochs when $R\le$0.7~mas; 7 - mean size of the knot, $\langle a\rangle$, over the epochs and its standard deviation; and 8 - type of knot (see above). 

\begin{deluxetable*}{lrrrrrrr}
\singlespace
\tablecolumns{8}
\tablecaption{\small\bf Jet Structure \label{Parm}}
\tabletypesize{\footnotesize}
\tablehead{
\colhead{Source}&\colhead{Knot}&\colhead{$N$}&\colhead{$<S>$}&\colhead{$<R>$}&\colhead{$<\Theta>$}&\colhead{$<a>$}&\colhead{$Type$}\\
\colhead{}&\colhead{}&\colhead{}&\colhead{Jy}&\colhead{mas}&\colhead{deg}&\colhead{mas}&\colhead{}\\
\colhead{(1)}&\colhead{(2)}&\colhead{(3)}&\colhead{(4)}&\colhead{(5)}&\colhead{(6)}&\colhead{(7)}&\colhead{(8)}
 }
\startdata
0219+428&$A0$&34&0.26$\pm$0.08&0.0&$-$171.0$\pm$7.6&0.04$\pm$0.03&Core \\
&$A1$&32&0.08$\pm$0.04&0.17$\pm$0.04&$-$160.1$\pm$10.3&0.08$\pm$0.04&St \\
&$A2$&15&0.04$\pm$0.01&0.39$\pm$0.04&$-$161.8$\pm$4.1&0.15$\pm$0.04&St \\
&$A3$&27&0.03$\pm$0.01&0.61$\pm$0.07&$-$169.7$\pm$4.6&0.19$\pm$0.09&St \\
&$B1$&6&0.02$\pm$0.01&1.03$\pm$0.21&$-$175.7$\pm$1.4&0.27$\pm$0.07&M \\
&$B2$&10&0.02$\pm$0.01&1.06$\pm$0.24&$-$172.3$\pm$6.9&0.38$\pm$0.08&M \\
&$A4$&28&0.03$\pm$0.01&2.34$\pm$0.07&$-$177.4$\pm$1.1&0.44$\pm$0.16&St \\
0235+164&$A0$&70&1.38$\pm$0.94&0.0&\nodata&0.05$\pm$0.02&Core \\
&$B1$&6&0.16$\pm$0.25&0.40$\pm$0.19&$-$17.7$\pm$6.4&0.21$\pm$0.09&M \\
&$B2$&15&1.03$\pm$0.92&0.22$\pm$0.11&163.1$\pm$12.6&0.15$\pm$0.05&M \\
&$B3$&32&0.30$\pm$0.11&0.18$\pm$0.04&165.8$\pm$4.7&0.13$\pm$0.05&M \\
0316$+$413&$A0$&24&2.67$\pm$0.92&0.0&170.3$\pm$11.0&0.12$\pm$0.03&Core \\
&$C1$&19&1.21$\pm$0.71&2.31$\pm$0.17&175.5$\pm$3.2&0.21$\pm$0.10&M \\
&$C2$&24&4.11$\pm$2.49&2.07$\pm$0.13&175.6$\pm$3.0&0.38$\pm$0.11&M \\
&$C3$&20&0.89$\pm$0.66&1.67$\pm$0.08&167.7$\pm$5.5&0.14$\pm$0.08&M \\
&$C4$&24&1.77$\pm$0.53&1.66$\pm$0.07&$-$160.0$\pm$4.1&0.72$\pm$0.12&M \\
&$C5$&21&0.96$\pm$0.69&1.55$\pm$0.11&172.7$\pm$1.3&0.17$\pm$0.10&M \\
&$C6$&21&0.54$\pm$0.36&0.99$\pm$0.22&159.5$\pm$4.5&0.23$\pm$0.14&M \\
&$C7$&24&0.94$\pm$0.42&0.43$\pm$0.13&159.3$\pm$9.5&0.20$\pm$0.11&M \\
&$C8$&11&1.00$\pm$0.54&0.16$\pm$0.03&149.7$\pm$17.7&0.12$\pm$0.08&M \\
0336$-$019&$A0$&40&0.99$\pm$0.40&0.0&80.1$\pm$3.6&0.06$\pm$0.03&Core \\
&$A1$&24&0.25$\pm$0.14&0.14$\pm$0.03&75.2$\pm$16.5&0.09$\pm$0.06&St \\
&$B1$&11&0.08$\pm$0.03&0.75$\pm$0.29&79.8$\pm$2.4&0.40$\pm$0.13&M \\
&$B2$&12&0.37$\pm$0.27&0.32$\pm$0.26&83.6$\pm$5.7&0.16$\pm$0.08&M \\
&$B3$&15&0.84$\pm$0.32&0.26$\pm$0.15&78.2$\pm$8.7&0.14$\pm$0.06&M \\
&$A2$&7&0.10$\pm$0.05&1.91$\pm$0.12&68.7$\pm$3.7&0.77$\pm$0.42&St \\
\enddata
\vspace{3mm}
$^*$ - the direction of the inner jet for 0235+164 is uncertain, since features detected in the jet
have a range of PA exceeding 180$^\circ$. The table is available entirely in a machine-readable format in the online journal (Jorstad et al. 2017, ApJ, 846, 98).
\end{deluxetable*} 
\subsection{Total Intensity Images}  
Figure~\ref{maps} presents a single image of each source in our sample at an epoch when the most prominent
features of the jet can be seen. The features are marked on the images according to the model
parameters obtained for that epoch. 
Table~\ref{MapParm} gives the parameters of the images as follows: 1 - name 
of the source, 2 - epoch of observation, 3 - total intensity peak of the map, $I_{\rm peak}$, 4 -
lowest contour shown, $I_{\rm low}^{\rm cnt}$, 5 - size of the restoring beam, and 6 - global total intensity peak 
of the sequence of images presented in a set of figures (Fig. SET 2, ), $I_{\rm peak}^{\rm seq}$.
\begin{figure*}
\plotone{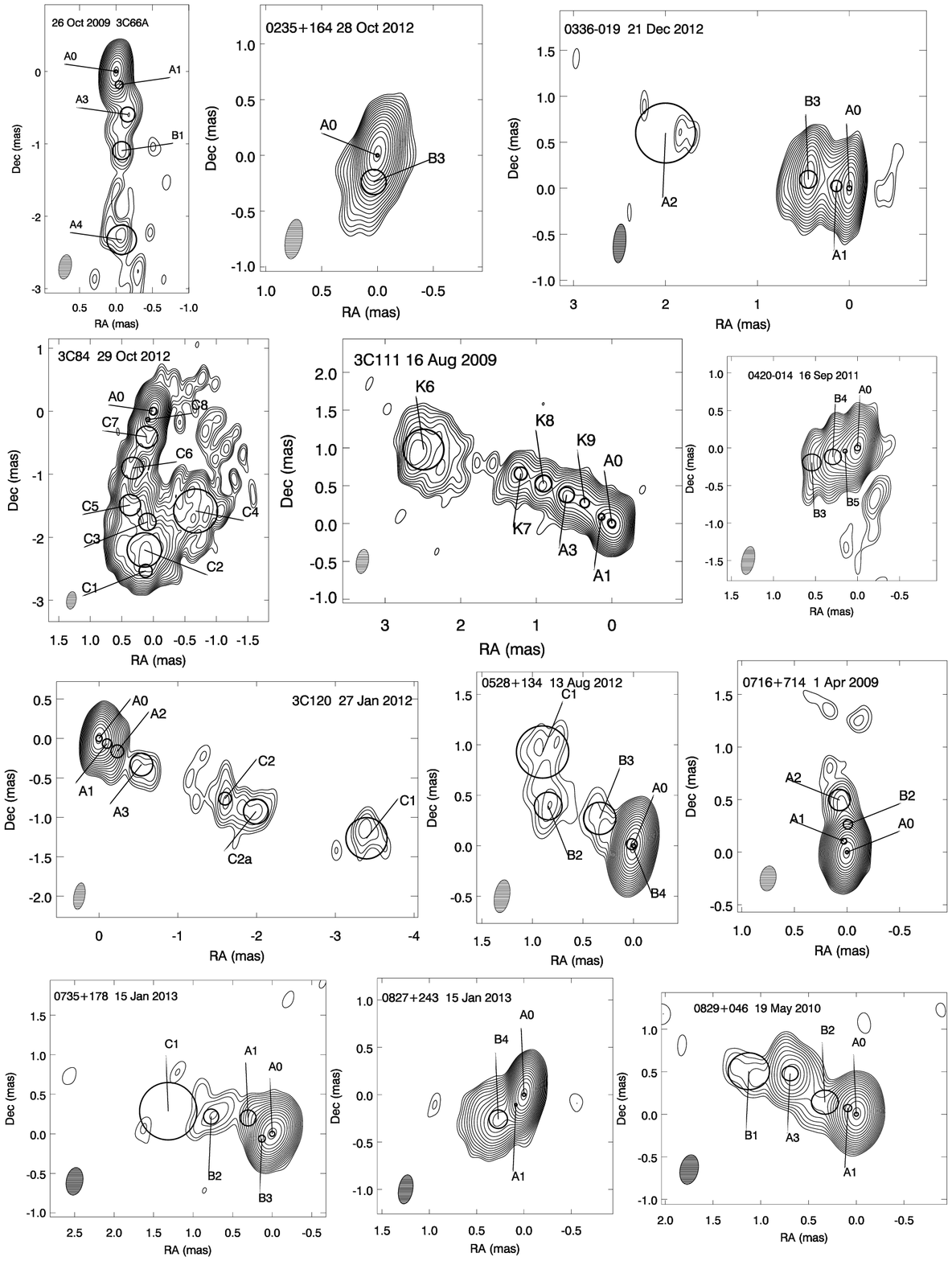}
\caption{Total intensity images at 43~GHz of the VLBA-BU-BLAZAR sample with an uniform weighting. Parameters of the images are given in Table~\ref{MapParm}. Contours decrease by a factor of $\sqrt{2}$
with respect to the peak flux density given in Table~\ref{MapParm}. Black circles on the images indicate positions and FWHM sizes of components according to the model fits.}
\label{maps}
\end{figure*}
\begin{figure*}
\figurenum{1}
\plotone{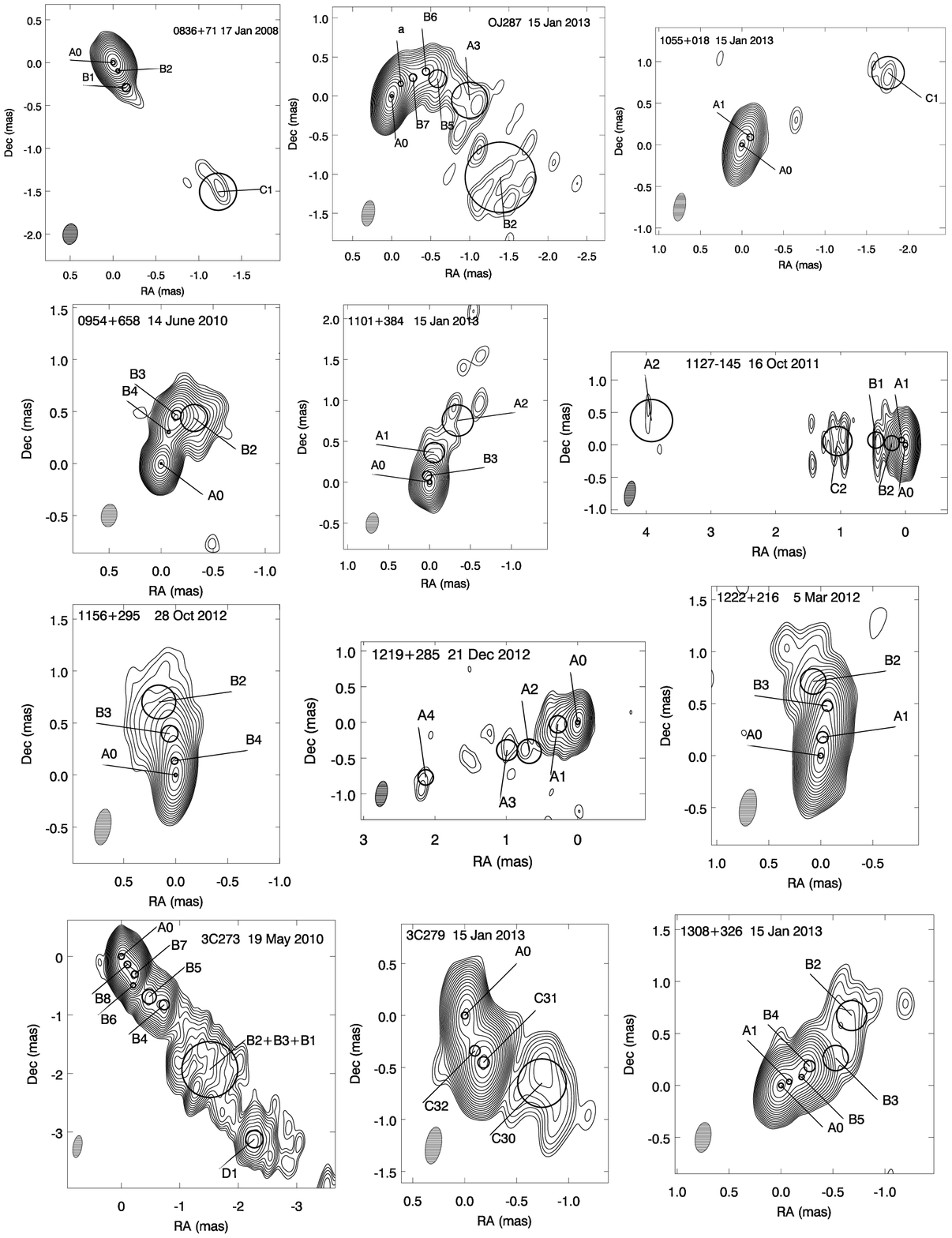}
\caption{Continued}
\end{figure*}
\begin{figure*}
\figurenum{1}
\plotone{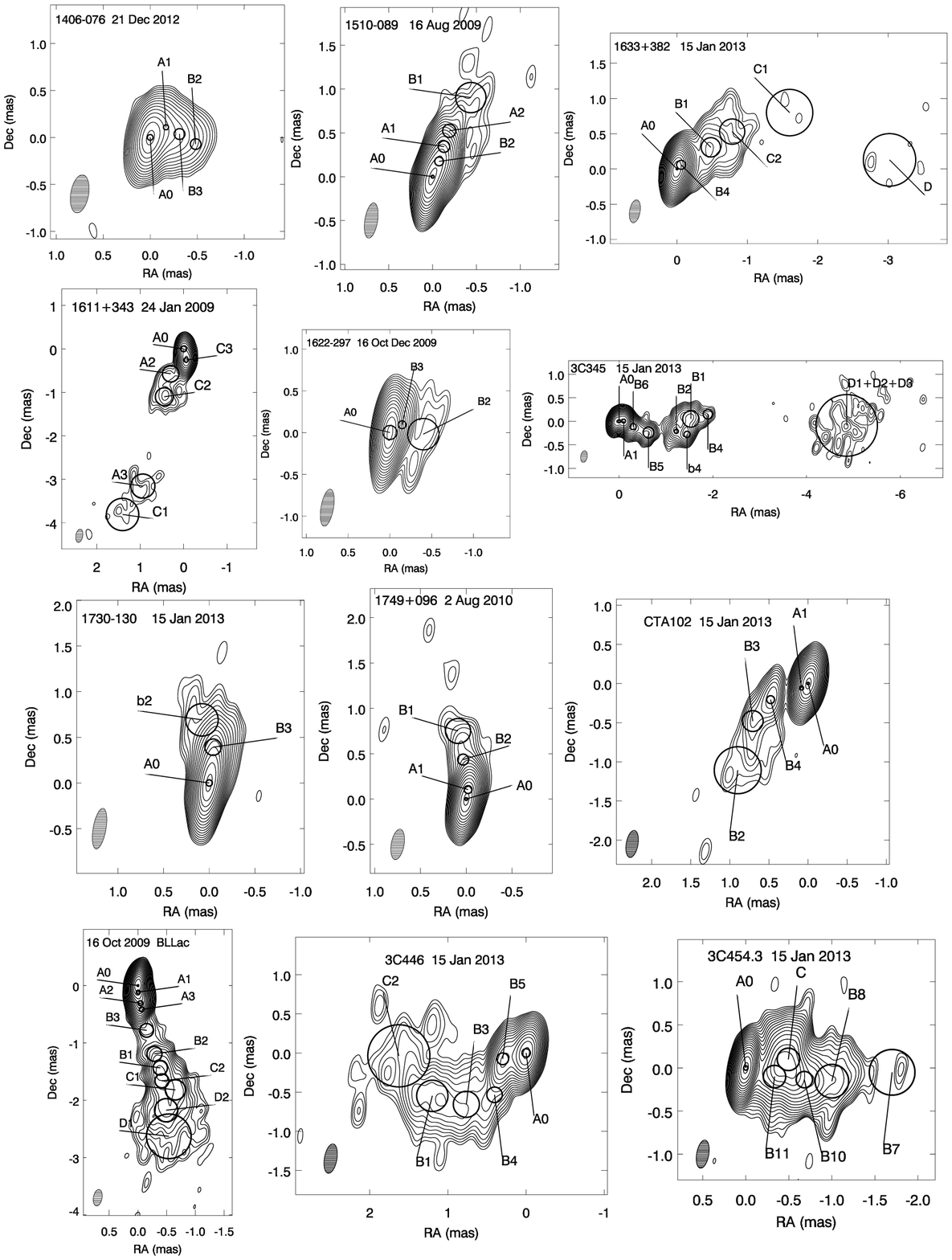}
\caption{Continued}
\end{figure*}

\begin{figure*}
\plotone{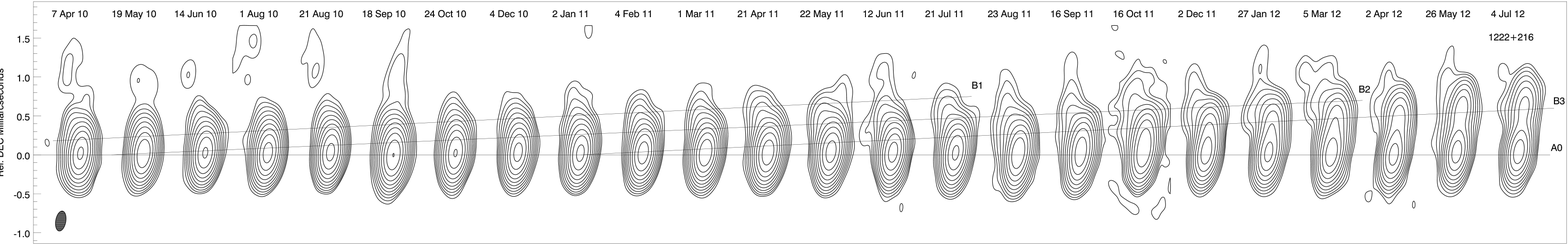}
\caption{Sample sequence of VLBA images at 43~GHz of the quasar 1222+216 convolved with the same beam as in
Figure~\ref{maps} and plotted using the global maximum over all of the epochs presented with contours decreasing by a factor of 2; straight lines across the images show positions of the core $A0$ and moving knots. Figures of sequences of images for all sources in the VLBA-BU-BLAZAR sample 
are available in the  online journal (Jorstad et al. 2017, ApJ, 846, 98). The parameters of the sequences are given in Table~\ref{MapParm}.}
\label{SETvlba}
\end{figure*}

\clearpage
\begin{deluxetable*}{lcccrc}
\singlespace
\tablecolumns{6}
\tablecaption{\small\bf Parameters of the VLBA Images in Figure~\ref{maps} \label{MapParm}}
\tabletypesize{\footnotesize}
\tablehead{
    \colhead{Source}&\colhead{Epoch}&\colhead{$I_{\rm peak}$}&\colhead{$I_{\rm low}^{\rm cnt}$}&\colhead{Beam Size}&\colhead{$I_{\rm peak}^{\rm seq}$} \\
\colhead{}&\colhead{}&\colhead{mJy~beam$^{-1}$}&\colhead{mJy~beam$^{-1}$}&\colhead{mas$\times$mas, deg}&\colhead{mJy~beam$^{-1}$}\\
\colhead{(1)}&\colhead{(2)}&\colhead{(3)}&\colhead{(4)}&\colhead{(5)}&\colhead{(6)}
}
\startdata
3C66A&2009/10/26&477&4.1&0.17$\times$0.34, $-$10&477\\
0235+164&2012/10/28&1151&4.1&0.15$\times$0.36, $-$10&5416\\
3C84&2012/10/29&3982&14.1&0.15$\times$0.28, $-$10&3072\\
0336$-$019&2012/12/21&1116&5.6&0.16$\times$0.38, $-$10&1817\\
3C111&2012/08/16&495&2.5&0.16$\times$0.32, $-$10&1791\\
0420$-$014&2011/09/16&2123&5.3&0.15$\times$0.38, $-$10&4745\\
3C120&2012/01/27&670&3.35&0.14$\times$0.34, $-$10&1795\\
0528+134&2012/08/13&1662&4.15&0.15$\times$0.33, $-$10&1185\\
0716+714&2009/04/01&11591&2.80&0.15$\times$0.24, $-$10&3392\\
0735+178&2013/01/15&456&2.28&0.21$\times$0.32, $-$10&542\\
0827+243&2013/01/15&1093&1.36&0.15$\times$0.31, $-$10&2899\\
0829+046&2010/05/19&260&1.30&0.19$\times$0.35, $-$10&853\\
0836+710&2008/01/17&753&3.77&0.16$\times$0.33, $-$10&1939\\
OJ287&2013/01/15&2792&4.91&0.17$\times$0.24, $-$10&3323\\
0954+658&2010/06/14&405&1.43&0.15$\times$0.24, $-$10&1078\\
1055+018&2013/01/15&1588&7.94&0.14$\times$0.34, $-$10&3793\\
1101+384&2013/01/15&333&1.67&0.15$\times$0.24, $-$10&308\\
1127$-$145&2011/10/16&1047&5.24&0.16$\times$0.39, $-$10&3173\\
1156+295&2012/10/28&549&3.88&0.15$\times$0.30, $-$10&2121\\
1219+285&2012/12/21&276&2.76&0.15$\times$0.24, $-$10&273\\
1222+216&2012/03/05&975&1.71&0.16$\times$0.33, $-$10&2341\\
3C273&2010/05/19&4290&7.55&0.15$\times$0.38, $-$10&9998\\
3C279&2013/01/15&9968&17.5&0.15$\times$0.36, $-$10&20193\\
1308+326&2013/01/15&550&1.38&0.15$\times$0.29, $-$10&2235\\
1406$-$076&2012/12/21&448&3.17&0.19$\times$0.41, $-$10&650\\
1510$-$089&2009/08/16&1186&2.96&0.19$\times$0.41, $-$10&5482\\
1611+343&2009/01/24&454&3.21&0.16$\times$0.31, $-$10&2195\\
1622$-$297&2009/10/16&1113&3.94&0.16$\times$0.58, $-$10&2248\\
1633+382&2013/01/15&2779&2.78&0.15$\times$0.33, $-$10&3572\\
3C345&2013/01/15&1088&2.72&0.14$\times$0.25, $-$10&4031\\
1730$-$130&2013/01/15&1899&6.72&0.14$\times$0.42, $-$10&4418\\
1749+096&2010/08/02&1528&3.82&0.14$\times$0.34, $-$10&5573\\
BLLac&2009/10/16&2389&2.39&0.15$\times$0.29, $-$10&5924\\
3C446&2013/01/15&629&2.23&0.16$\times$0.38, $-$10&3740\\
CTA102&2013/01/15&2709&6.77&0.15$\times$0.35, $-$10&3741\\
3C454.3&2013/01/15&1091&1.36&0.14$\times$0.33, $-$10&21835\\
\enddata
\end{deluxetable*}
We have constructed a sequence of images for each source that illustrates the evolution of the brightest knots during the period of monitoring presented here. An example of a sequence for the quasar 1222+216 is given in Figure~\ref{SETvlba}. The images in the sequences are convolved with the same beam as for a single epoch, as presented in Figure~\ref{maps}, and contours are plotted relative to the global maximum over epochs, given in Table~\ref{MapParm}.

\subsection{Observed Brightness Temperatures of Jet Features}\label{STB}
\begin{figure*}
\plottwo{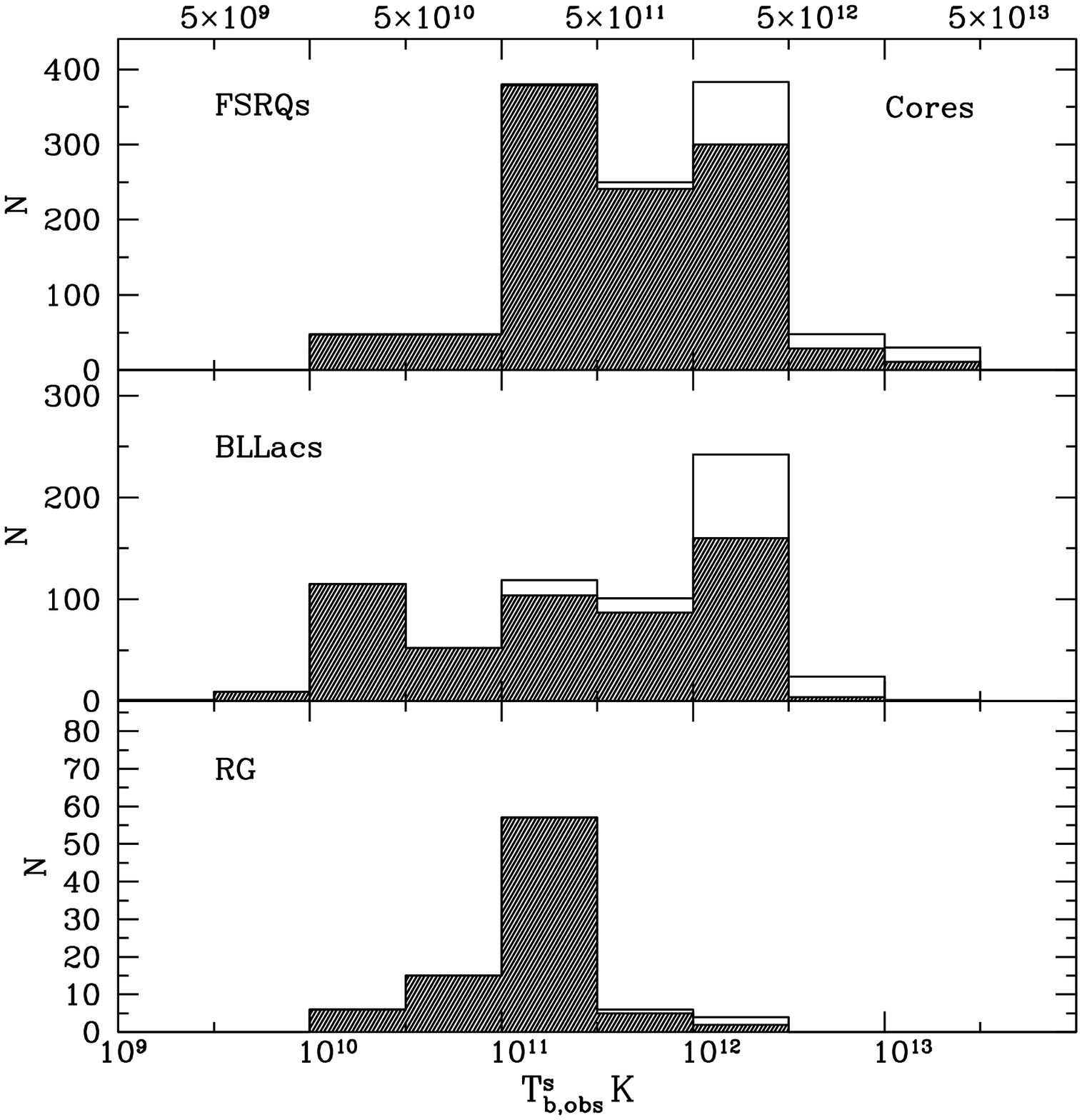}{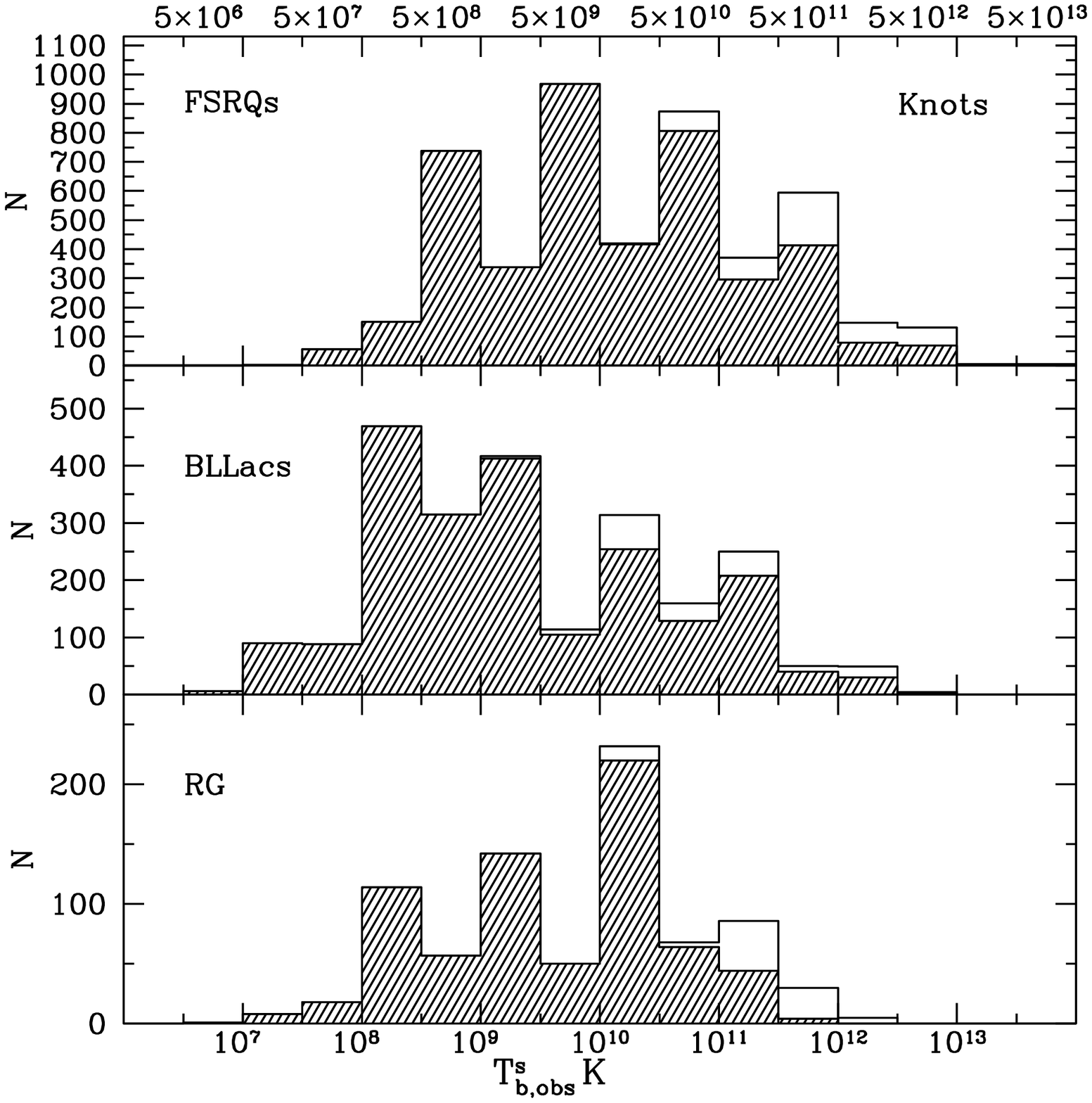}
\caption{{\it Left:} Distributions of brightness temperatures in the host galaxy frame of the cores at all epochs in the FSRQs ({\it top}), 
BLLacs ({\it middle}), and RGs ({\it bottom}). {\it Right:} Distributions of brightness temperatures of knots different from the core 
for the different sub-classes. Unshaded parts of the distributions represent lower limits to $T_{\rm b,obs}^{\rm s}$. The scale at the x axis is neither linear nor logarithmic.}
\label{figTB3}
\end{figure*}

We have calculated observed brightness temperatures, $T_{\rm b,obs}$, for all knots detected in the jets of our sample. 
Values of  $T_{\rm b,obs}$ are listed in Table~\ref{Model}. Note that 10.1\% of $T_{\rm b,obs}$ values are lower limits or uncertain because for these knots the model fit yields a size less than 0.02~mas, which is too small to be significantly resolved on the longest baselines. Lower limits to $T_{\rm b,obs}$ for such knots are calculated using $a$=0.02~mas (corresponding to 1/5 of our best resolution) and marked by the letter $L$ in Table \ref{Model}.
Figure~\ref{figTB3}, {\it left} shows the brightness temperatures of the cores, $T_{\rm b,obs}^{\rm s}=T_{\rm b,obs}(1+z)$ in the host galaxy frame, at all epochs in the different sub-classes. The values of $T_{\rm b,obs}^{\rm s}$ of the cores range from 
5$\times$10$^{9}$~K to 5$\times$10$^{13}$~K. The Kolmogorov-Smirnov (K-S) test gives a probability of 78.6\% that the distributions of $T_{\rm b,obs}^{\rm s}$ of the FSRQ and BLLacs cores are drawn from the same population. The K-S statistic, $KS$, equals 0.250, which specifies the maximum deviation between the cumulative distributions of the data. The distributions of both the FSRQs and BLLacs have a high probability (93.4\% and 81.4\%, respectively) of being different from the distribution of $T_{\rm b,obs}^{\rm s}$ of the RG cores. Note that the K-S test does not make assumptions about the underlying 
distribution of the data. The distribution of $T_{\rm b,obs}^{\rm s}$ in the RG cores has a well defined peak at 10$^{11}$--5$\times$10$^{11}$~K. The distribution of $T_{\rm b,obs}^{\rm s}$ of the FSRQ cores has a bimodal shape, while the distribution of $T_{\rm b,obs}^{\rm s}$ of the BLLac cores peaks at 10$^{12}$--5$\times$10$^{12}$~K, with similar percentages of brightness temperatures exceeding 10$^{12}$~K in the FSRQs and BLLacs (38.8\% and 40.2\%, respectively) and much smaller number of 4.5\% in the RGs. The distributions of $T_{\rm b,obs}^{\rm s}$ in the FSRQ and BLLac cores are similar to the distributions obtained by \cite{YY05} for the highest measured brightness temperatures of cores in their full sample, where the distribution of quasars has two peaks, with the highest brightness temperature peak around 5$\times$10$^{12}$~K being less prominent, while the distribution of $T_{\rm b,obs}^{\rm s}$ in BLLacs maximizes at this temperature.  
 
Figure~\ref{figTB3}, {\it right} shows the distributions of brightness temperatures in the host galaxy frame of jet knots different
from the cores. Values of $T_{\rm b,obs}^{\rm s}$ are lower than those of the cores, with peaks at 5$\times$10$^{9}$--10$^{10}$,
10$^{8}$--5$\times$10$^{8}$, and 10$^{10}$--5$\times$10$^{10}$ for the FSRQ, BLLac, and RG knots, respectively. The K-S test gives a low probability of 34.8\% that the distributions of $T_{\rm b,obs}^{\rm s}$ of jet knots in the FSRQs and BLLacs are similar. The distribution for FSRQ jet knots is shifted to higher values of $T_{\rm b,obs}^{\rm s}$ with respect to that of the BLLacs, with 26\% and 4.5\% of $T_{\rm b,obs}^{\rm s}>$5$\times$10$^{11}$~K in FSRQ and BLLac knots, respectively. According to the K-S test, the distribution of $T_{\rm b,obs}^{\rm s}$ of knots in the RGs is different from those of the FSRQs and BLLacs, with
a significantly higher value of $T_{\rm b,obs}^{\rm s}$ of the peak with respect to the peak of the distribution for BLLacs. 
Therefore, the BLLacs possess cores as intense as those in the FSRQs and significantly more intense cores than cores in the RGs, but less intense knots in the extended jet than features in the FSRQs and RGs. In addition, 7.5\% of all knots have $T_{\rm b,obs}<$10$^{8}$~K. The latter are waek diffuse features, located beyond 1~mas and with size $\geq0.5$~mas.
 
The brightness temperature of unbeamed, incoherent synchrotron emission produced by relativistic electrons in energy equipartition with the magnetic field is $T_{\rm b,eq}\sim 5\times$10$^{10}$~K \citep{RH94}. This is close to values of $T_{\rm b,obs}$ measured in the jet knots of the RGs at the majority of epochs. Values of $T_{\rm b,obs}\sim5\times$10$^{11}$-5$\times$10$^{12}$~K in the FSRQs and BLLacs can be explained as equipartition brightness temperatures amplified by relativistic boosting with Doppler factor $\delta$, $T_{\rm b,obs}\propto\delta T_{\rm b,int}$, since $\delta$ can be as high as $\sim$50 in blazars \citep[e.g., ][]{Hovatta09}. However, $T_{\rm b,obs}>$10$^{13}$~K is difficult to reconcile with incoherent synchrotron emission under equipartition conditions. It is possible that equipartition is violated during high activity states in a blazar \citep{Homan09L}. In fact, the values of $T_{\rm b,obs}>$10$^{13}$~K in our sample are observed in the cores or features nearest to the core in the blazars 0716+714, OJ287, 3C~273, 3C~279, 1749+096, BL~Lac, and 3C~454.3 during major outbursts; short discussions of these events can be found in Appendix \ref{Notes}. A very compact structure as small as 26~$\mu$as with $T_{\rm b,obs}>$10$^{13}$~K was recently detected in the quasar 3C~273 with a VLBI array including the space-based {\it RadioAstron} antenna \citep{YY16}. However later observations of the quasar with a similar high angular resolution when 3C~273 was in a low-activity state found the value of $T_{\rm b,obs}$ to be an order of magnitude lower than the equipartition value \citep{Bruni17}. 
These findings support the presence of very bright and compact features in blazar jets and possible dominance of the energy density of radiating particles over the magnetic field in the VLBI core at some epochs, which should be connected with 
active states of the source.       

\section{Velocities in the Inner Parsec-Scale Jet \label{Move}} 
We have identified 290 distinct emission features in the parsec-scale jets of the sources in our sample and measured apparent 
speeds of 252 these components. This excludes the cores, for which we assume no proper motion, and two knots that appeared on 
images of 0528+134 ($B4$) and 0735+178 ($B3$) during the last three epochs considered here, without measurable motion.  Out of these 
252 features, 54 components satisfy the criteria for a quasi-stationary feature listed above, with three sources having such features upstream 
of the core. All objects in our sample exhibit superluminal apparent speeds in the parsec-scale jet, 
except the radio galaxy 3C~84 and BL~Lac object Mkn421. Table~\ref{beta} presents the results of calculations of the jet kinematics as follows: 
1 - name of the source; 2 - designation of the component; 3 - order of polynomial used to fit the motion, $l$; 4 - proper motion, $\mu$, 
in mas~yr$^{-1}$; 5 - direction of motion, $\Phi$, in degrees; 6 - acceleration along the jet direction, $\mu_\parallel$, in  mas~yr$^{-2}$; 
7 - acceleration perpendicular to the jet direction, $\mu_\perp$, in  mas~yr$^{-2}$;
8 - apparent speed in units of the speed of light, $\beta_{\rm app}$; and 9 - epoch of ejection, $T_\circ$. 

\begin{figure*}
\plottwo{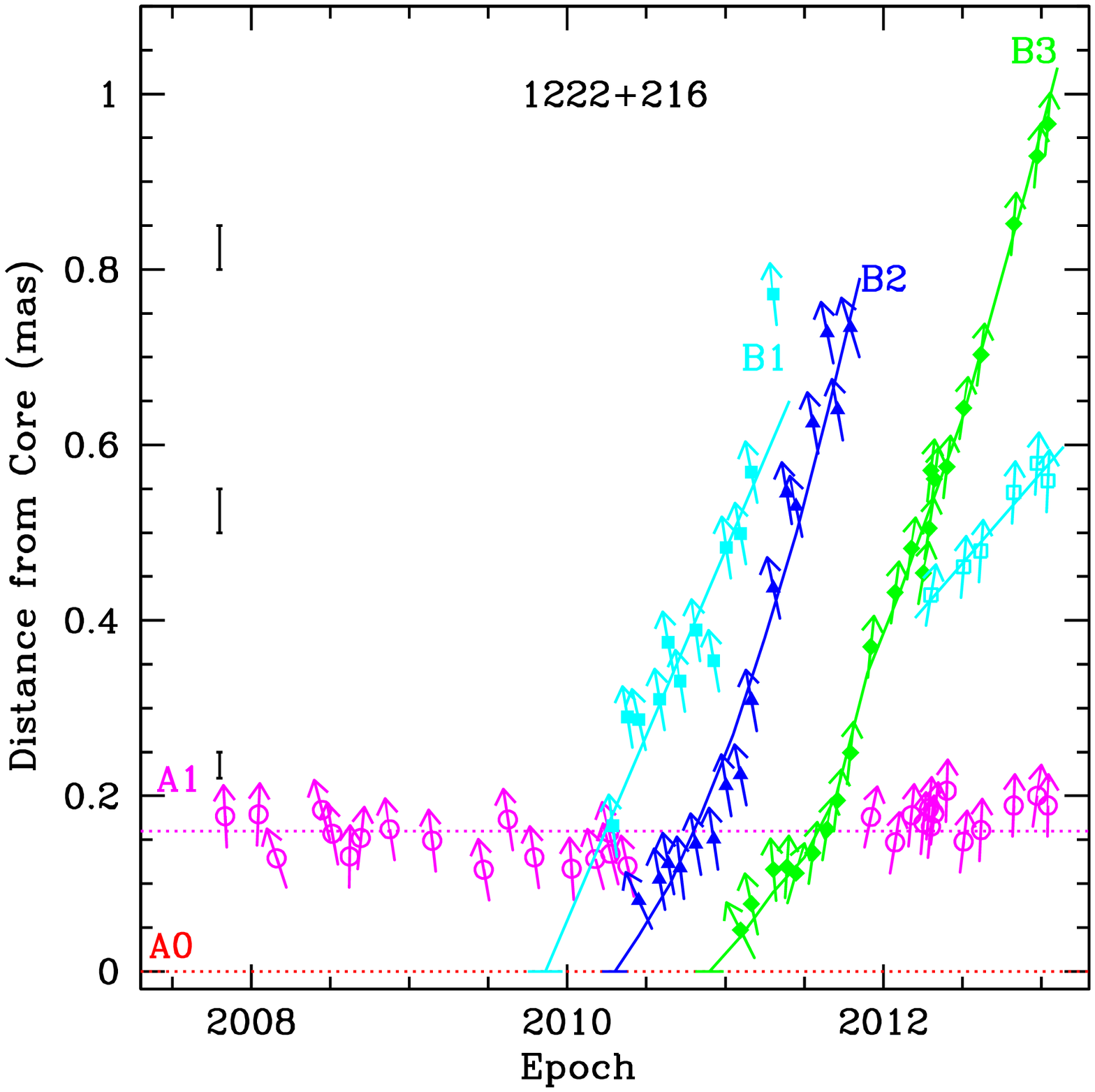}{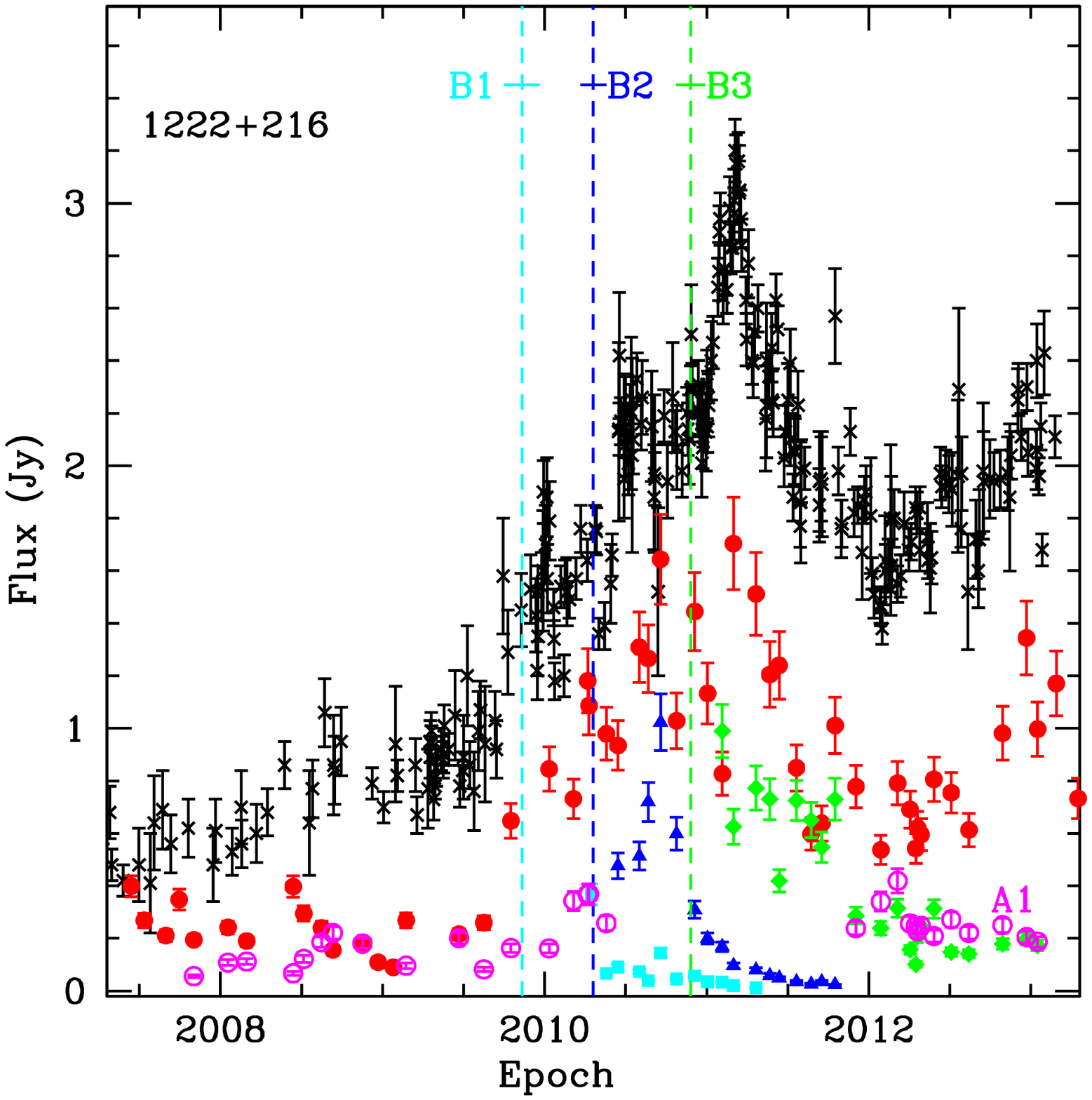}
\caption{{\it Left:} Separation vs.\ time of knots in the jet of the quasar 1222+216 from the VLBA-BU-BLAZAR sample. 
The vectors show the PA of each knot with respect to the core at the corresponding epoch. The solid lines or curves (depending on parameter $l$
in Table~\ref{beta}) represent polynomial fits to the motion, while dotted red and magenta lines mark the position of the core, 
$A0$ and stationary feature, $A1$, respectively. The vertical 
black line segments show approximate 1$\sigma$ positional uncertainties based on $T_{\rm b,obs}$. {\it Right:} The light curves 
of the core, $A0$ (filled circles, red), and jet components, and the light curve of the entire source at 37~GHz (black crosses); 
dashed lines indicate epochs of ejection of moving knots. Symbols and colors correspond to the same knot in the left and right plots. Plots for all sources in the sample are available in the online journal (Jotstad et al. 2017, ApJ, 846, 98). 
}
\label{SETevol}
\end{figure*}

For each source we construct a plot displaying the separation of moving knots from the core, usually within 1~mas of the core, 
and position of stationary features other than the core, if detected (Fig.\ SET 4). Figure~\ref{SETevol}, {\it left} presents an example of such plots 
for the quasar 1222+216. In addition, we have compiled light curves of the core and brightest knots in the jet for each 
source, an example of which is shown in Figure~\ref{SETevol}, {\it right} for the quasar 1222+216 as well. The  
light curves also include flux density measurements at the 14~m radio telescope of the Mets\"{a}hovi Radio Observatory (Aalto Univ., Finland) 
at 37~GHz \citep{TV94}, if available for the object. In Figure~\ref{SETevol}, {\it right} 
we mark derived epochs of ejection of knots to determine whether the appearance of a new knot can be associated with a millimeter-wave outburst or other event. 
We discuss such comparisons in the notes on individual sources in Appendix~\ref{Notes}.
 
\begin{deluxetable*}{rrrrrrrrl}
\singlespace
\tablecolumns{9}
\tablecaption{\small\bf Velocity and Acceleration in Jets \label{beta}}
\tabletypesize{\footnotesize}        
\tablehead{
\colhead{Source}&\colhead{Knot}&\colhead{$l$}&\colhead{$<\mu>$}&\colhead{$<\Phi>$}
&\colhead{$\dot{\mu}_\parallel$}&\colhead{$\dot{\mu}_\perp$}&\colhead{$<\beta_{\rm app}>$}&\colhead{$T_\circ$}\\
\colhead{}&\colhead{}&\colhead{}&\colhead{mas~yr$^{-1}$}&\colhead{deg}&\colhead{mas~yr$^{-2}$}&\colhead{mas~yr$^{-2}$}
&\colhead{c}&\colhead{}\\
\colhead{(1)}&\colhead{(2)}&\colhead{(3)}&\colhead{(4)}&\colhead{(5)}&\colhead{(6)}&\colhead{(7)}&\colhead{(8)}&\colhead{(9)}
}
\startdata 
0219+428&$A1$&1&0.008$\pm$0.005&$-$164.9$\pm$20.2&\nodata&\nodata&0.20$\pm$0.14&\nodata \\
&$A2$&1&0.019$\pm$0.012&$-$101.7$\pm$8.3&\nodata&\nodata&0.51$\pm$0.32&\nodata \\
&$A3$&1&0.028$\pm$0.010&149.3$\pm$10.5&\nodata&\nodata&0.76$\pm$0.26&\nodata \\
&$A4$&1&0.011$\pm$0.009&$-$136.8$\pm$9.6&\nodata&\nodata&0.29$\pm$0.22&\nodata \\
&$B1$&1&1.043$\pm$0.052&$-$176.8$\pm$0.6&\nodata&\nodata&28.20$\pm$1.41&2008.80$\pm$0.23 \\
&$B2$&1&0.472$\pm$0.019&$-$164.1$\pm$0.8&\nodata&\nodata&12.76$\pm$0.53&2009.42$\pm$0.29 \\
0235+164&$B1$&1&0.524$\pm$0.033&$-$14.1$\pm$1.2&\nodata&\nodata&26.27$\pm$1.67&2007.44$\pm$0.10 \\
&$B2$&1&0.267$\pm$0.029&128.8$\pm$1.6&\nodata&\nodata&13.39$\pm$1.47&2008.30$\pm$0.09 \\
&$B3$&1&0.062$\pm$0.006&$-$170.7$\pm$0.2&\nodata&\nodata&3.08$\pm$0.31&2008.80$\pm$0.55 \\
0316+413&$C1$&1&0.292$\pm$0.013&150.1$\pm$0.7&\nodata&\nodata&0.35$\pm$0.02&\nodata \\
&$C2$&1&0.183$\pm$0.010&$-$137.4$\pm$0.6&\nodata&\nodata&0.22$\pm$0.02&\nodata \\
&$C3$&1&0.185$\pm$0.013&70.6$\pm$0.5&\nodata&\nodata&0.22$\pm$0.02&\nodata \\
&$C4$&1&0.064$\pm$0.010&$-$142.6$\pm$0.6&\nodata&\nodata&0.08$\pm$0.02&\nodata \\
&$C5$&1&0.144$\pm$0.010&159.1$\pm$0.5&\nodata&\nodata&0.17$\pm$0.01&\nodata \\
&$C6$&1&0.298$\pm$0.009&166.0$\pm$0.3&\nodata&\nodata&0.36$\pm$0.01&2008.5$\pm$0.8 \\
&$C7$&1&0.153$\pm$0.007&176.0$\pm$0.1&\nodata&\nodata&0.18$\pm$0.01&(2009.0$\pm$1.5) \\
&$C8$&1&0.093$\pm$0.029&$-$161.4$\pm$1.3&\nodata&\nodata&0.11$\pm$0.03&2011.1$\pm$0.7 \\
0336$-$019&$A1$&1&0.010$\pm$0.006&$-$6.3$\pm$17.4&\nodata&\nodata&0.46$\pm$0.33&\nodata \\
&$A2$&1&0.023$\pm$0.118&\nodata&\nodata&\nodata&1.07$\pm$5.49&\nodata \\
&$B1$&1&0.625$\pm$0.023&79.6$\pm$0.5&\nodata&\nodata&29.08$\pm$1.09&2007.34$\pm$0.05 \\
&$B2$&2&0.482$\pm$0.018&81.0$\pm$0.4&0.498$\pm$0.030&$-$0.056$\pm$0.040&22.44$\pm$0.85&2008.40$\pm$0.06 \\
&$B2^*$&1&0.676$\pm$0.027&80.3$\pm$0.8&\nodata&\nodata&31.42$\pm$1.26&2009.17$\pm$0.05 \\
&$B3$&2&0.190$\pm$0.012&70.1$\pm$0.4&0.117$\pm$0.013&0.054$\pm$0.016&8.84$\pm$0.55&2010.13$\pm$0.22 \\
&$B3^*$&1&0.339$\pm$0.042&75.1$\pm$1.3&\nodata&\nodata&15.79$\pm$1.95&2011.64$\pm$0.08 \\
\enddata
\vspace{3mm}
Bracketed parameter $l=4$ designates cases when the $\chi^2$ criterion is not reached 
for fitting the trajectory of a knot with polynomials of order from 0 to 4. Bracketed parameter $T_\circ$ designates 
cases when the time of ejection is calculated based on the best-fit polynomial for the $R$ coordinate. 
The table is available entirely in a machine-readable format in the online journal (Jorstad et al. 2017, ApJ, 846, 98). 
\end{deluxetable*}

\subsection{Properties of Moving Features} 

\begin{figure*}
\plottwo{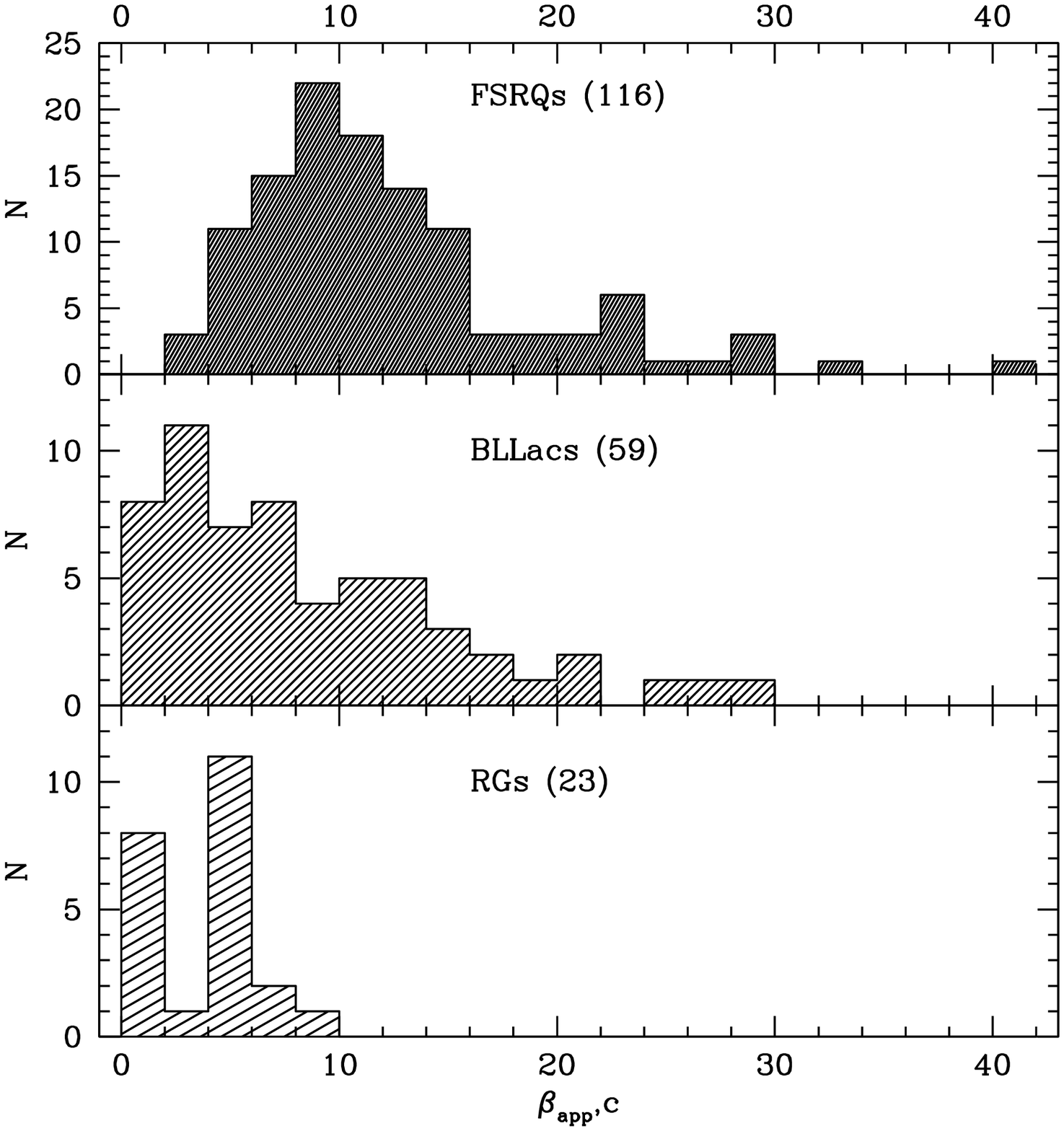}{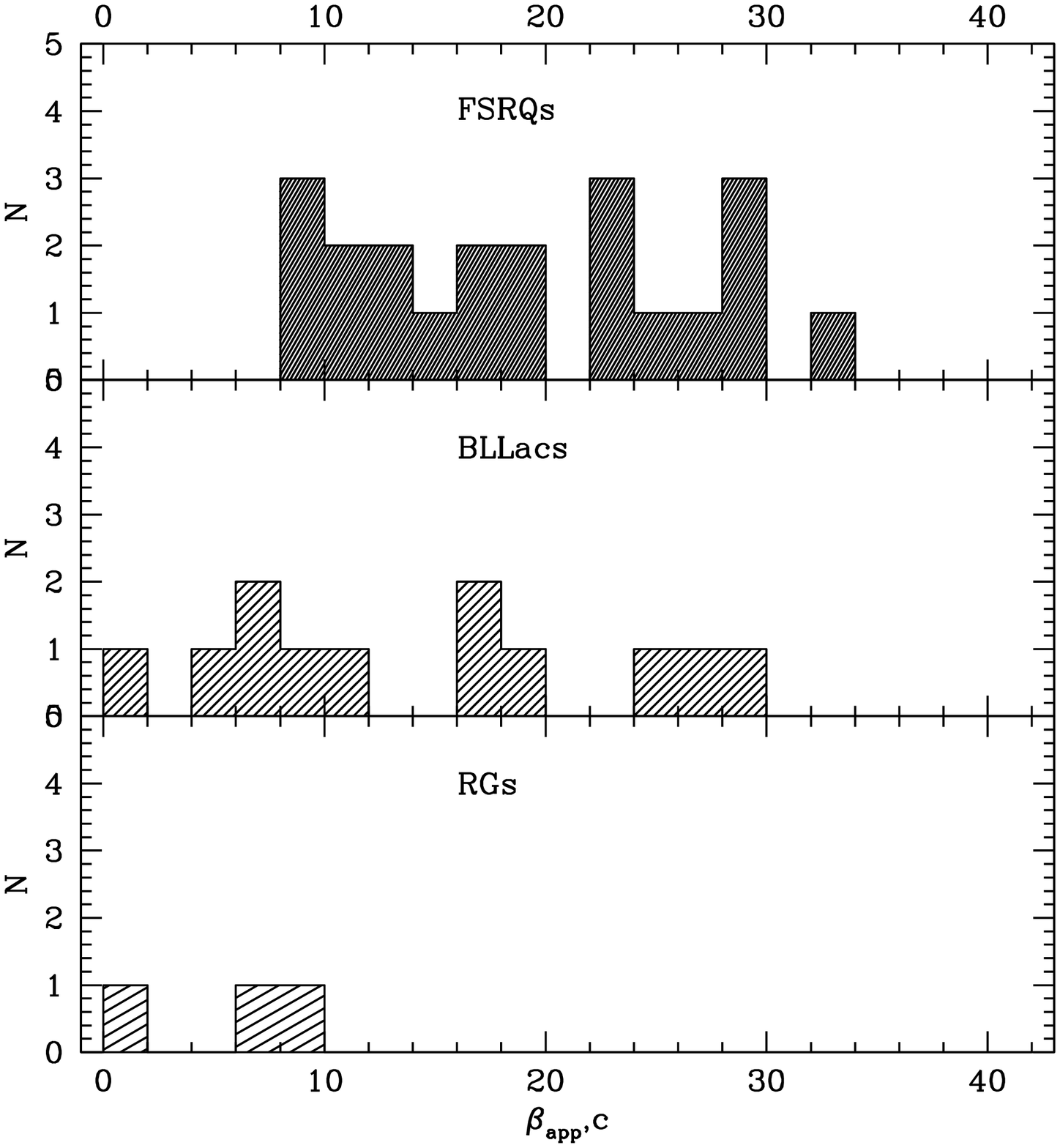}
\caption{{\it Left:} Distributions of apparent speeds of all moving knots detected in the FSRQs (top), BLLacs (middle), and RGs(bottom). {\it Right:} Distributions of maximum apparent speeds of moving knots detected in the different sub-classes.} \label{Speeds}
\end{figure*}

We derive statistically significant proper motions ($\mu>2\sigma_\mu$) in 198 of the jet components. Figure~\ref{Speeds}, {\it left} presents 
distributions of apparent speeds of these features separately for the FSRQs, BLLacs, and RGs. The distributions of apparent speeds in the FSRQs and 
BLLacs cover a wide range of $\beta_{\rm app}$ from 2 to $40c$ and peak at 8-$10c$ and 2-$4c$, respectively, with at least one knot in 13 different FSRQs and 7 different BLLacs represented in these peak intervals. There is one FSRQ knot with a speed exceeding $40c$: $B1$ in 0528+134, with $\beta_{\rm app}\sim$80~c (see Appendix~\ref{Notes}); we include it in the 40-$42c$ bin in order to limit the size of the figure. The maximum at 4-$6c$ in the distribution of apparent speeds of the RGs is mainly from knots in the radio galaxy 3C~111. The K-S test gives a probability of 49.5\% ($KS$=0.240) that the apparent speed distribution of the FSRQs is different from that of the BLLacs, therefore the difference between the distributions is uncertain. The distributions of both the FSRQs and BLLacs have a high probability (99.8\% and 96.4\%, respectively) of being different from the distribution of $\beta_{\rm app}$ of the RGs. 

We have identified the highest reliable speed in each jet. This is identified as the highest apparent speed 
measured in a source  for knots observed at least at 6 epochs, with standard deviation of the average size of the knot 
less than 
$\langle a\rangle/2$. We exclude from consideration apparent speeds of trailing features, and we do not follow these rules if a source 
has only one moving feature detected in the jet. (Note that we detect only a single moving feature in one source, 1219+285, and only two moving knots in three of the other BLLacs.)
Figure~\ref{Speeds}, {\it right} presents the distributions of the highest apparent speeds 
for FSRQs, BLLacs, and RGs. The maximum speeds of the FSRQs and BLLacs cover the same wide range as for entire sample, displayed in Figure~\ref{Speeds}, {\it left}, but without distinct maxima. 
According to the K-S statistic, the distributions of the FSRQs and BLLacs are not different from each other at 70\% probability. 
This implies that apparent speeds of the parsec-scale jets in $\gamma$-ray bright FSRQs and 
BLLacs could be drawn from the same parent population. Although we also plot the distributions for the RGs, the sample is too small to draw any statistical conclusions.

\begin{figure}
\plotone{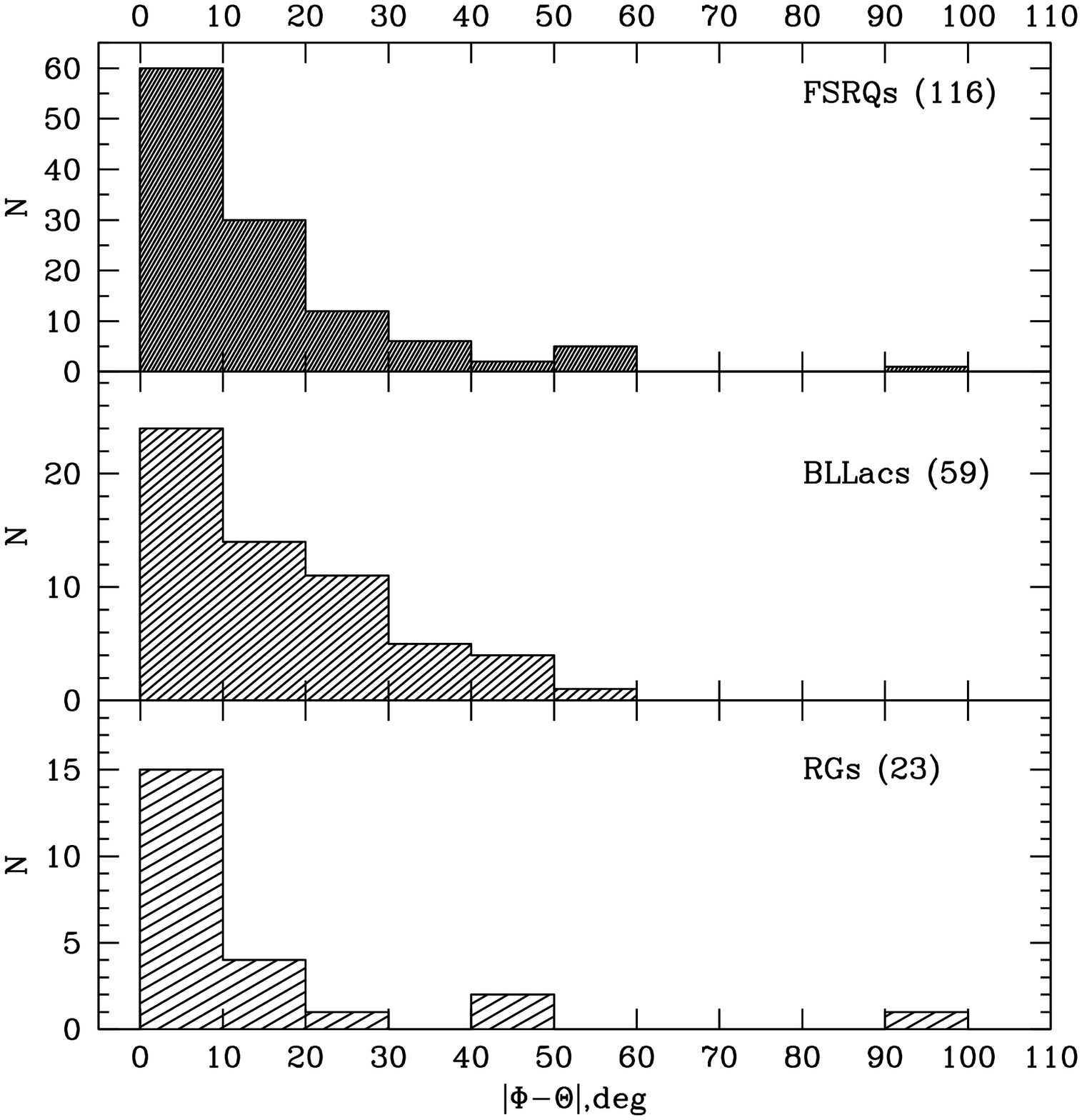}
\caption{Distributions of differences between the velocity vector and the line from the core to the average position of moving knots in the FSRQs (top), BLLacs (middle), and RGs (bottom).} \label{Phi}
\end{figure}

We analyze the direction of the velocity vector of each knot, relative to the line from the core to the average position of the knot over 
epochs (the positional line), by calculating the difference between the direction $\Phi$ of the apparent velocity, given in Table~\ref{beta}, 
and the average position angle of the knot, $\langle\Theta\rangle$, given in Table~\ref{Parm}.  Figure~\ref{Phi} plots the distributions of 
values of $|\Phi-\langle\Theta\rangle|$ for the FSRQs, BLLacs, and RGs. In all three sub-classes the most prominent groups of knots
are those with velocity vectors aligned with the positional lines within 10$^\circ$ (51.7\%, 40.7\%, and 65.2\% of the knots in FSRQs, BLLacs, and RGs, 
respectively). The BLLacs have the smallest percentage of knots with nearly unidirectional motion, and the largest value of 
$\langle\sigma(\Phi)\rangle$ [where $\langle\sigma(\Phi)\rangle=\sum_{i=1}^{N_k}\sigma(\Phi_{\rm i})/N_k$, with $N_k$ the number of knots in the sub-class], 
which is 0.9$^\circ$, 1.3$^\circ$, and 1.0$^\circ$ for FSRQs, BLLacs, and RGs, respectively. BL Lacs also have the largest value of 
$\langle\sigma(\langle\Theta\rangle)\rangle$ [where $\langle\sigma(\langle\Theta\rangle)\rangle=\sum_{i=1}^{N_k}\sigma(\langle\Theta\rangle_{\rm i})/N_k$],
which is 7.5$^\circ$, 8.5$^\circ$, and 4.2$^\circ$ for FSRQs, BLLacs, and RGs, respectively. According to Figure~\ref{Phi}, the largest deviations of the direction of the velocity vector from the positional line of moving knots, which range from 30$^\circ$ to 60$^\circ$, are observed in 12.1\% of 
the FSRQs, 16.9\% of the BLLacs, and 13.0\% of the RGs. There are two exceptions: knots $C1$ in the quasar 0836+710 and $C3$ in the radio galaxy 3C~84, which have velocity vectors perpendicular to their positional lines. Large values of $|\Phi-\langle\Theta\rangle|$ imply 
that either the jet is broader than the size of individual components, and/or knots possess complex trajectories, 
deviating substantially from a straight line. We note that the two knots with motion transverse to the jet are observed near a bend in the jet (see Appendix~\ref{Notes}).

\begin{figure*}
\plottwo{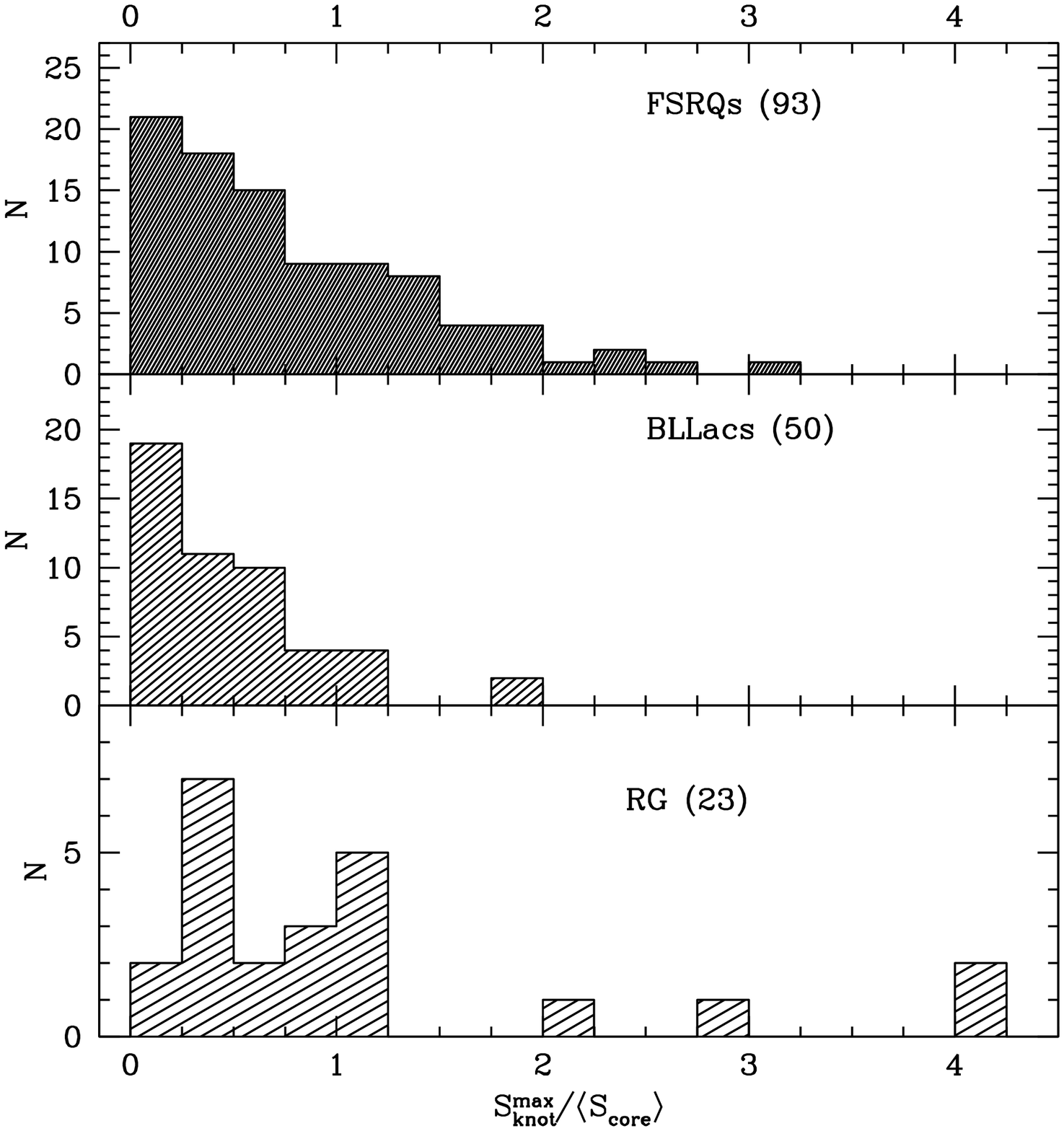}{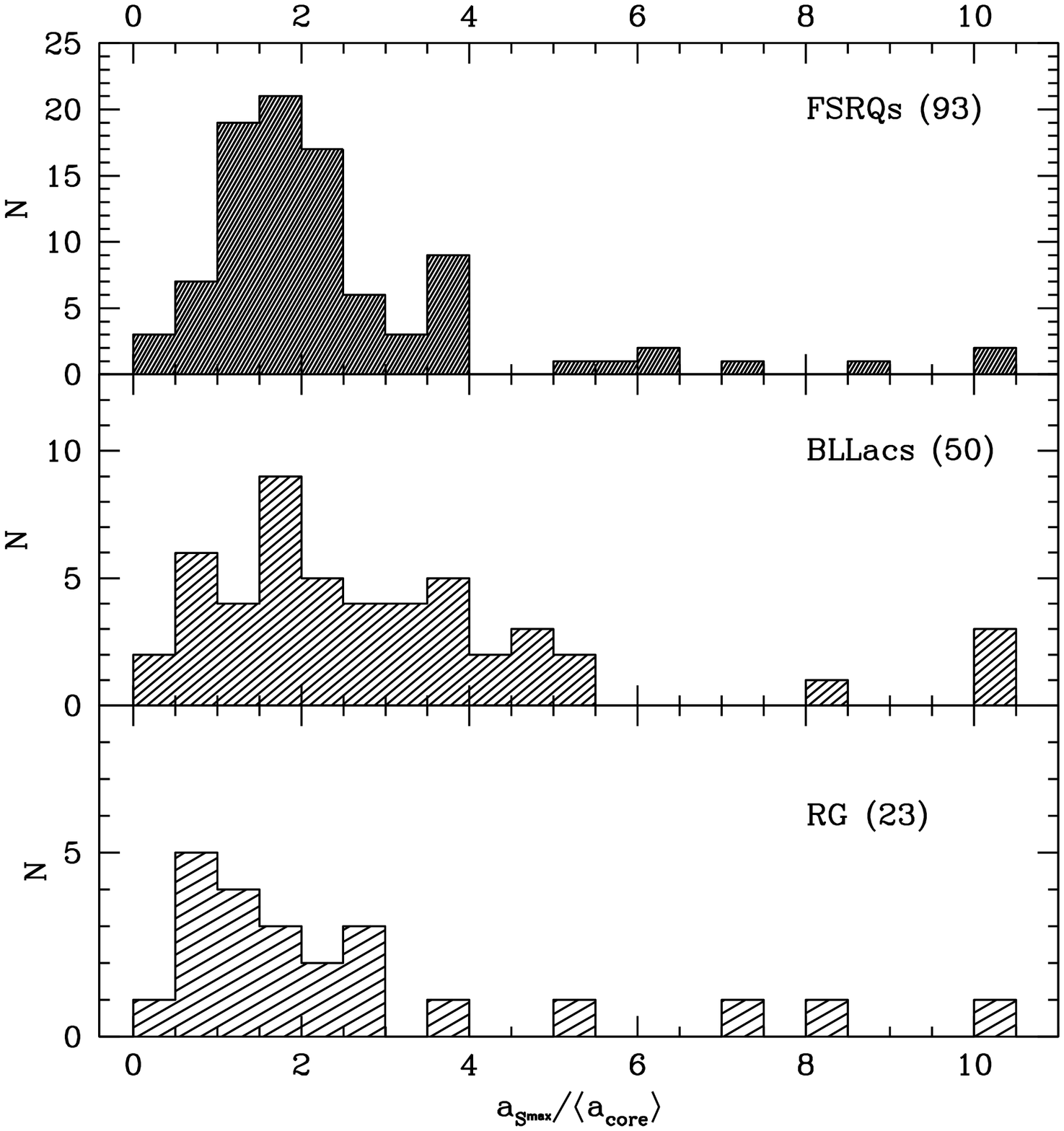}
\caption{{\it Left:} Distributions of maximum flux densities of moving knots normalized by the average core flux densities in the FSRQs (top), BLLacs (middle), and RGs (bottom). {\it Right:} Distributions of sizes of moving knots measured at the time of
maximum flux density and normalized by the average core size in the FSRQs (top), BLLacs (middle), and RGs (bottom)} \label{Size}
\end{figure*}

The emission of knots at 43~GHz is incoherent synchrotron emission, with a flux that depends on the density of relativistic electrons, represented by
coefficient $N_\circ$ of the energy distribution, magnetic field strength, $B$, spectral index, $\alpha$, size of the emission region $a$, and Doppler factor, $\delta$. We have investigated whether disturbances, observed in the jets as knots of enhanced brightness, 
have different flux densities relative to the respective cores across the three sub-classes.
Figure~\ref{Size}, {\it left} presents distributions of maximum flux 
densities (as measured according to the model fit) relative to the average flux densities of the cores for moving knots in the FSRQs, BLLacs, and RGs. We have included in the analysis only knots with average angular distances from the core $<1$~mas. The K-S test yields a small probability of 16\% ($KS$=0.351) that the distributions of the FSRQs and BLLacs are similar. Although both distributions peak at values $S_{\rm knot}^{\rm max}\le\langle S_{\rm core}\rangle/4$, there is a significant difference in the percentage of knots whose maximum flux densities are higher than the mean flux density of the core, 32\% in the FSRQs and 12\% in the BLLacs. This result agrees with the finding
that the brightness temperatures of knots other than the core in the BLLacs are lower than $T_{\rm b,obs}$ of such
features in the FSRQs, while the $T_{\rm b,obs}$ values of the BLLac cores are higher than those in the FSRQs. 
 
It is often assumed that the particles and magnetic field in moving knots in the parsec-scale jet are in energy equipartition. This implies that a higher value of $B$ should be associated with a larger value of $N_\circ$. If we assume that the magnetic field strength of a knot near the core (where the maximum flux of a knot is usually observed) is not significantly different from that of 
the core, and that the spectral indices of knots are similar across the sub-classes (although see \citealt{Hovatta14}), the main parameters affecting the distributions 
should be the Doppler factor and the size of the emission region. We compute and discuss values of $\delta$ of moving knots 
in \S~\ref{Physics} and analyze there whether the derived Doppler factors can explain the differences in the distributions shown in 
Figure~\ref{Size}, {\it left}. Here we address the question of whether the size of moving knots can play a role in these 
differences. Figure~\ref{Size}, {\it right} displays the distributions of relative angular sizes of moving knots (ratios of the size 
of the knot at maximum flux density to the average size of the core) for the FSRQs, BLLacs, and RGs. The samples are the same as 
those used to construct the distributions of relative flux densities. The K-S test gives a probability of 84\% that the distributions 
of relative sizes for the FSRQs and BLLacs are the same ($KS$=0.174), implying that this parameter should not play a major role in 
the differences of the relative flux distributions. 
It is interesting to note that, for both the FSRQs and BLLacs, the most common size of a knot at the peak of the flux density is about twice the core size.

\subsection{Acceleration and Deceleration}\label{SAccel}

Each of the apparent speeds of the 198 features discussed above represents the average apparent speed at the mid-point in time of a 
knot's  trajectory. This speed is equal to the apparent speed for the majority of knots, which move ballistically. However, 31.3\% of knots exhibit non-ballistic motion with a statistically significant change of the apparent speed: 42 features in 16 FSRQs, 15 features in 7 BLLacs, and 5 features in 2 RGs. Out of these, 19 knots have trajectories 
that require polynomials of 3rd or 4th order to fit the motion. For the latter, Table~\ref{beta} gives the average values of 
acceleration or deceleration. For a given knot, we cannot distinguish whether the acceleration is connected with an intrinsic change in speed 
(and therefore, in the Lorentz factor, $\Gamma$) or with a change of the intrinsic viewing angle, $\Theta_\circ$. However, \cite{Homan09} have proposed a statistical approach to resolve this ambiguity. They have estimated that, in a flux limited sample of beamed jets, if the observed accelerations are caused by changes in the jet 
direction instead of changes in Lorentz factors, the observed relative parallel acceleration should not 
exceed $\sim60\%$ of the observed relative perpendicular acceleration, averaged over the sample. 

\begin{figure*}
\plottwo{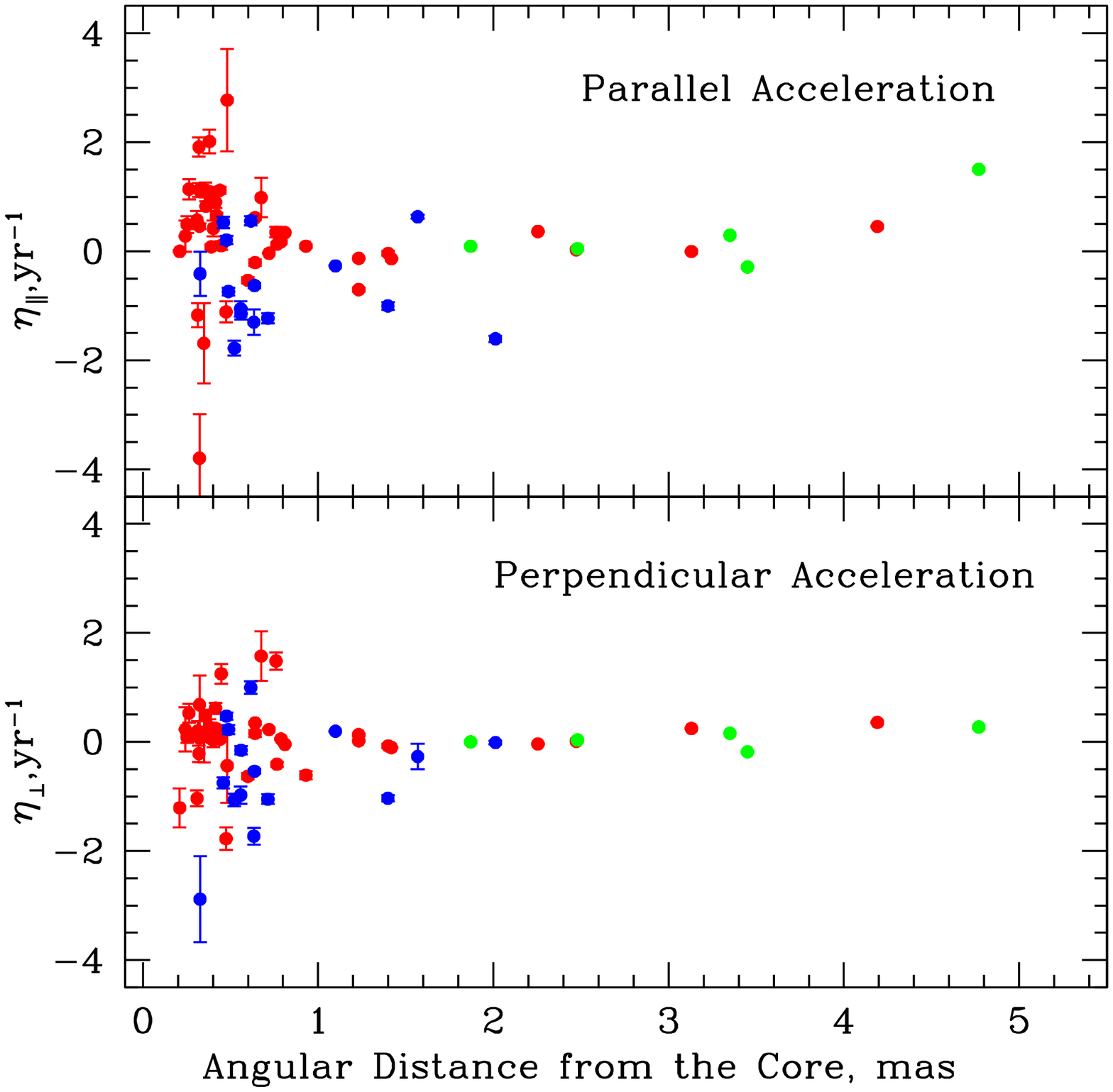}{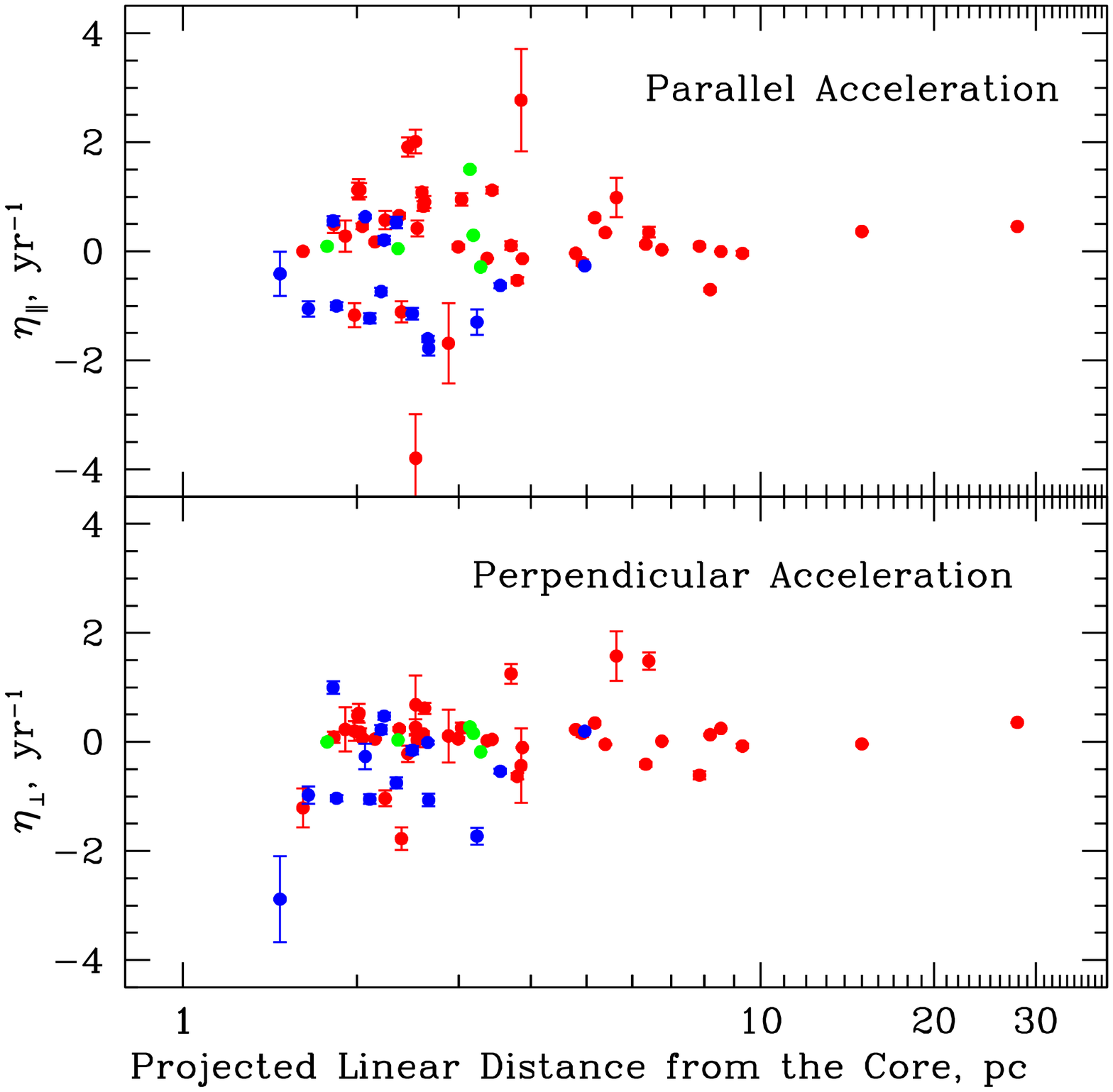}
\caption{{\it Left:} Relative parallel (top) and perpendicular (bottom) accelerations of moving knots in the FSRQs (red), BLLacs (blue) and RGs (green) vs.\ their average angular distances. {\it Right:} The same values vs.\ the average projected linear distances.} \label{Accel}
\end{figure*}

We follow the formalism in \cite{Homan09} and compute the relative accelerations parallel to the jet, 
$\eta_\parallel = (1+z)\dot{\mu}_\parallel/\mu$, and perpendicular to the jet, $\eta_\perp = (1+z)\dot{\mu}_\perp/\mu$. 
Figure~\ref{Accel}, {\it left} shows values of relative accelerations with respect to the average angular distance of each knot, 
while Figure~\ref{Accel}, {\it right} plots the same values as functions of the average projected linear distance. Table~\ref{TAccel} gives 
the averaged values of the relative parallel and perpendicular accelerations, weighted by their uncertainties, for the entire sample and 
separately for the FSRQs and BLLacs. It is clear from Table~\ref{TAccel} and  Figure~\ref{Accel} that there are similarities and differences
in the acceleration properties of the sources. First, independent of whether the entire sample or different sub-classes is considered, 
the parallel acceleration is larger by at least a factor of $\sim$2 than the perpendicular acceleration. This result agrees very well 
with that reported by \cite{Homan15} using the MOJAVE sample. Second, the parallel and perpendicular accelerations
for the entire sample are significantly less than those found in the MOJAVE sample, while $\langle\eta_\parallel\rangle$ and 
$\langle\eta_\perp\rangle$ for the FSRQs are similar to the median values of the corresponding accelerations calculated 
in the MOJAVE sample. The latter can be explained by a difference in the behavior of the FSRQs and BLLacs in our sample:
while the majority of knots in the FSRQ jets accelerate, Table~\ref{TAccel} and  Figure~\ref{Accel} indicate that the majority of knots in the BLLac jets decelerate.   

\begin{deluxetable}{lr}
\singlespace
\tablecolumns{2}
\tablecaption{\small\bf Acceleration Statistics \label{TAccel}}
\tabletypesize{\footnotesize}
\tablehead{
\colhead{Characteristic}&\colhead{Values}}
\startdata
Number of knots& 62 \\
Number of knots for FSRQs& 42\\
Number of knots for BLLacs& 15 \\
Number of knots for RGs& 5 \\
$\langle\eta_\parallel\rangle$ for all sample,yr$^{-1}$&0.034\\ 
$\langle\eta_\perp\rangle$ for all sample, yr$^{-1}$& 0.012 \\
$\langle\eta_\parallel\rangle$ for FSRQs, yr$^{-1}$& 0.117 \\
$\langle\eta_\perp\rangle$ for FSRQs, yr$^{-1}$ &  0.066 \\
$\langle\eta_\parallel\rangle$ for BLLacs, yr$^{-1}$& $-$0.421 \\  
$\langle\eta_\perp\rangle$ for BLLacs, yr$^{-1}$ & $-$0.147 \\
\enddata
\end{deluxetable}

\begin{figure}
\plotone{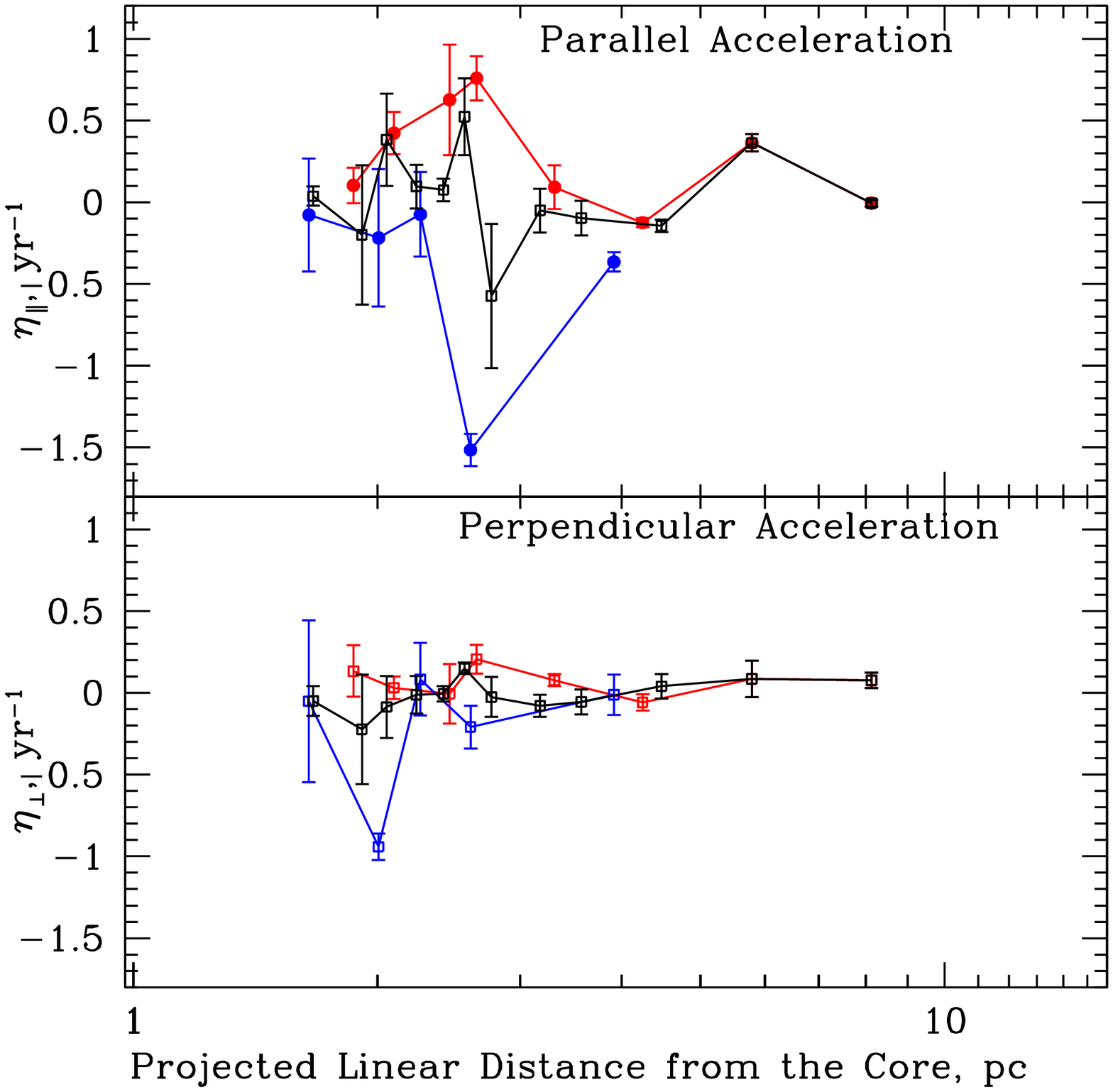}
\caption{{\it Top:} Binned values of the relative parallel accelerations for knots in all sources (black), FSRQs alone (red), and BLLacs alone (blue) vs.\ average projected linear distances of these knots. 
{\it Bottom:} The same for the relative perpendicular accelerations.} \label{AccelL}
\end{figure}

To determine the location where the acceleration/deleceration occurs, we have sorted the values of $\eta_\parallel$
and $\eta_\perp$ according to increasing linear {\emph projected} distances from the core and calculated weighted (by the uncertainties) average values 
for each 5 successive sorted values for all sources and for FSRQs alone, and 3 successive values for the BLLacs alone. These values 
of $\langle\eta_\parallel\rangle_{\rm bin}$ and $\langle\eta_\perp\rangle_{\rm bin}$ are plotted in Figure~\ref{AccelL} versus 
average distance for each bin. In the case of the entire sample and FSRQs alone, we ignore knots beyond 10~pc, since there are only 
two such knots in the sample. Figure~\ref{AccelL} shows that the most dramatic changes in the motion 
of jet components occur within 3~pc of the core. Main features in the curves of the parallel acceleration (a positive global maximum 
for the FSRQs and a negative global minimum for the BLLacs) are located between 2 and 3~pc from the core. Note that 10 measurements 
corresponding to the two highest points in the FSRQ curve include data from 9 different quasars, while the global minimum in the BLLac 
curve consists of measurements in 3 different sources. For the FSRQs all values of $\langle\eta_\parallel\rangle_{\rm bin}$ are 
positive within 5~pc of the core and values of $\langle\eta_\parallel\rangle_{\rm bin}$ between 2 and 3~pc exceed those 
of $\langle\eta_\perp\rangle_{\rm bin}$ by more than a factor of 3. In general, 
$|\langle\eta_\parallel\rangle_{\rm bin}|>|\langle\eta_\perp\rangle_{\rm bin}|$ for all FSRQ bins, except for the last one at $\sim8$~pc. 
For the BLLacs, all values of $\langle\eta_\parallel\rangle_{\rm bin}$ are negative, although we track the deceleration only up to 4~pc. 
However, the BLLac sample also possesses a global minimum in $\langle\eta_\perp\rangle_{\rm bin}$ at $\sim$2~pc, 
which exceeds in magnitude the corresponding value of $\langle\eta_\parallel\rangle_{\rm bin}$ by a factor of $\sim$4, although only 
2 sources (BL~Lac and OJ049) have contributed to this bin. Therefore, we conclude that the jets of the quasars and, perhaps, radio galaxies 
in our sample exhibit an intrinsic acceleration connected with an increase of the Lorentz factor, while the jets of the BLLacs undergo an intrinsic deceleration (a decrease of $\Gamma$) within 4~pc (projected) of the core. In addition, there is a hint that, very close to the core, jets of the BLLacs experience strong curvature, which could result in deceleration after a knot executes the bend. Although this result needs to be confirmed by higher-resolution imaging (e.g., at 86 GHz), it is supported by the finding in \S~\ref{STB} that the observed brightness temperatures of the cores of the BLLacs tend to be higher, and the $T_{\rm b,obs}$ values of jet features lower than those of the FSRQs and RGs. Deceleration of knots in the vicinity of the core could be a reason for the corresponding decrease in intensity. 

\subsection{Properties of Quasi-Stationary Features}

\begin{figure*}
\plottwo{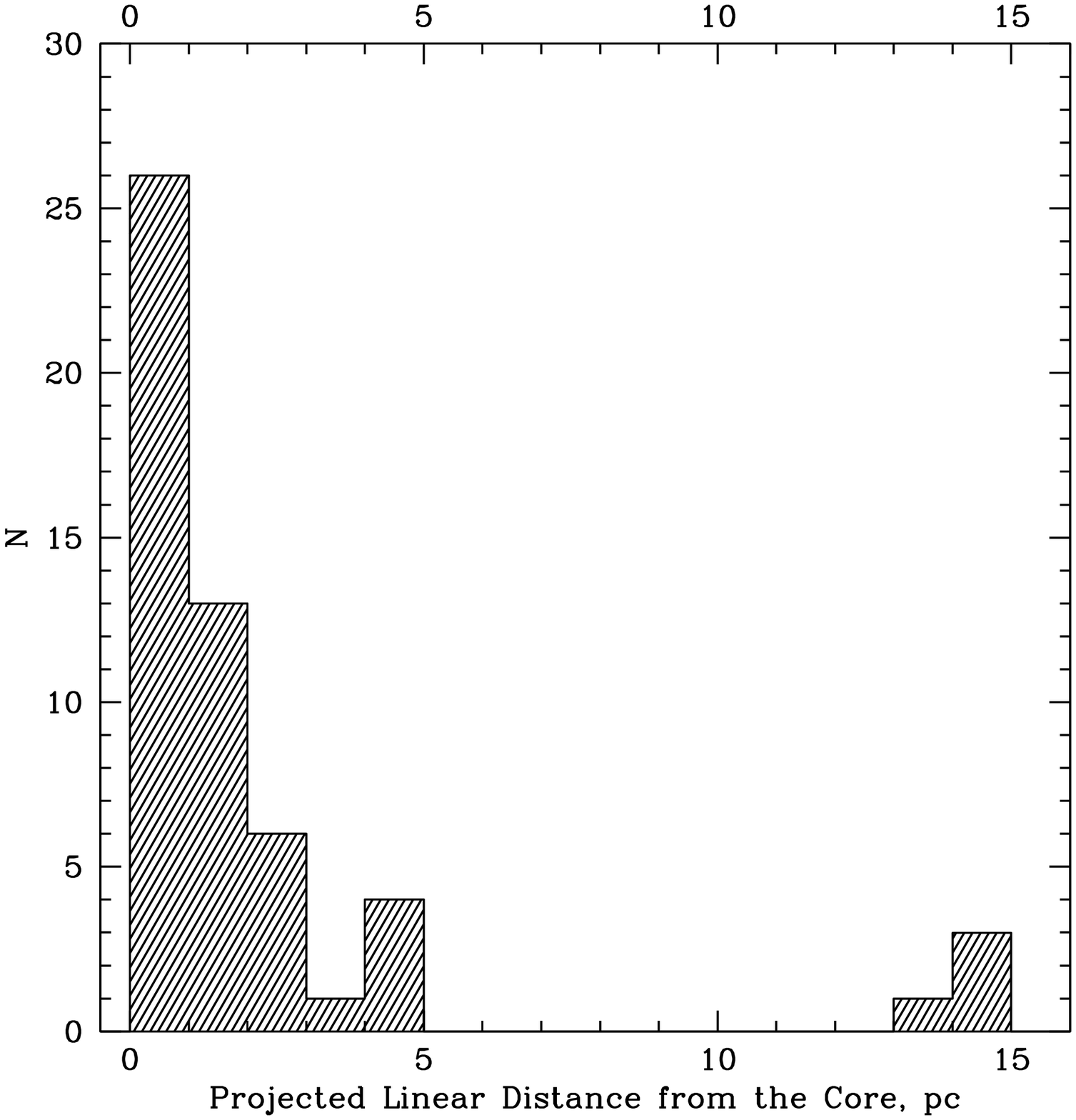}{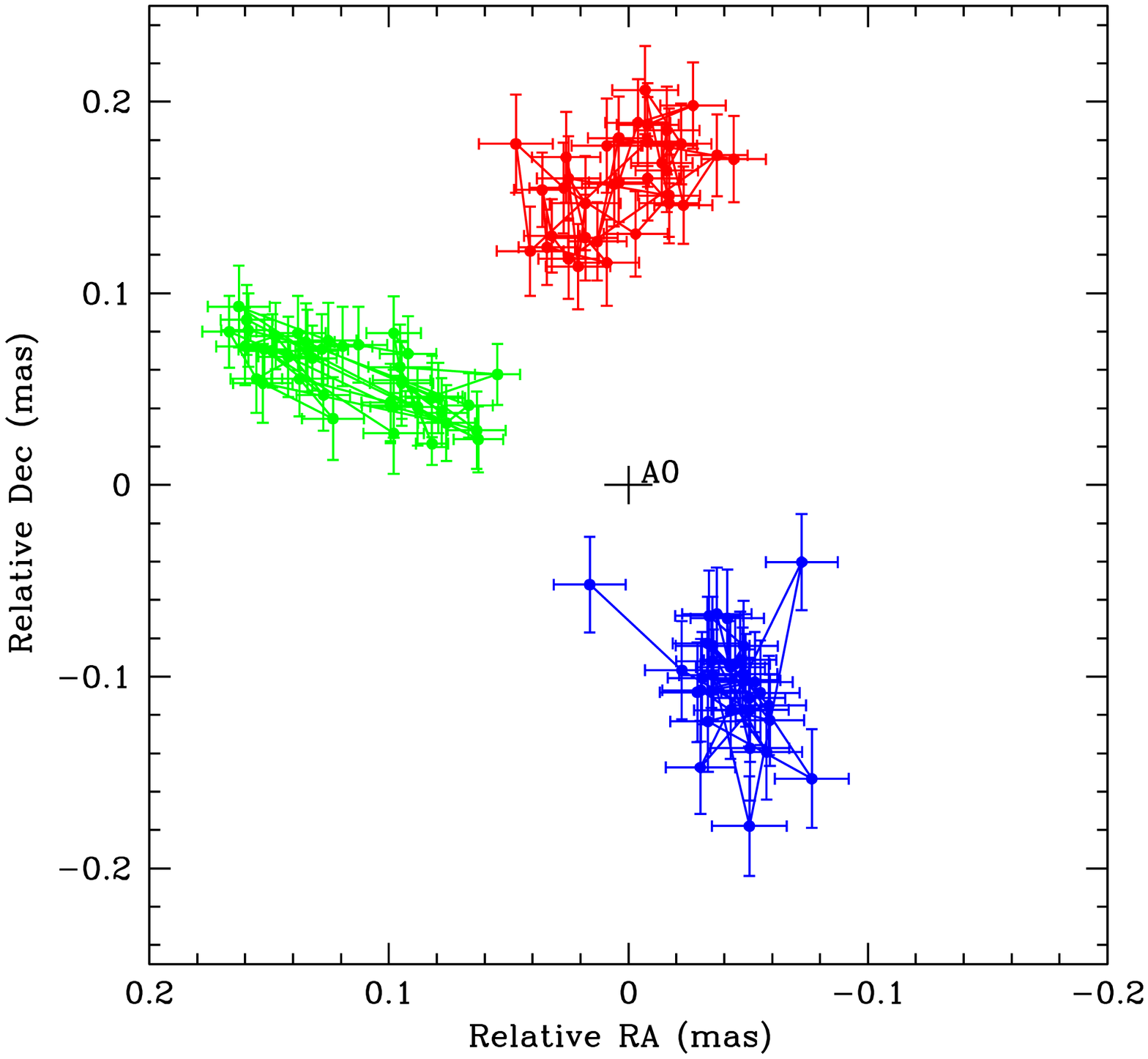}
\caption{{\it Left:} Distribution of projected linear distances of quasi-stationary features in our sample. {\it Right:} Trajectories of stationary features $A1$ in 1222+216 (red), 3C~66A (blue), and 3C~111 (green) with respect to the core, $A0$ (black cross at (0,0) position).} \label{StTraj}
\end{figure*}

\begin{figure*}
\plottwo{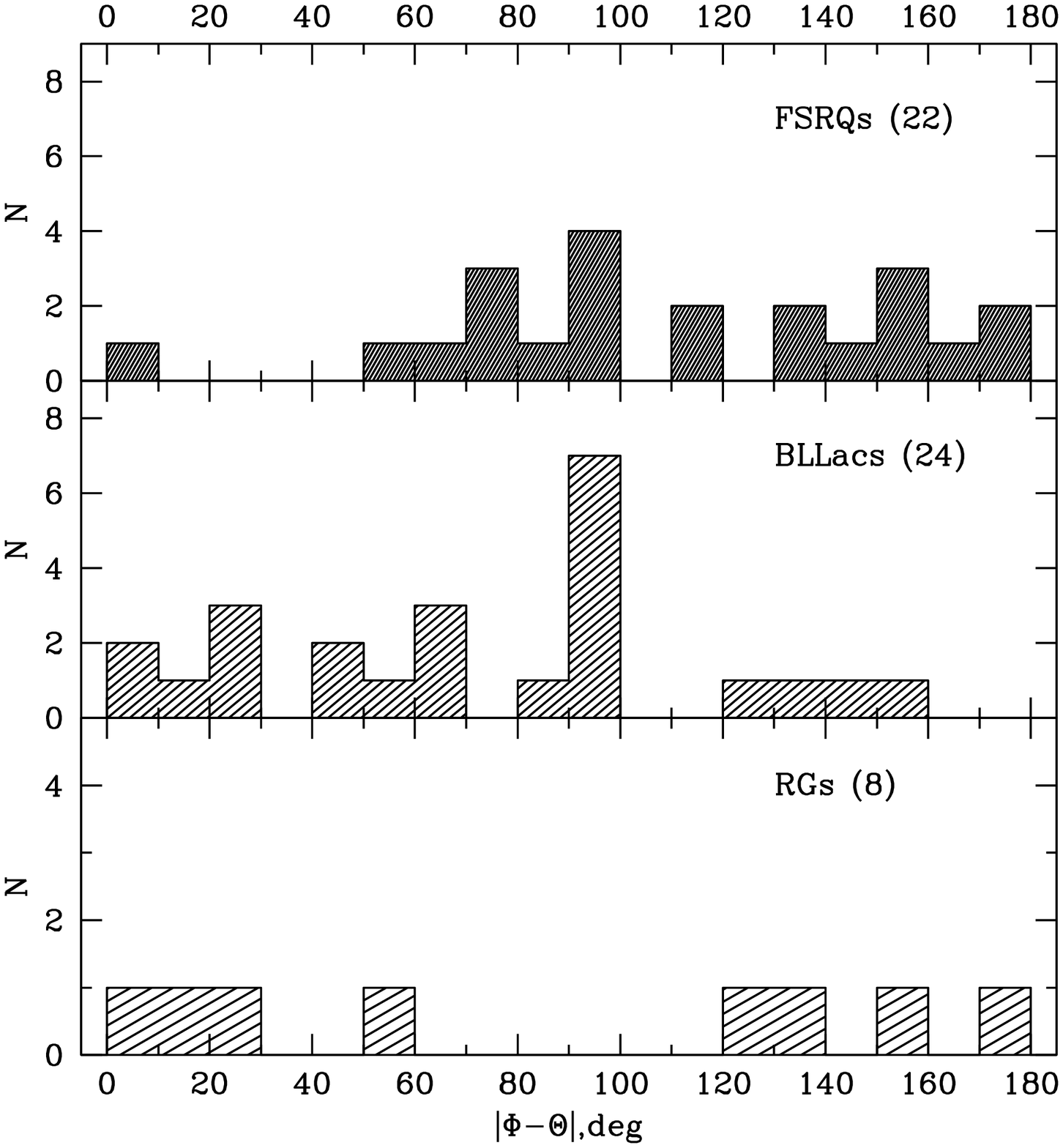}{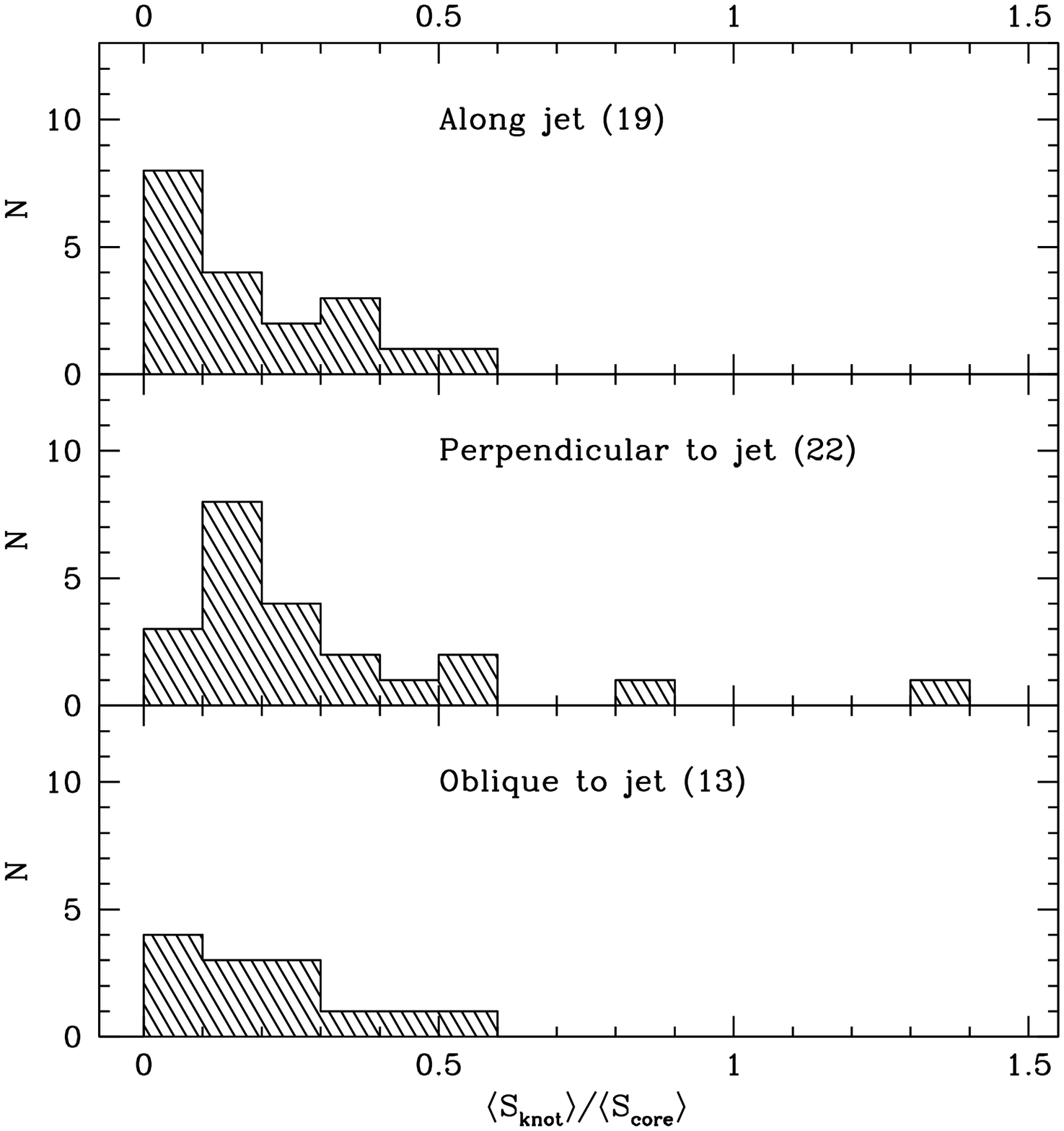}
\caption{{\it Left:} Distributions of differences between the direction of preferred motion and the line from the core to the average position of quasi-stationary knots in the FSRQs (top), BLLacs (middle), and RGs (bottom). {\it Right:} Distributions of relative flux densities of quasi-stationary features for different preferred directions of oscillations - along (top), transverse (middle), and oblique (bottom) to the jet.} 
\label{StPhi}
\end{figure*}

We have identified 54 features (24 in BLLacs, 22 in FSRQs, and 8 in RGs) with $\mu<2\sigma_\mu$, measured over 10 or more epochs, 
that we classify as quasi-stationary features. All BLLacs in our sample except 0235+164 possess at least one such stationary feature in addition 
to the core, with some of them containing three stationary knots within 1~mas of the core. The latter case is similar to the radio galaxies 3C~111 
and 3C~120. Figure~\ref{StTraj}, {\it left} shows the distribution of average projected linear distances of all stationary knots in our sample. 
The distribution has a prominent peak at projected distances $<$1~pc from the core, indicating that these stationary features tend to form in the 
vicinity of the core. There is no difference between the distributions of locations of stationary features in the FSRQs, BLLacs, and RGs with respect to the distribution shown in Figure~\ref{StTraj}, {\it left} according to the K-S test 
($KS$=0.111, 0.103, and 0.115 for FSRQs, BLLacs, and RGs, respectively). Although we classify such features as stationary knots, their 
locations tend to fluctuate about particular positions in the jet. Figure~\ref{StTraj}, {\it right} plots examples of trajectories of stationary features 
in each sub-class. One can see that, despite rather chaotic motion, the loci of (X,Y) positions for a given stationary feature form a pattern that indicates a preferred direction of the fluctuations, $\Phi$. We compare this direction with that of the line from the core to the average position of 
the stationary feature, $\langle\Theta\rangle$. Figure~\ref{StPhi}, {\it left} shows the distribution 
of the difference between these two values, separately for the FSRQs, BLLacs, and RGs. 

According to Figure~\ref{StPhi},~{\it left} stationary features have all possible directions of fluctuations -- along, 
perpendicular, or oblique to the jet axis, which supports the idea that they are different from knots classified as moving features. We separate the entire range of $|\Phi-\langle\Theta\rangle|$ into three categories: 1) fluctuation along the jet, 
$|\Phi-\langle\Theta\rangle|\le$30$^\circ$ or $|\Phi-\langle\Theta\rangle|>$150$^\circ$, 2) transverse fluctuation, 60$^\circ<|\Phi-\langle\Theta\rangle|\le$120$^\circ$, and 
3) oblique fluctuation, 30$<|\Phi-\langle\Theta\rangle|\le$60$^\circ$ or 120$<|\Phi-\langle\Theta\rangle|\le$150$^\circ$. The distribution for stationary features in the RGs
shows that knot positions fluctuate either along (62.5\%) or oblique (37.5\%) to the jet,
although the sample is small. In the quasars the positions of 50\% of stationary features vary with a preferable direction perpendicular to the jet, and 27.3\% along the jet. In the BLLacs 45.8\% and 29.1\% of stationary knots oscillate perpendicular
and along the jet, respectively. We have analyzed the difference in the relative flux density of stationary features in 
the different sub-classes and categories. According to the K-S test, the distributions of the average flux densities of stationary knots normalized by the average flux density of the core for the FSRQs and BLLac are essentially the same, and
different from the distribution of those for the RGs with a probability of $\sim$70\%. In the case of the categories (Figure~\ref{StPhi},~{\it right}) the KS test gives a probability of $\sim$65\% and 68\% that the distribution of relative flux densities for stationary knots oscillating perpendicular to the jet are different from those with fluctuations along and oblique to the jet,
respectively, with the former tending to be brighter than the latter. 

\subsection{Comparison of Jet Kinematics with Results of the MOJAVE Survey}

\begin{figure*}
\plottwo{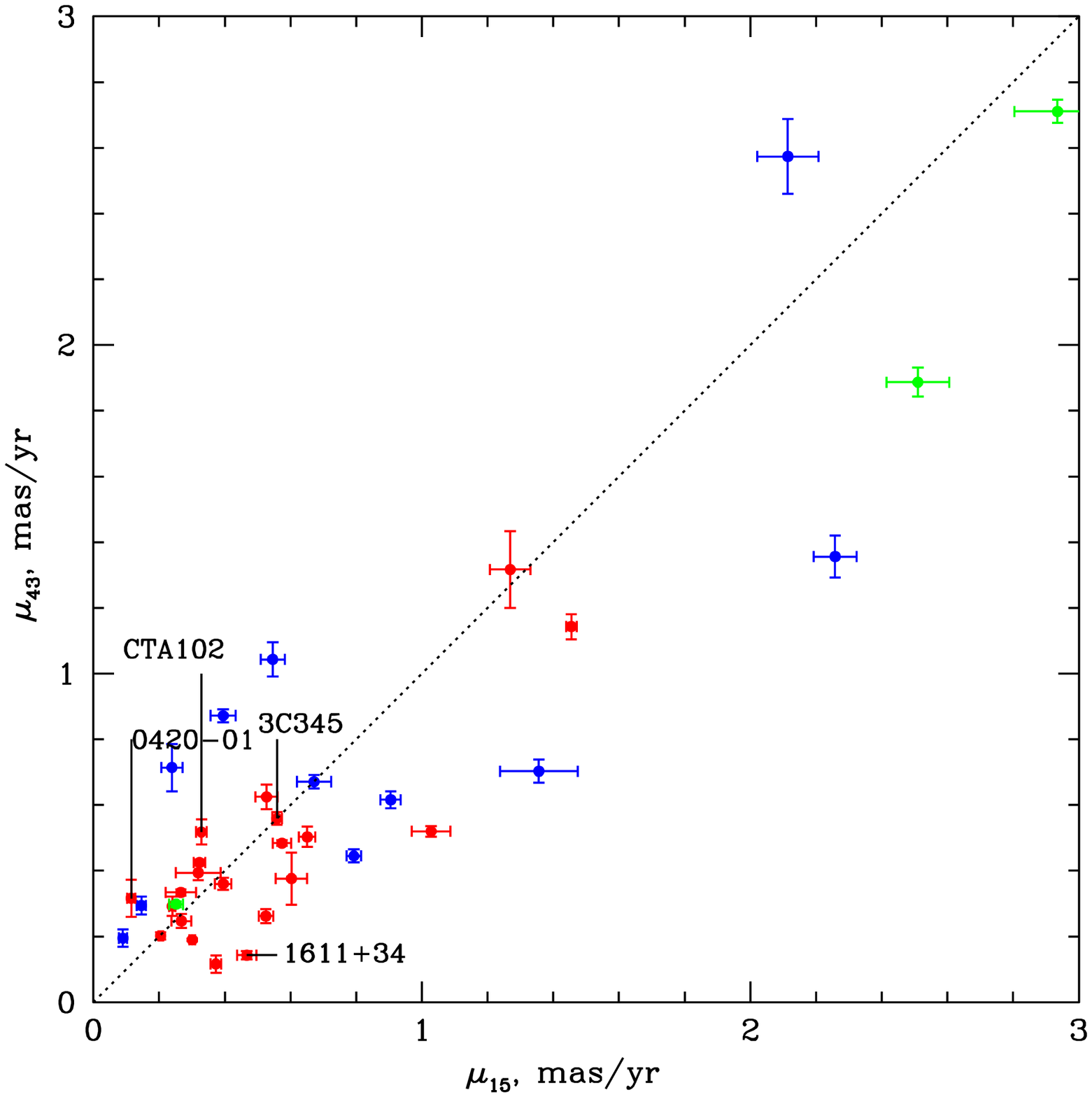}{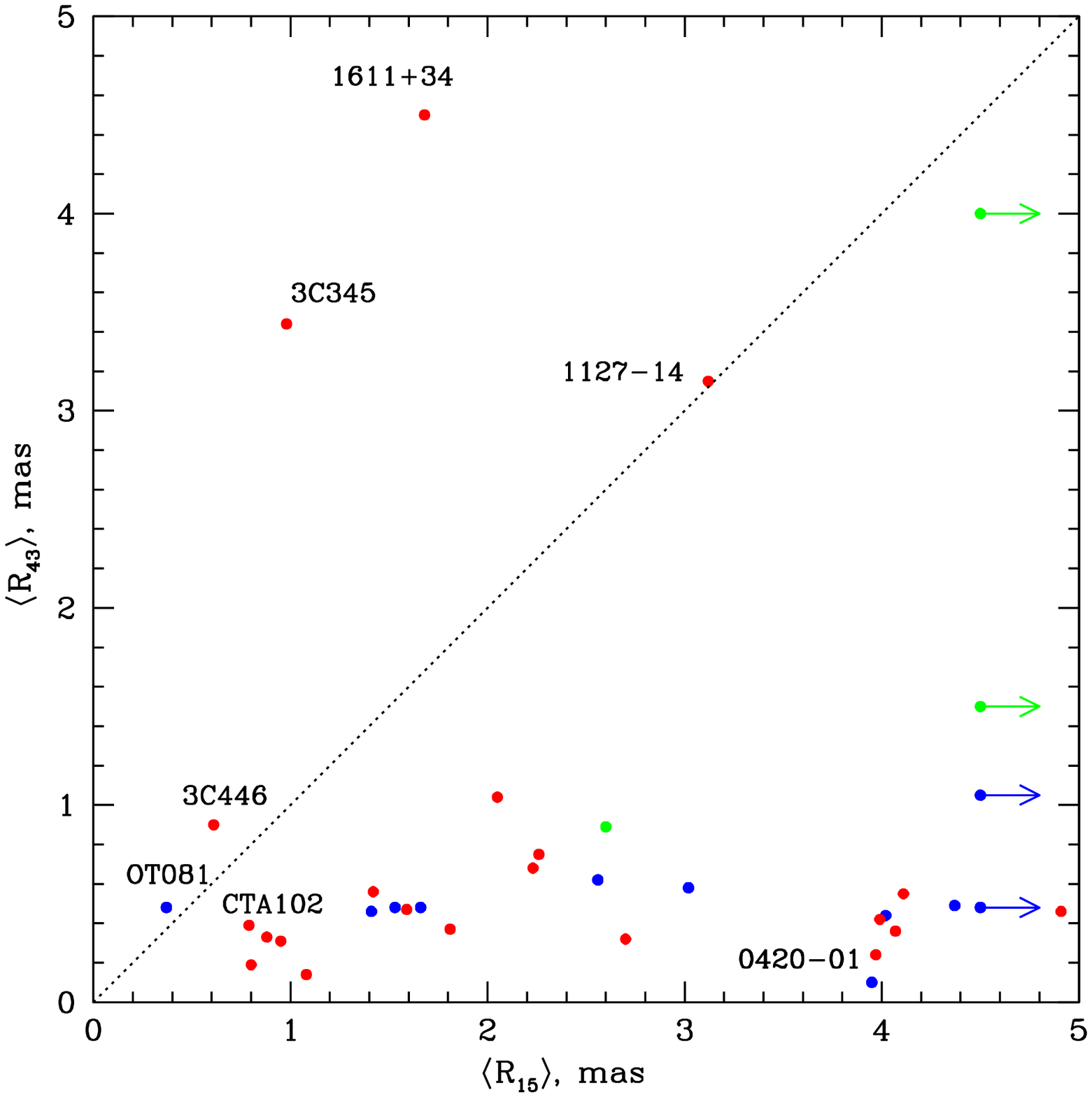}
\caption{{\it Left:} Maximum proper motion at 43~GHz vs.\ that at 15~GHz \citep[from][]{LIST16} of jet features of each of the FSRQs (red),
BLLacs (blue), and RGs (green) in our sample. {\it Right:} The average angular distances from the core
of features with the maximum proper motion at 43~GHz vs.\ that at 15~GHz.} 
\label{CompSpeed}
\end{figure*}

The MOJAVE survey monitors all sources from our sample with the VLBA at 15 GHz. The results of the jet kinematics 
presented by \cite{LIST16} cover the period of time discussed here. However, the jet features observed during this period at 15 and 43~GHz are most likely different disturbances, except perhaps the brightest knots at 43~GHz, since the majority of features in our sample are detected within 1~mas of the core (Table~\ref{Parm}). In contrast, most of the features that appear in the MOJAVE images are tracked beyond 1~mas from the core. Nevertheless, a comparison of jet kinematics at different frequencies is important for extending the description of their general properties to a wider range of distances from the central engine. We use the maximum proper motion reported by \cite{LIST16} for each source in our sample to compare with the corresponding maximum proper motion of each source, as given in Table~\ref{beta}. This allows us to avoid uncertainties in the redshifts of some sources and differences in the cosmological parameters applied. The comparison sample consists of 35 sources, excluding the BL~Lac object 0235+164 owing to its compactness at 15~GHz. 
Figure~\ref{CompSpeed}, {\it left} shows relationship  between $\mu^{\rm max}$ at 15~GHz and 43~GHz. 

Analysis of the proper motions reveals that 9 sources in the sample (26\%) have $\mu^{\rm max}_{15}=\mu^{\rm max}_{43}$ within the 1$\sigma$ uncertainty. According to Monte-Carlo simulations, this level of correspondence can occur by chance with a probability of $\sim$33\%. Figure~\ref{CompSpeed}, {\it right} plots average angular distances of features 
with maximum proper motions at 43 and 15~GHz. Except for several quasars and the BLLac object OT~081 marked on the plot, such knots are observed significantly farther from the core at 15~GHz than the fastest knots detected at 43~GHz, as expected given the factor of $\sim 3$ coarser resolution. The proper motions of the BL Lac object OT~081 and the quasar 1127$-$145 were measured almost at the same distances at both frequencies. This quasar has the same  $\mu^{\rm max}$ at 15 and 43~GHz, while for OT081 $\mu^{\rm max}_{43}>\mu^{\rm max}_{15}$ by a factor of 2. There are 15
sources in the sample for which the differences between the proper motions are statistically significant, $>2[\sigma(\mu_{15})+\sigma(\mu_{43})]$, and $\mu^{\rm max}$ is larger when the distance is farther from the core. This has a low probability, $\sim5\%$, of occurring by chance. The chance probability is lower, $\sim$3\% (11 sources out of 24), if we exclude the BLLacs. The latter supports our results discussed in \S~\ref{SAccel}, which are consistent with the finding of \cite{Homan15} that FSRQs, and most likely RGs, with detected $\gamma$-ray emission accelerate outward within several parsecs of the core. Note that the two quasars that deviate the most from the behavior indicated above, 
0420$-$014 and CTA~102 (see Figure~\ref{CompSpeed}), exhibit strong curvature of the jet at $\sim$0.5 and 2~mas from the core, respectively (see Appendix~\ref{Notes}). 
The situation is different for the BLLacs: although the fastest knots (besides that in OT~081) are 
observed at 15~GHz significantly farther from the core than those at 43~GHz, acceleration and deceleration with distance from the core occur in roughly the same number of cases. The latter neither supports nor argues against our finding in \S~\ref{SAccel} that features in the BLLacs jets tend to decelerate very close to the core. 

\section{Physical Parameters of the Jets}
We use the apparent velocities, trajectories, and light curves derived for moving features to compute physical parameters in the parsec-scale 
jets of the sources in our sample. These parameters include Doppler, $\delta$, and Lorentz, $\Gamma$, factors, viewing angle, 
$\Theta_\circ$, and half opening angle, $\theta_\circ$. We have restricted the analysis to the most reliable features, meaning those identified at 
least at 6 epochs and with ejection times within the period of VLBA monitoring reported here.
This results in a 
{\it reliable sample} that consists of 84, 46, and 12 moving knots in the FSRQs, BLLacs, and RGs, respectively.  

\subsection{Timescale of Variability and Doppler Factor}\label{Physics}

We use the formalism developed by J05 to derive the variability Doppler factor, $\delta_{\rm var}$,  of 
superluminal jet features. J05 have shown that the flux density of the majority of superluminal knots observed at 43~GHz decreases faster than the knot expands, which implies that the decay is caused by 
radiative losses rather than adiabatic cooling. Since the decline is also shorter than the
light-travel time without corrections for relativistic effects, it is likely that the observed 
timescale, $\tau_{\rm var}$, is governed by the Doppler-adjusted light-crossing time. In this case,
\begin{equation} 
\delta_{\rm var}=\frac{15.8\;s\;D}{\tau_{\rm var}\;(1+z)}, \label{EDop}
\end{equation} 
where $D$ is the luminosity distance in Gpc and $\tau_{\rm var}$ is measured in yr. Here 
$s$ is the angular size of the knot in mas, equal to $1.6a$ for a Gaussian model with FWHM=$a$ if the 
actual shape of the knot is similar to a uniform face-on disk.
Since we have a better time sampling of VLBA observations than in J05, we propose a different
approach to estimate the critical parameters $\tau_{\rm var}$ and $s$ than that used in J05,
allowing us to derive more robust values of $\delta_{\rm var}$.

\begin{figure*}
\plottwo{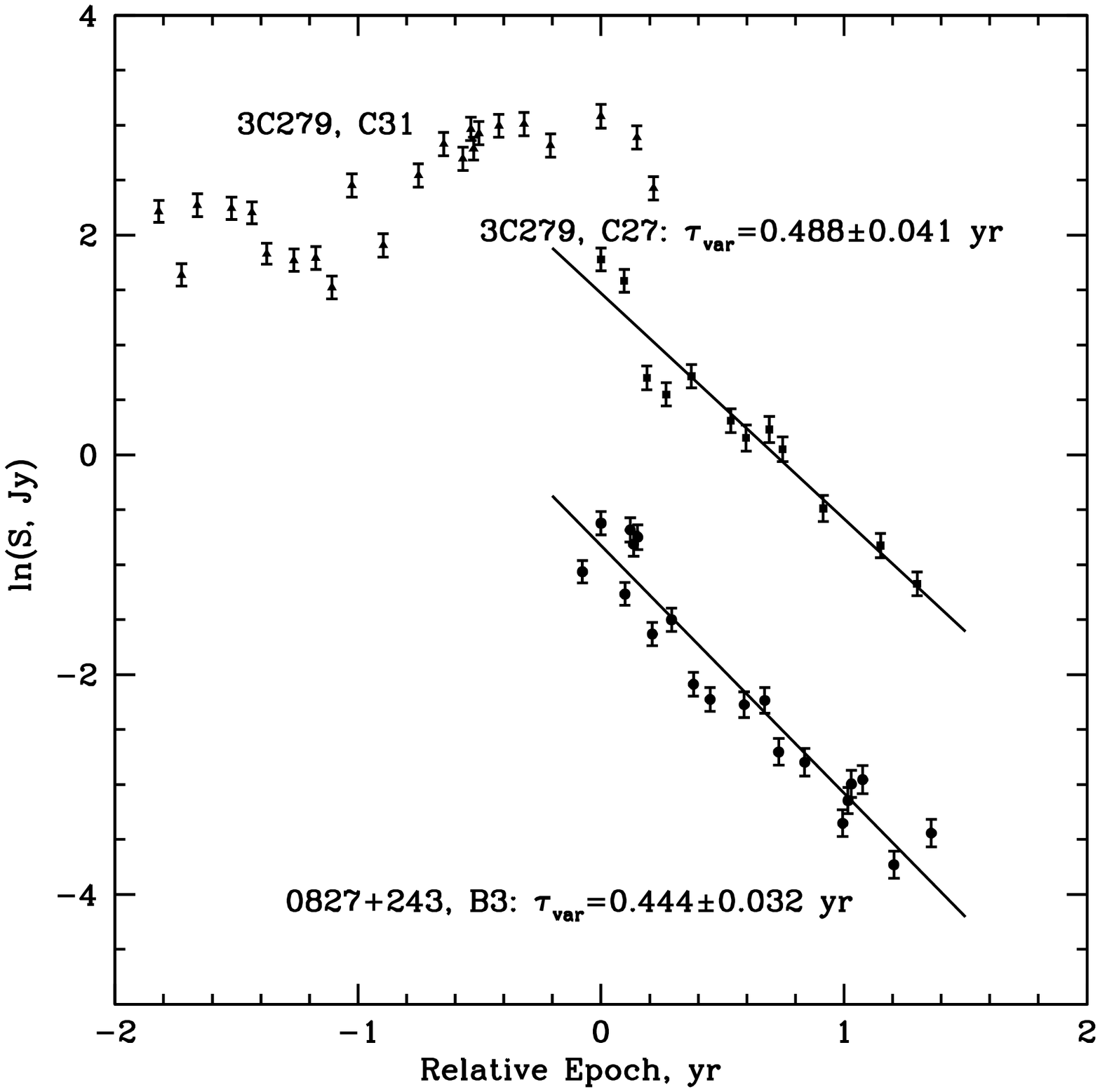}{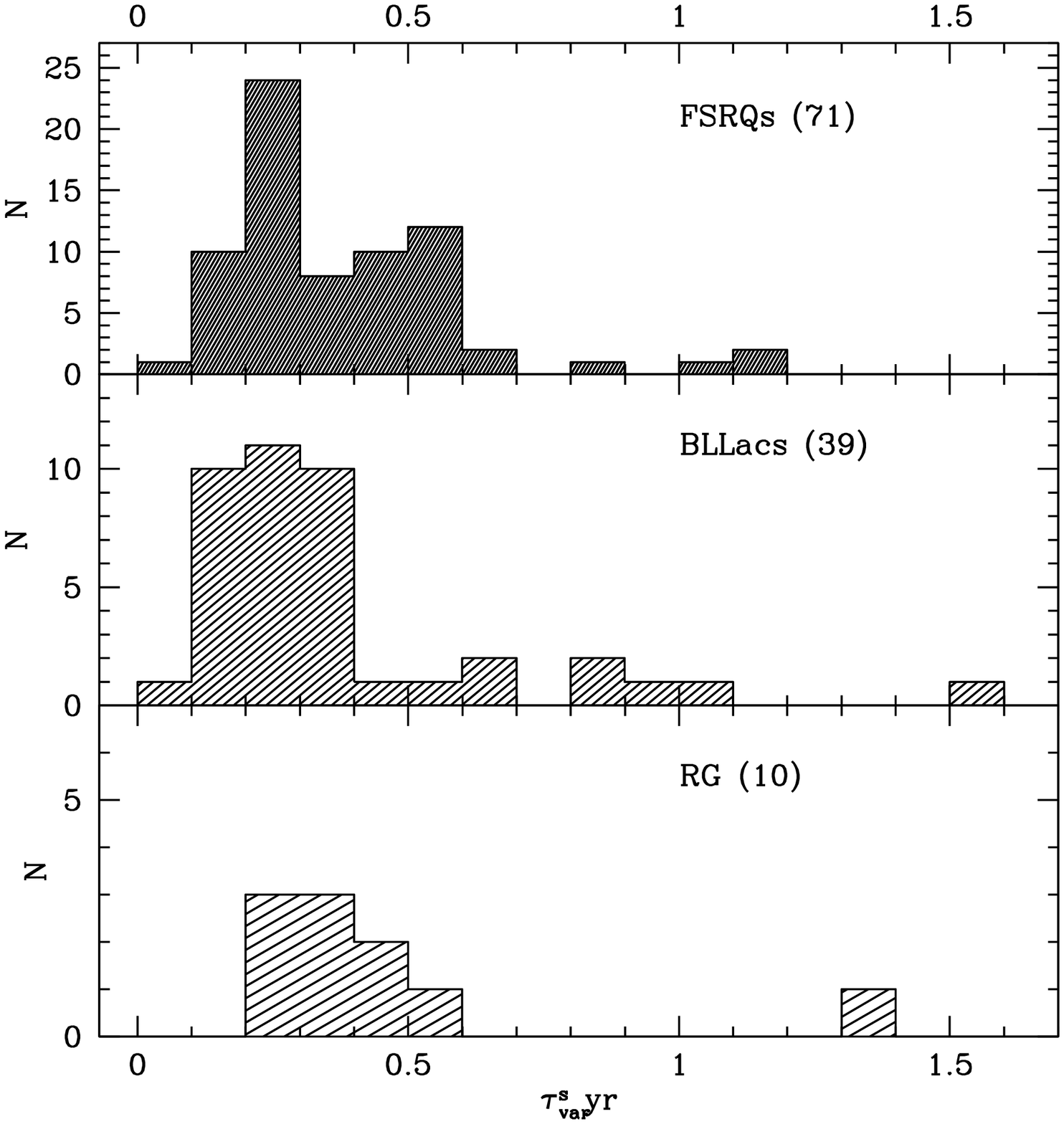}
\caption{{\it Left:} Light curves of moving knots $B3$ in 0827+243 (circles), and $C27$ (squares) and $C31$ (triangles) in 3C~279. The solid lines represent least-square fits to the flux density 
variations of $B3$ and $C27$. {\it Right:} Distributions of timescales of variability of moving knots in the host galaxy frame for the FSRQs (top), BLLacs (middle), and RGs (bottom).} 
\label{TauV}
\end{figure*}

The majority of AGN flares at millimeter wavelengths can be modeled by profiles with an exponential rise and decay in the form of 
$\ln S(t)\propto t$ \citep{TV94,LIST01}. As shown by \cite{SAV02}, a millimeter-wave outburst tends to be 
closely connected with the emergence of a new knot moving down the jet, while the evolution of the
total flux flare is similar to that of the knot associated with the flare. Usually, 
the flux of the knot fades as it separates from the core, so that we observe the decay branch 
of a knot's light curve. We determine the maximum flux density, $S_{\rm max}$, in the light curve of each knot in the reliable sample, and 
approximate the flux decay as an exponential function, $\ln (S(t)/S_{o}) = k(t-t_{\rm max})$, 
where $t_{\rm max}\ge t$ is the epoch corresponding to $S_{\rm max}$, $S_{o}$ is the
flux density of a least-squares fit to the light curve decay at time $t_{\rm max}$, and
$k$ is the slope of the fit (see Figure \ref{TauV}, {\it left}). The timescale of 
variability is then $\tau_{\rm var}= \mid$1/k$\mid$. We require at least 3 measurements of the 
decay branch when the knot is located within 1~mas of the core. 
This use of flux densities measured at multiple epochs represents an improvement in accuracy
over the method applied by J05 \citep[see also][]{BJO74}, which used only two points
on the light curve to derive $\tau_{\rm var}$.
In general, for the majority of knots $\tau_{\rm var}$ can be determined with an accuracy of
$\sim$10\%, as shown in 
Figure~\ref{TauV}, {\it left} for knots $C27$ in 3C~279 and $B3$ in OJ~248.  However, the light
curves of some knots, e.g., that of $C31$ in 3C~279 in the same figure,
have no well-determined maximum. This leads to significant uncertainties 
in the timescale of variability, rendering the knot unfit for deriving physical parameters.
Table~\ref{PhysParm} gives the values of the timescale of variability that we consider sufficiently 
accurate ($\sigma_{\rm \tau_{\rm var}}<\tau_{\rm var}/2)$ to use in calculations of the physical
parameters of the corresponding knots (71 knots 
in 21 FSRQs, 39 knots in 11 BLLacs, and 10 knots in two RGs, 9 of which are in 3C~111). 

To compare variability timescales of different sub-classes, we transfer variability timescales in the observer frame, $\tau_{\rm var}$, into variability timescales in the host galaxy frame, $\tau_{\rm var}^{\rm s}$, as 
$\tau_{\rm var}^{\rm s}=\tau_{\rm var}/(1+z)$.  Figure~\ref{TauV}, {\it right} shows distributions of values of $\tau_{\rm var}^{\rm s}$ for the FSRQs, BLLacs, and RGs. The K-S test gives a maximum cumulative difference between the distributions of the FSRQs and BLLacs equal 
to $KS$=0.167, with a probability of 94.5\% that the distributions are similar. The distributions of the FSRQs and BLLacs peak at a timescale of $\sim3$ months. Both distributions, as well as the distribution of $\tau_{\rm var}^{\rm s}$ in the RGs, suggest that the most common 
timescale of variability of a superluminal knot in the AGN frame $<0.5$~yr (for 74.6\%, 75.7\%, and 80.0\% of knots in the FSRQs, BLLacs, and RGs, respectively).

We estimate the angular size of the knot, $a$, as the average over the epochs used to calculate
$\tau_{\rm var}$. Therefore, we replace $s$ in equation~\ref{EDop}
with $\langle s\rangle=1.6\langle a\rangle$, where 
$\langle a\rangle=\sum^{\rm n}_{\rm i=1} a_{\rm i}/n$, $a_{\rm i}$ is the FWHM size of the
knot at epoch $i$, and $n$ is the number of epochs in the light curve.
The values of the variability Doppler factors are given in Table~\ref{PhysParm} for each knot in the reliable sample.

\begin{figure*}
\plottwo{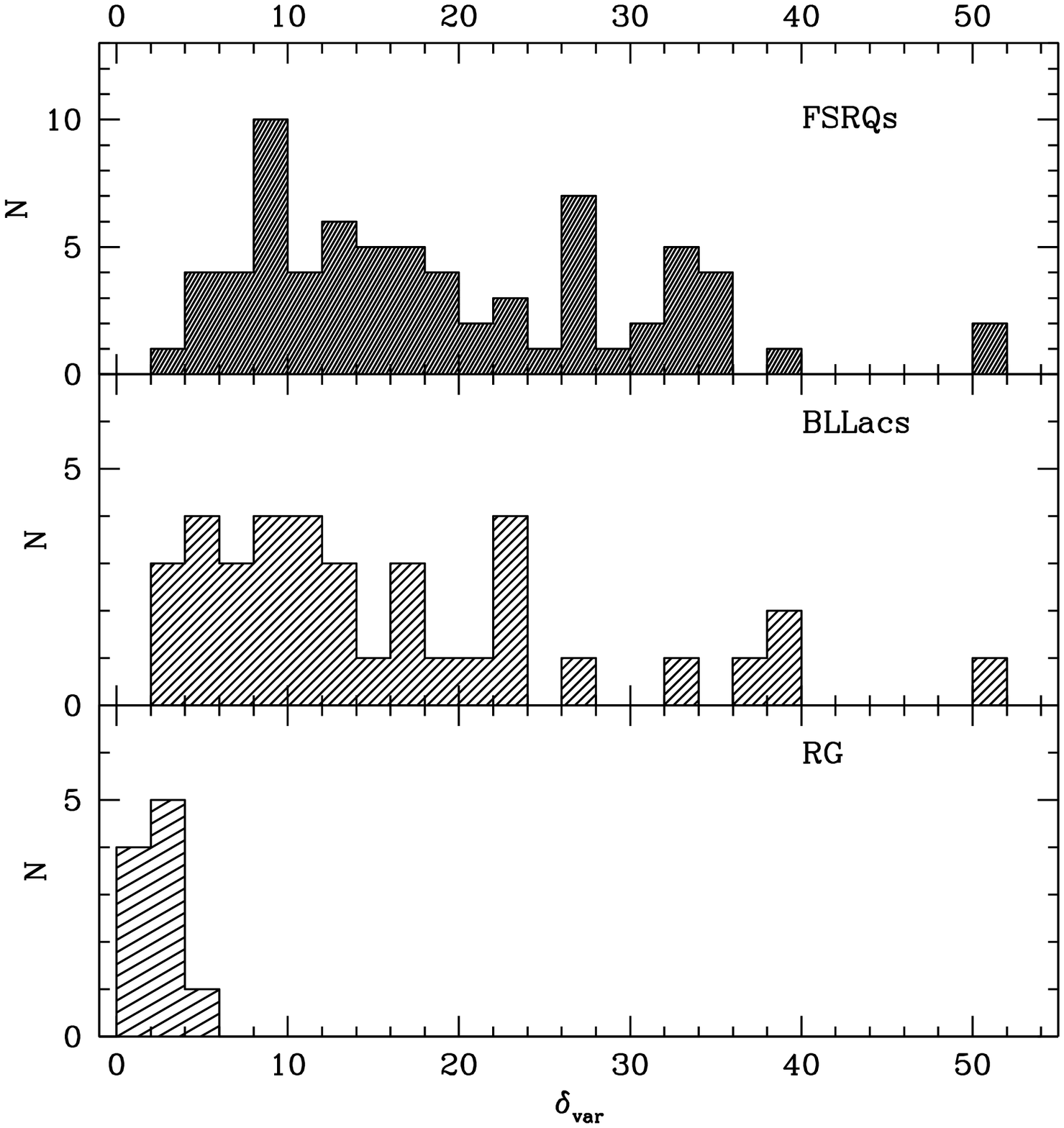}{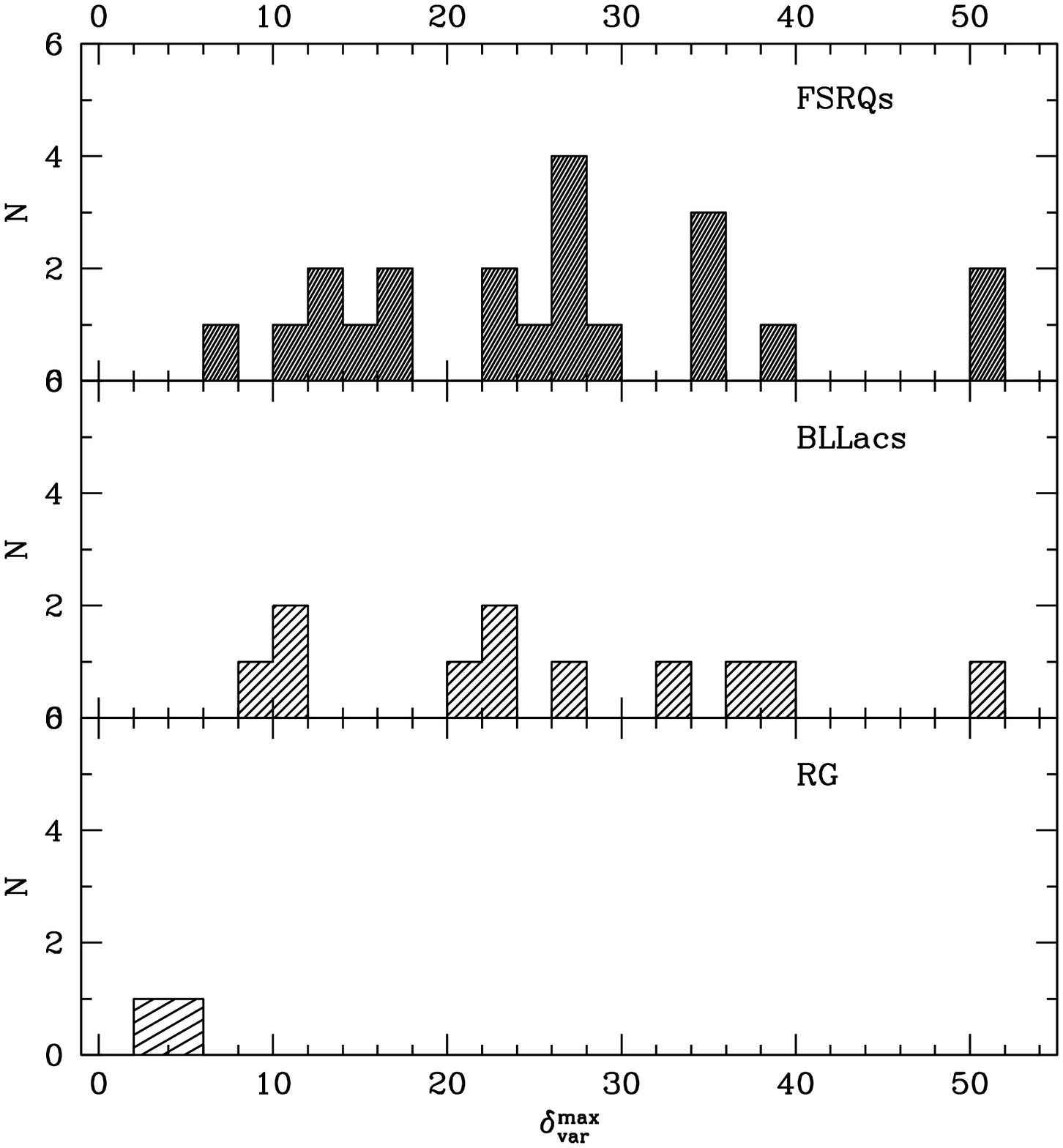}
\caption{{\it Left:} Distributions of variability Doppler factors in the FSRQs (top), BLLacs (middle), and 
RGs (bottom), derived from the light curves of superluminal knots. {\it Right:} Distributions of Doppler factors, with each source represented by the maximum value of
$\delta_{\rm var}$ of its knots. 
The interval 50$<\delta_{\rm var}\le$52  includes all cases with $\delta_{\rm var}\ge$50.} 
\label{Doppler}
\end{figure*}

Figure~\ref{Doppler}, {\it left} displays the distributions of Doppler factors derived for knots
in the FSRQs, BLLacs, and RGs. Values of $\delta_{\rm var}$ for knots in the FSRQs and BLLacs are 
distributed mostly between 2 and 40,  although there are 2 knots in the FSRQs and 1 knot in the BLLacs 
with $\delta_{\rm var}>$50. The K-S test gives a probability of 30\% that the distributions of Doppler factors in the FSRQs and BLLacs could be drawn from the same parent population. The distribution of the FSRQs possesses a more prominent ``tail'' of 
high Doppler factors than that of the BLLacs ($\sim$39\% of the FSRQ knots have $\delta_{\rm var}>$20, while such high Doppler factors occur in only $\sim$28\% of the BLLac knots). This is consistent with more intense (higher $T_{\rm b,obs}$) jet
features in the FSRQs (Figure~\ref{figTB3}, {\rm left}) and higher values of their relative flux density distributions
(Figure~\ref{Size}, {\rm left}) with respect to the properties of moving knots in the BLLacs.
The distribution for the RGs peaks at $\delta_{\rm var}\sim$3. Figure~\ref{Doppler}, {\it right} plots
the distributions of Doppler factors when each source is represented by the maximum value of $\delta_{\rm var}$.
The distribution for the FSRQs does not extend below $\delta_{\rm var}=6$ and peaks at $\delta_{\rm var}\sim$27, 
with the highest value ($\sim$60) found for 1510$-$089. The same distribution for the BLLacs ranges from $\sim8.5$ (0735+178)
to $\sim60$ (0235+164), without a significant peak and with a median of 23. The K-S test gives a 
higher probability, 72\%, that the distributions are similar.
 
\subsection{Lorentz Factor and Viewing Angle of Jet Components}

\begin{figure*}
\plottwo{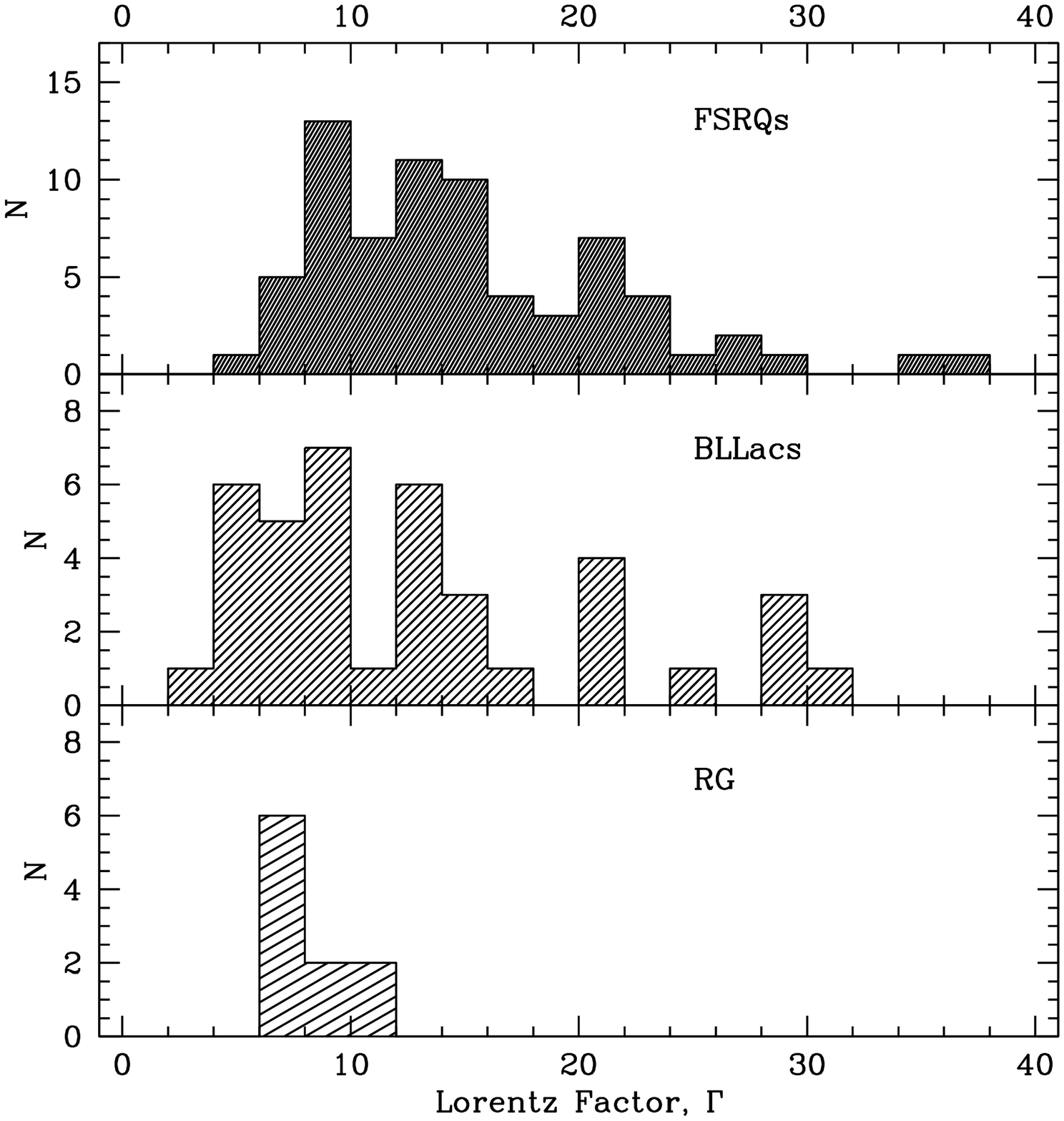}{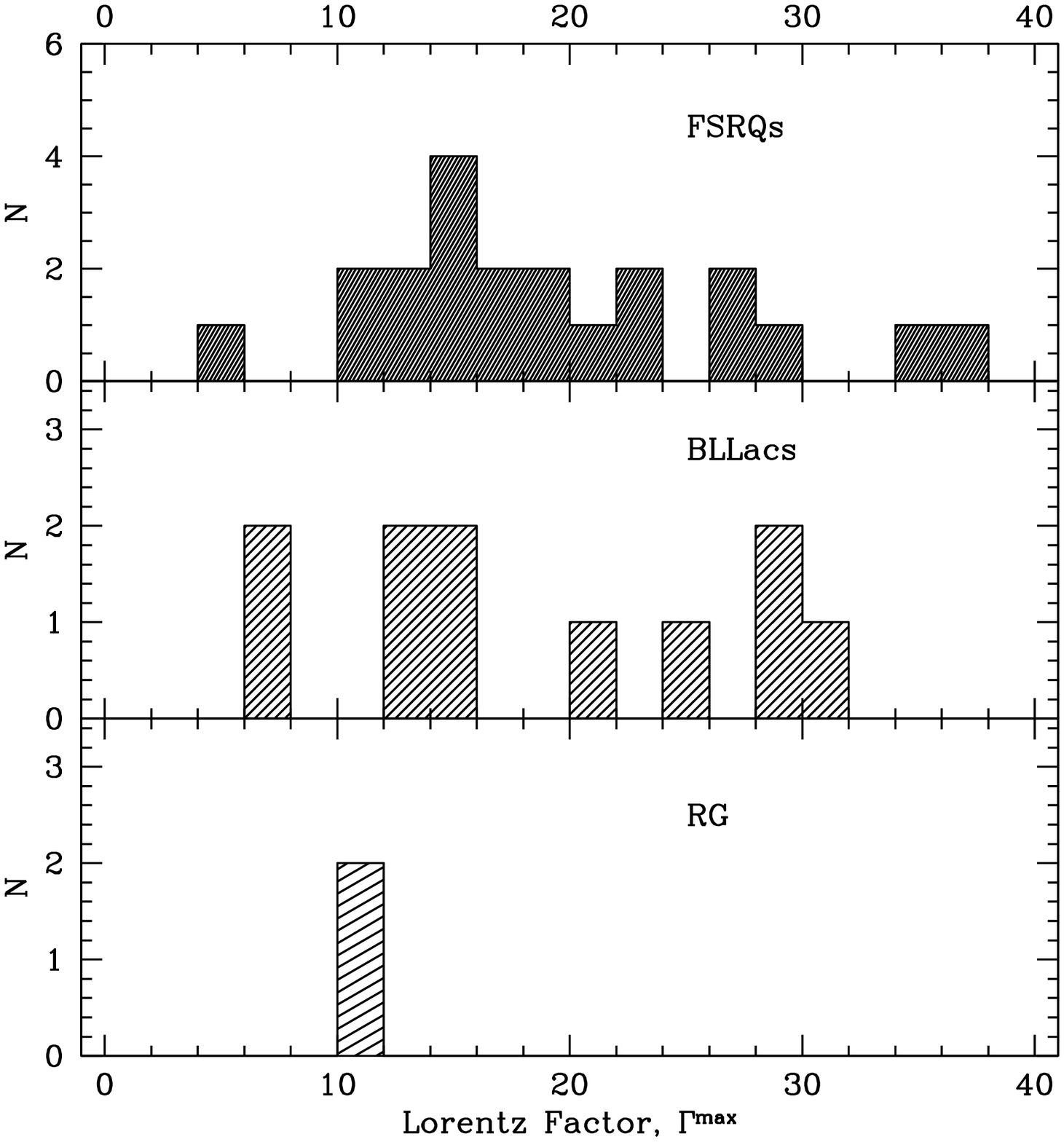}
\caption{{\it Left:} Distributions of Lorentz factors in the FSRQs (top), BLLacs (middle), and RGs (bottom), derived from apparent speeds and variability Doppler factors of superluminal knots. {\it Right:} Distributions of Lorentz factors, with each source represented by the maximum value of $\Gamma$ of its knots.} 
\label{Lorentz}
\end{figure*}

The apparent speed and Doppler factor of a jet component are functions of the Lorentz factor of the knot, $\Gamma$, and its position angle with respect to the line of sight, $\Theta_\circ$:
\begin{equation}\label{Eapp}
     \beta_{\rm app}=\frac{\beta\sin\Theta_\circ}{1-\beta\cos\Theta_\circ};~~~~~~~~~~
     \delta =\frac{1}{\Gamma(1-\beta\cos\Theta_\circ)},
\end{equation}
where $\beta=\sqrt{1-\Gamma^{-2}}$. Using trajectories and light curves of individual knots, we
derive $\beta_{\rm app}$ and $\delta_{\rm var}$, which allows us to solve for
$\Gamma$ and $\Theta_\circ$ independently as follows:
\begin{equation}\label{Egam}
     \Gamma =\frac{\beta_{\rm app}^2+\delta_{\rm var}^2 +1}{2\;\delta_{\rm var}};~~~~~~~
     \tan\;\Theta_\circ =\frac{2\;\beta_{\rm app}}{\beta_{\rm app}^2+\delta_{\rm var}^2 -1}.
\end{equation}
Table~\ref{PhysParm} presents the values of $\Gamma$ and $\Theta_\circ$ for all knots in the
reliable sample. 
Figure~\ref{Lorentz}, {\it left} shows distributions of Lorentz factors of the FSRQ, BLLac, and RG
knots. The distributions for the FSRQs and BLLacs have similar positions of peaks: a global peak
at $\Gamma\sim$9, a secondary peak at $\Gamma\sim$13, and a third peak at $\Gamma\sim$21. The K-S test gives a probability of $\sim$82\% that the samples are drawn from the same distribution. The distribution for knots in the RGs peaks at $\Gamma\sim$7, which is defined by the Lorentz factors of knots in the radio galaxy 3C~111. Figure~\ref{Lorentz},  {\it right} presents distributions of Lorentz factors when each source is represented by the maximum value of $\Gamma$ if several knots are observed in the same source. This distribution for the FSRQs ranges from 4 to 38, with a median of 17. The same distribution for the BLLacs
has a range of 6 to 32 and median of 15. The K-S test does not give a clear conclusion about the similarity of the distributions, since it yields a probability of 56\% that they are comparable. The radio galaxies 3C~111 and 3C~120 have
the same $\Gamma^{\rm max}\sim$11.

\begin{figure*}
\plottwo{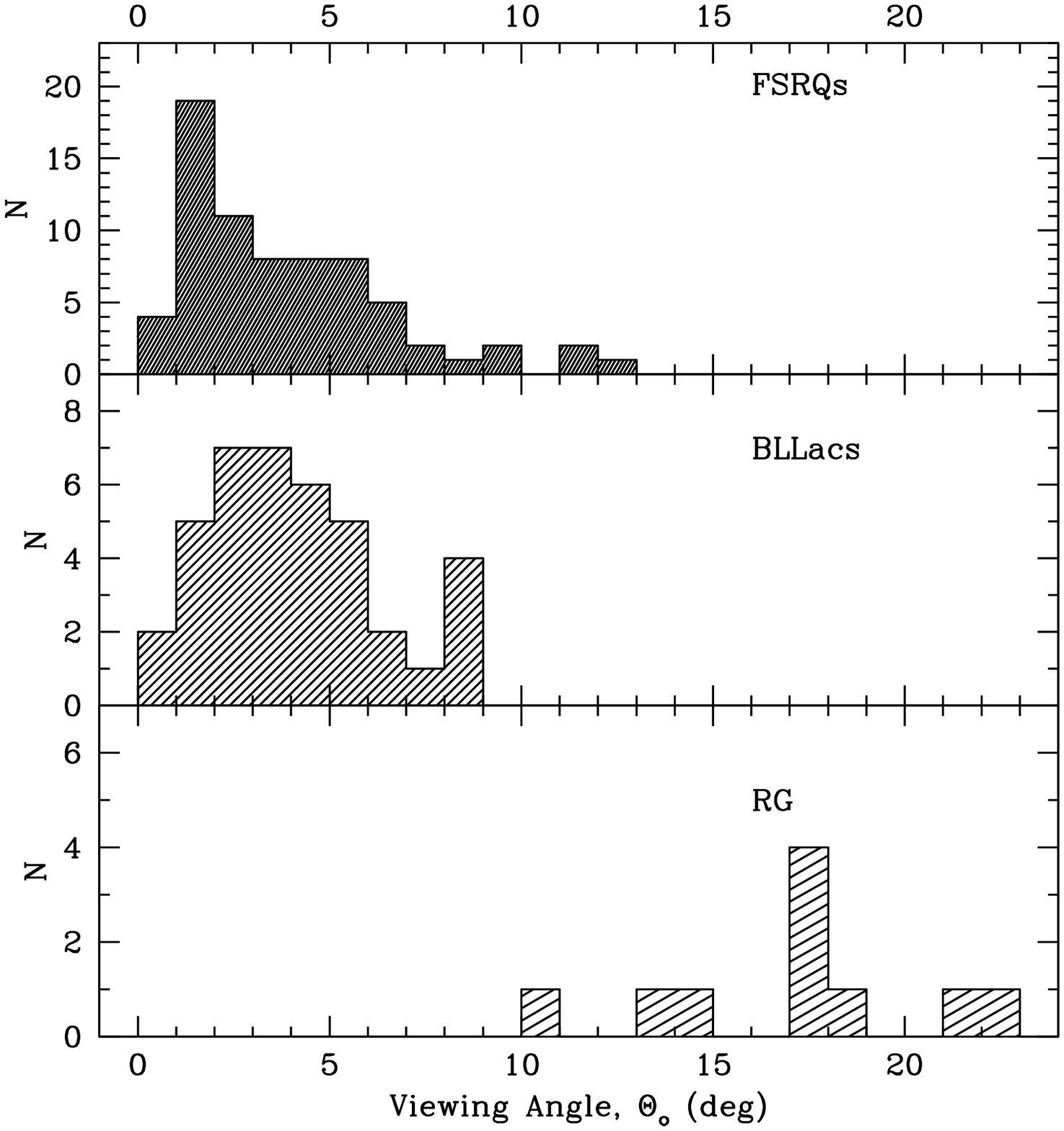}{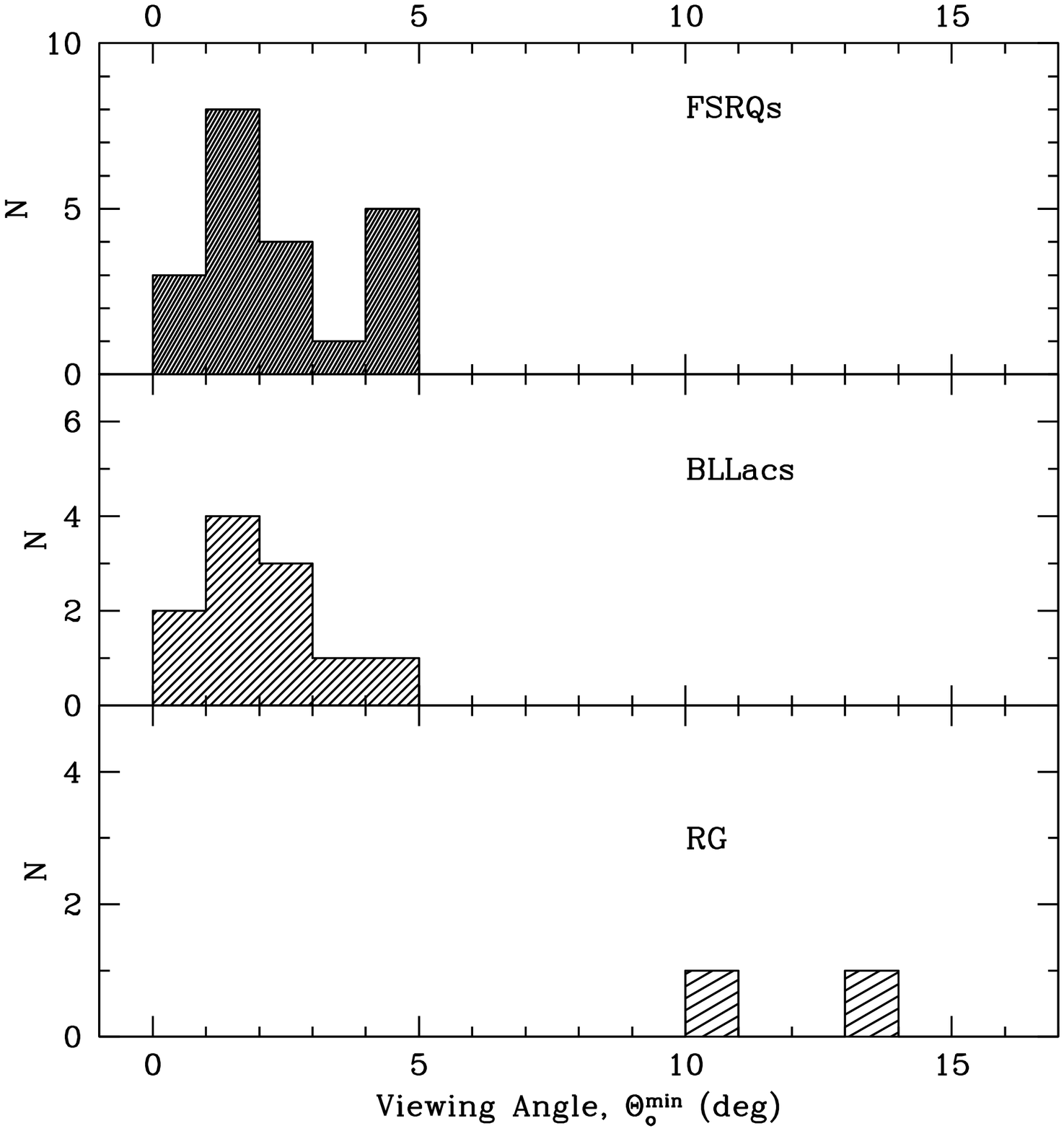}
\caption{{\it Left:} Distributions of viewing angles in the FSRQs (top), BLLacs (middle), and RGs (bottom), derived from apparent speeds and variability Doppler factors of superluminal knots. {\it Right:} Distributions of viewing angle, with each source represented by the lowest value
of $\Theta_\circ$ of its knots.} 
\label{ViewA}
\end{figure*}

Figure~\ref{ViewA}, {\it left} plots distributions of viewing angles for knots in the FSRQs, BLLacs, and RGs. The jets in the BLLacs appear to have larger values of $\Theta_\circ$ with respect to those of the FSRQs. The distribution for the FSRQs has a prominent maximum at $\Theta_\circ\sim$1.5$^\circ$, while the distribution for the BLLacs peaks between 2$^\circ$ and 3$^\circ$, although in both samples $\sim$70\% of the features have $\Theta_\circ<$5$^\circ$. The viewing angles of all knots in the RGs exceed 10$^\circ$. According to the K-S test, the probability that the distributions for the FSRQs and BLLacs are the same is $\sim$61\%. The distributions of the smallest viewing angles, if there are multiple reliable knots, for both the FSRQs and BLLacs jets range from nearly zero to
$5^\circ$ and peak around $1^\circ$ (Figure \ref{ViewA}, {\it right}). The probability provided by the K-S test increases to 99.9\% that the distributions are drawn from the same population. 

Table~\ref{PhysParm} presents the results of this analysis: 1 - name of the source; 2 - designation
of the knot;
3 - timescale of variability of the flux density of the knot, $\tau_{\rm var}$ and its uncertainty;
4 - average size of the knot, $\langle a\rangle$ and its standard deviation;
5 - number of measurements employed to calculate $\tau_{\rm var}$ and $\langle a\rangle$, $N_t$;
6 - variability Doppler factor of the knot, $\delta_{\rm var}$, and its uncertainty;
7 - bulk Lorentz factor of the knot, $\Gamma$ and its uncertainty; and
8 - intrinsic viewing angle of the knot, $\Theta_\circ$ and its uncertainty. 

\begin{deluxetable*}{rcrrrrrr}
\singlespace
\tablecolumns{8}
\tablecaption{\small\bf Physical Parameters of Jet Features \label{PhysParm}}
\tabletypesize{\footnotesize}        
\tablehead{
\colhead{Source}&\colhead{Knot}&\colhead{$\tau_{\rm var}$, yr}&\colhead{$\langle a\rangle$, mas}&\colhead{$N_t$}&\colhead{$\delta_{\rm var}$}&\colhead{$\Gamma$}&\colhead{$\Theta_\circ$, deg}\\
\colhead{(1)}&\colhead{(2)}&\colhead{(3)}&\colhead{(4)}&\colhead{(5)}&\colhead{(6)}&\colhead{(7)}&\colhead{(8)}
}
\startdata 
0219+428&  B1&    0.314$\pm$ 0.066&  0.27$\pm$ 0.07&    6&  37.0$\pm$  8.7&  29.3$\pm$  4.2&   1.5 $\pm$ 0.2\\
        &  B2&    1.281$\pm$ 0.477&  0.38$\pm$ 0.08&   10&  12.8$\pm$  3.7&  12.8$\pm$  2.1&   4.5$\pm$  0.7\\
0235+164&  B1&    0.420$\pm$ 0.171&  0.21$\pm$ 0.09&    6&  39.9$\pm$ 16.7&  28.6$\pm$  6.9&   1.3$\pm$  0.3\\
        &  B2&    0.201$\pm$ 0.014&  0.15$\pm$ 0.05&    9&  59.5$\pm$ 12.0&  31.3$\pm$  4.9&   0.4$\pm$  0.1\\
0336-019&  B1&    1.876$\pm$ 0.830&  0.40$\pm$ 0.13&   11&  15.8$\pm$  6.9&  34.7$\pm$  8.2&   3.0$\pm$  0.7\\
        &  B2&    0.490$\pm$ 0.073&  0.16$\pm$ 0.08&    8&  24.2$\pm$  7.8&  22.5$\pm$  4.1&   2.4$\pm$  0.4\\
        &  B3&    0.931$\pm$ 0.355&  0.14$\pm$ 0.06&    7&  11.1$\pm$  5.7&   9.1$\pm$  2.6&   5.0$\pm$  1.4\\             
\enddata
\vspace{3mm}
The table is available entirely in a machine-readable format in the online journal (Jorstad et al. 2017, ApJ, 846,98).
\end{deluxetable*}

\subsection{Average Jet Kinematic Parameters and Opening Angles of Jets}
We have estimated the average physical parameters, $\langle\delta\rangle, \langle\Gamma\rangle$, and $\langle\Theta_\circ\rangle$, for each jet in the reliable sample. For sources with multiple superluminal components, these are weighted averages of corresponding values of the components, with the weights inversely proportional to the uncertainties of the values. The uncertainties of the average parameters are calculated as weighted standard deviations of the average values. For sources with a single reliable component, the average physical parameters are equal to the parameters of that knot. Table~\ref{ParAve} lists the resulting average values, with the number of knots used to calculate the average values indicated. 

\begin{figure}
\plotone{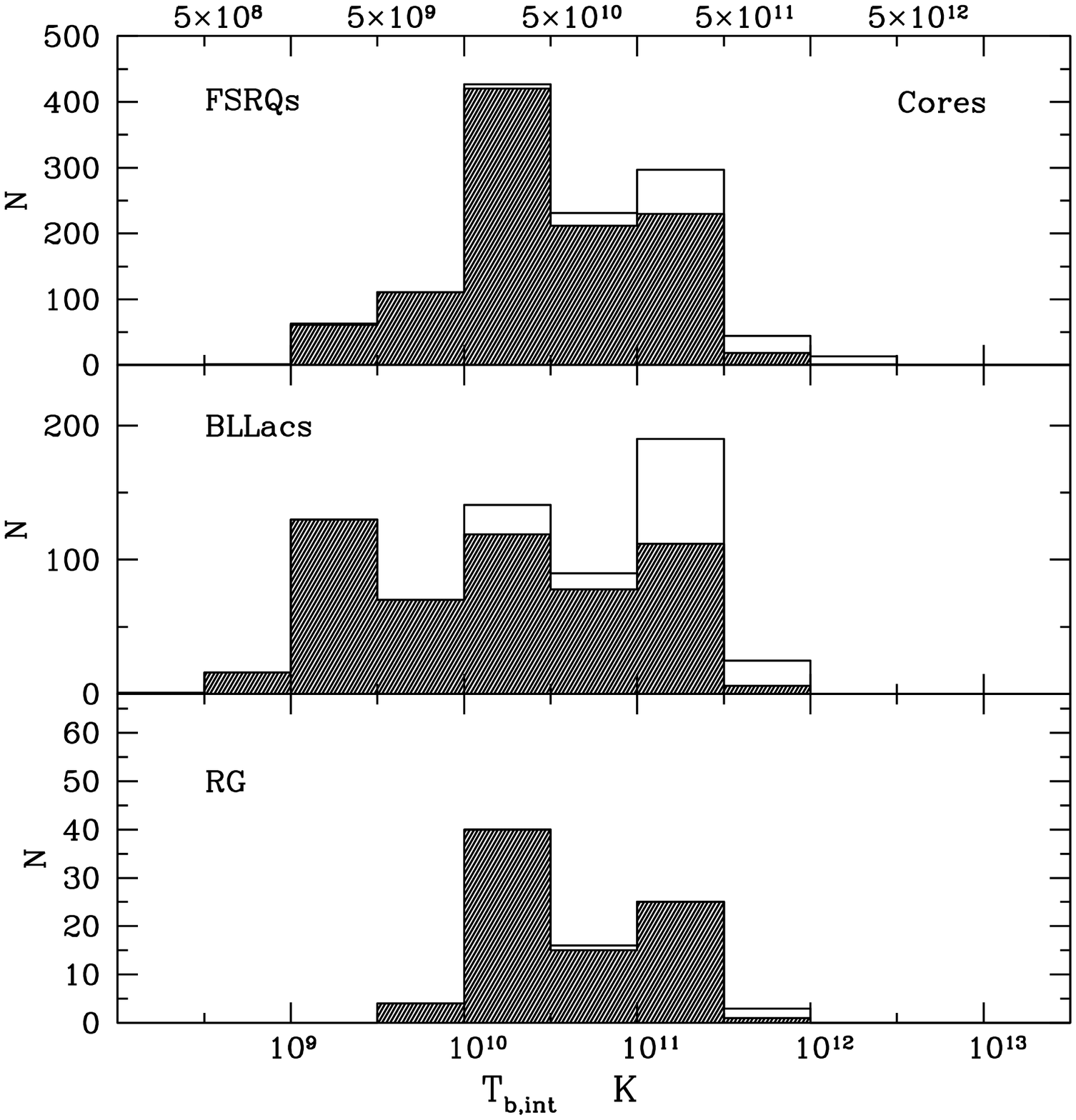}
\caption{{\it Left:} Distributions of intrinsic brightness temperatures of the cores at all epochs in the FSRQs ({\it top}), BLLacs ({\it middle}), 
and RGs ({\it bottom}); unshaded parts of the distributions represent lower limits to $T_{\rm b,int}$; the scale at the x axis is neither linear nor 
logarithmic.}
\label{figTBI}
\end{figure}

We have calculated values of the intrinsic brightness temperature of cores, $T_{\rm b,int}=T_{\rm b,obs}(1+z)^{1+\alpha}/\delta^{1+\alpha}$, where $\alpha$ is the spectral
index ($S_\nu\propto\nu^{-\alpha}$), using $\langle\delta\rangle$ as $\delta$ and $\alpha\approx$0. Figure~\ref{figTBI}
shows the distributions of $T_{\rm b,int}^{\rm core}$ for different sub-classes at all epochs. The distributions 
for the FSRQs and RGs peak at $T_{\rm b,int}\sim T_{\rm b,eq}$, with a small percentage of $T_{\rm b,int}^{\rm core}<10^{10}$
(14.7\% in the FSRQs and 4.5\% in the RGs). Although the distribution of $T_{\rm b,int}^{\rm core}$ for the BLLacs peaks at a higher temperature than that of the FSRQ and RG distributions, it possesses the largest percentage of cores with $T_{\rm b,int}^{\rm core}<10^{10}$ (32.7\%). The derived maximum intrinsic brightness temperature of the core for each source is given in Table~\ref{ParAve}. Except for several sources, $T_{\rm b,int}^{\rm max}$
of the cores exceeds $T_{\rm b,eq}$ by a factor of 10, with several extreme cases, for which $T_{\rm b,int}^{\rm max}>$100$T_{\rm b,eq}$: 3C~273, BLLac, 3C~454.3, and 3C~111. All these cases are associated with strong multi-wavelength activities of the sources
(see \S Appendix~\ref{Notes}). The intrinsic brightness temperature in the unresolved core 
of BL~Lacertae $>$3$\times$10$^{12}$~K was obtained by \cite{JL16} with VLBI including {\it RadioAstron} space observations at 22~GHz, which supports our findings of very high intrinsic brightness temperatures in VLBI cores. However, these extreme brightness temperatures are most likely transient events rather than persistent conditions according to Figure~\ref{figTBI}. The results presented in Figure~\ref{figTBI} suggest that cores of the RG maintain the equipartition conditions most of the time, while for the FSRQs and BLLacs $\sim$30\% of $T_{\rm b,int}^{\rm core}>5\times 10^{11}$~K. This argues in favor of the presence of very bright and compact features in blazar jets and possible departure from equipartition of energy between the magnetic field and radiating particles in the VLBI core at some epochs.

We have determined the intrinsic opening semi-angle of the jet, $\theta_\circ$, for each source in the reliable sample in two different manners. Both methods assume that, over the period of monitoring presented here,
we have detected a sufficient number of features in each jet, the combination of which covers the entire opening angle of the jet. The first method, A, is based on the relation between the intrinsic and projected opening angles used by J05:
\begin{equation}\label{EopA}
  \theta_\circ=\theta_{\rm p}\sin\Theta_\circ.   
\end{equation} 

We employ the standard deviation of position angles, $\Theta$, of all knots in the jet with respect to the average projected jet direction, $\Theta_{\rm A0}$, to represent the projected opening semi-angle of the jet as $\theta_{\rm p}$=2$\sigma\langle\Theta\rangle_{\rm A0}$, and use the mean of the uncertainties of $\langle\Theta\rangle$ of the knots as the uncertainty of $\theta_{\rm p}$. Values of $\sigma\Theta_{\rm A0}$ are given in Table~\ref{Parm} for all sources in the sample except 0235+164, for which the projected direction of the jet is undetermined. The second method, B, is applied only to the jets with multiple reliable knots, for which Table~\ref{PhysParm} lists a range of viewing angles. We choose the minimum and maximum viewing angles for each jet, $\Theta_\circ^{\rm min}$ and $\Theta_\circ^{\rm max}$, respectively, and calculate the intrinsic opening semi-angle as $\theta_\circ=(\Theta_\circ^{\rm max}-\Theta_\circ^{\rm min})/2$.  Table~\ref{ParAve} gives the physical jet parameters for each source in our sample with reliable superluminal knots: 1 - name of the source; 2 - luminosity distance, $D$; 3 - projected opening angle
$\theta_{\rm p}$, and its uncertainty; 3 - Doppler factor,
$\langle\delta\rangle$ and its standard deviation; 4 - Lorentz factor, $\langle\Gamma\rangle$, and its standard deviation; 5 - viewing angle, $\langle\Theta_\circ\rangle$, and its standard deviation;
6 - maximum intrinsic brightness temperature of the core, $T_{\rm b,int}^{\rm core}$, letter ``L'' near the value designates
a lower limit;
7 - projected opening semi-angle, $\theta_{\rm p}$, and its uncertainty;
8 - opening semi-angle of the jet, $\theta_\circ^{\rm A}$, derived using method A, and its uncertainty; 9 - opening semi-angle of the jet, $\theta_\circ^{\rm B}$, derived using method B, and its uncertainty; 
and 10 - number of knots, $N_k$, used in the calculation of the average parameters; in the case of $N_k=$1 the parameters of the jet corresponds to the parameters of a knot. 

\begin{deluxetable*}{lcrrrrrrrr}
\singlespace
\tablecolumns{10}
\tablecaption{\small\bf Physical Parameters of Jets\label{ParAve}}
\tabletypesize{\footnotesize}        
\tablehead{
\colhead{Source}&\colhead{D}&\colhead{$\langle\delta\rangle$}&\colhead{$\langle\Gamma\rangle$}&\colhead{$\langle\Theta_\circ\rangle$}&\colhead{T$_{\rm b,int}^{\rm core}$}&\colhead{$\theta_{\rm p}$}&\colhead{$\theta_\circ^{\rm A}$}&\colhead{$\theta_\circ^{\rm B}$}&\colhead{$N_k$}\\
\colhead{}&\colhead{Gpc}&\colhead{}&\colhead{}&\colhead{deg}&\colhead{$10^{10}$K}&\colhead{deg}&\colhead{deg}&\colhead{deg}&\colhead{}\\
\colhead{(1)}&\colhead{(2)}&\colhead{(3)}&\colhead{(4)}&\colhead{(5)}&\colhead{(6)}&\colhead{(7)}&\colhead{(8)}&\colhead{(9)}&\colhead{(10)}
}
\startdata 
0219+428&  2.458&  16.5$\pm$  5.2&  16.1$\pm$  4.5&   1.7$\pm$  0.5&4.24L&15.2$\pm$5.5&0.5$\pm$  0.4&  1.5$\pm$  0.5&2\\
0235+164&  6.121&  52.8$\pm$  8.4&  30.4$\pm$  3.1&   0.5$\pm$  0.3&9.27L&\nodata& \nodata& 0.5$\pm$  0.4&2 \\
0316+413$^*$&  0.077&$\sim$11  &$\sim$7&$\sim$4.4&6.20&22.1$\pm$6.4&$\sim$1.7&\nodata&0 \\  
0336$-$019&  5.422&  15.7$\pm$  4.9&  14.4$\pm$  6.0&   2.7$\pm$  0.3&15.7L&7.2$\pm$5.3&0.3$\pm$  0.2&  1.3$\pm$  0.9& 3\\
0415+379&   0.215&  2.0$\pm$  0.5&   7.7$\pm$  0.7&  16.7$\pm$  2.3&126L&8.0$\pm$3.6&2.3$\pm$  0.6&    4.4$\pm$  3.1&9\\
0420$-$014& 5.928&  16.9$\pm$  7.4&  13.6$\pm$  3.0&   1.5$\pm$  0.3&78.9&23.4$\pm$7.2&0.6$\pm$  0.2&  2.2$\pm$  1.1&6\\
0430+052&   0.145&  4.5$\pm$  2.0&  10.7$\pm$  2.4&  10.4$\pm$  2.3&29.0&6.6$\pm$4.4&1.2$\pm$  0.5&    \nodata&1 \\
0528+134&  16.109&  19.5$\pm$  4.3&  12.6$\pm$  3.1&   1.7$\pm$  0.5&80.7&22.8$\pm$7.1&0.7$\pm$  0.4&    0.9$\pm$  0.5&3\\
0716+714&  1.553&  15.6$\pm$  4.0&  14.0$\pm$  3.7&   3.0$\pm$  0.4&65.8&26.6$\pm$9.1&1.4$\pm$  0.3&    1.8$\pm$  1.0&6\\
0735+178&  2.327&   8.5$\pm$  4.0&   5.6$\pm$  1.4&   5.7$\pm$  1.4&8.23&52.2$\pm$9.2&5.2$\pm$  1.1&    \nodata&1\\
0827+243&  6.129& 22.8$\pm$ 8.5&  15.4$\pm$  4.1&   1.2$\pm$  0.2&29.7&24.0$\pm$9.1&0.6$\pm$  0.3 &   1.4$\pm$  0.6& 4 \\
0829+046&  0.840&  13.3$\pm$  1.3&   8.1$\pm$  1.7&   1.4$\pm$  0.4&9.45&7.9$\pm$7.4&0.2$\pm$  0.3&    0.8$\pm$  0.6&2\\
0836+710& 17.185&   15.6$\pm$  1.5&  17.0$\pm$  2.2&   2.9$\pm$  0.9&15.8&6.8$\pm$4.6&0.3$\pm$  0.2&    1.0$\pm$  0.6&3\\
0851+202&  1.589&   8.7$\pm$  2.1&   6.1$\pm$  1.2&   2.6$\pm$  1.4&76.8&33.0$\pm$9.9&1.5$\pm$  0.7&    3.8$\pm$  1.0&7\\
0954+658&  1.969&   8.6$\pm$  2.5&  11.4$\pm$  3.1&   1.5$\pm$  0.7&28.4&21.0$\pm$9.1&0.5$\pm$  0.6&    3.0$\pm$  0.8&7\\
1055+018&  5.722&  20.0$\pm$  8.0&  15.0$\pm$  5.2&   2.7$\pm$  0.9&41.0&24.8$\pm$8.8&1.2$\pm$  0.6 &   \nodata&1 \\
1101+384$^*$&0.131&$\sim$24&$\sim$13&$\sim$1.0&1.50&55.2$\pm$11.0&$\sim$0.9&\nodata&0 \\     
1127$-$145&  8.142& 20.7$\pm$  1.9&  12.6$\pm$  2.0&   1.5$\pm$  0.3&22.9&15.4$\pm$4.0&0.4$\pm$  0.2&   0.4$\pm$  0.5&2 \\
1156+295&  4.446&  11.8$\pm$  2.8&  10.4$\pm$  2.7&   1.0$\pm$  0.2&39.3&13.6$\pm$6.1&0.2$\pm$  0.2&     2.5$\pm$  0.5&3 \\
1219+285& 0.470&   9.6$\pm$  2.3&   6.0$\pm$  0.9&   4.8$\pm$  0.7&6.13&9.2$\pm$3.9&0.8$\pm$  0.5&     \nodata&1 \\
1222+216& 2.379&    7.4$\pm$  2.1&  13.9$\pm$  2.1&   5.6$\pm$  1.0&46.0&16.2$\pm$7.2&1.6$\pm$  0.6 &    1.5$\pm$  1.0&3\\
1226+023&  0.755&  4.3$\pm$  1.3&   8.5$\pm$  2.2&   6.4$\pm$  2.4&325&6.6$\pm$3.5  &0.7$\pm$  0.5&     4.0$\pm$  1.3&8\\
1253$-$055& 3.080&   18.3$\pm$  1.9&  13.3$\pm$  0.6&   1.9$\pm$  0.6&82.9&47.4$\pm$10.4&1.6$\pm$  1.0&   2.3$\pm$  2.2&3 \\
1308+326&   6.591& 20.9$\pm$  1.2&  13.2$\pm$  1.1&   1.9$\pm$  0.4&8.91&58.4$\pm$8.3&1.9$\pm$0.8&     0.6$\pm$  0.4&2\\
1406$-$076&  10.854&  11.1$\pm$  1.6&  9.7$\pm$  2.1&4.8$\pm$  0.6&20.7&16.4$\pm$5.1   &1.4$\pm$  0.4&   0.7$\pm$  1.3&2\\
1510$-$089&  1.919& 35.3$\pm$  4.6&  22.5$\pm$  3.3& 1.2$\pm$  0.3&14.1&11.4$\pm$4.3   &0.3$\pm$  0.2&   0.5$\pm$  0.3&5\\
1611+343&  10.009&   7.5$\pm$  2.6&   4.1$\pm$  1.0&   4.0$\pm$  1.0&12.3&20.8$\pm$5.8 &1.5$\pm$  0.7&     \nodata&1 \\
1622$-$297& 5.133&   9.5$\pm$  1.6&   7.8$\pm$  1.5&   4.9$\pm$  1.1&21.8&30.8$\pm$13.6&1.0$\pm$  0.4&   1.4$\pm$  2.1&3 \\
1633+382&  13.775&  12.4$\pm$  3.7&   8.2$\pm$  1.0&   2.6$\pm$  1.0&72.3&41.2$\pm$5.2 &1.9$\pm$ 1.4&     2.1$\pm$  0.9&3 \\
1641+399&  3.480& 11.9$\pm$  3.0&  10.8$\pm$  0.8&   3.7$\pm$  1.0&42.6&18.6$\pm$5.0   &1.2$\pm$  0.6 &    2.1$\pm$  0.9&4 \\
1730$-$130& 5.817&    7.9$\pm$  1.1&  8.9$\pm$  2.5&   3.0$\pm$  1.2&83.3&16.2$\pm$5.3 &0.8$\pm$  0.7&   3.2$\pm$ 1.4&3 \\
1749+096&  1.685&  17.7$\pm$  7.7&  11.0$\pm$  3.6&   2.4$\pm$  1.0&45.0&26.2$\pm$11.3 &1.1$\pm$  0.7&     1.8$\pm$  1.0&4 \\
2200+420&  0.311&   7.5$\pm$  1.3&   5.6$\pm$  1.4&   5.1$\pm$  2.3&98.1L&6.0$\pm$3.7 &0.5$\pm$  0.5&     3.4$\pm$  0.9&6 \\
2223$-$052&  10.053& 13.6$\pm$  3.1&  10.0$\pm$  1.4&   3.3$\pm$  0.4&41.7&22.0$\pm$10.7 &1.3$\pm$  0.3&   1.2$\pm$  1.1&2 \\
2230+114&   6.911& 30.5$\pm$  3.3&  21.7$\pm$  1.3&   1.6$\pm$  0.4&38.7&23.8$\pm$8.3&0.7$\pm$  0.3 &    0.5$\pm$  0.3&4 \\
2251+158&  5.477&  24.4$\pm$  3.7&  13.8$\pm$  1.4 &  0.7$\pm$  0.4&165&22.6$\pm$9.4&0.4$\pm$  0.3 &    0.9$\pm$  0.6&4 \\
\enddata
\vspace{3mm}
$^*$ - estimates of parameters are discussed in \S~\ref{subclass}.
\end{deluxetable*}

Comparison between $\theta_\circ^{\rm A}$ and $\theta_\circ^{\rm B}$ values shows that, in general, the values are consistent 
within the uncertainties, especially for sources with $N_k\ge$3, although there are several cases (e.g., 1156+295, 3C273, and BLLac) for which
$\theta_\circ^{\rm B}>\theta_\circ^{\rm A}$.
This may result from $\theta_\circ^{\rm A}$ being
underestimated, a possibility that we plan to test through further observations.
We compare the results given in Table~\ref{ParAve} with those obtained by J05 for 13 sources
common to both samples. The parameters are in good agreement within the uncertainties, except for 0528+134 and OJ~287, for which Table 11 in J05 lists higher values of $\langle\delta\rangle$ and $\langle\Gamma\rangle$. These discrepancies are very interesting, since the observations occurred during quite different stages of jet activity. Throughout the J05 epochs, 0528+134 was significantly more active than during the more recent monitoring reported here, while OJ~287 became more active after its
inner jet executed a dramatic change in projected direction around 2005 \citep{IVAN12}. Such changes
illustrate the need for continued monitoring of blazars with the VLBA to sample a range of kinematic
parameters during different activity states at different wavelengths to understand their range of variability in structure and kinematics.

\begin{figure*}
\plottwo{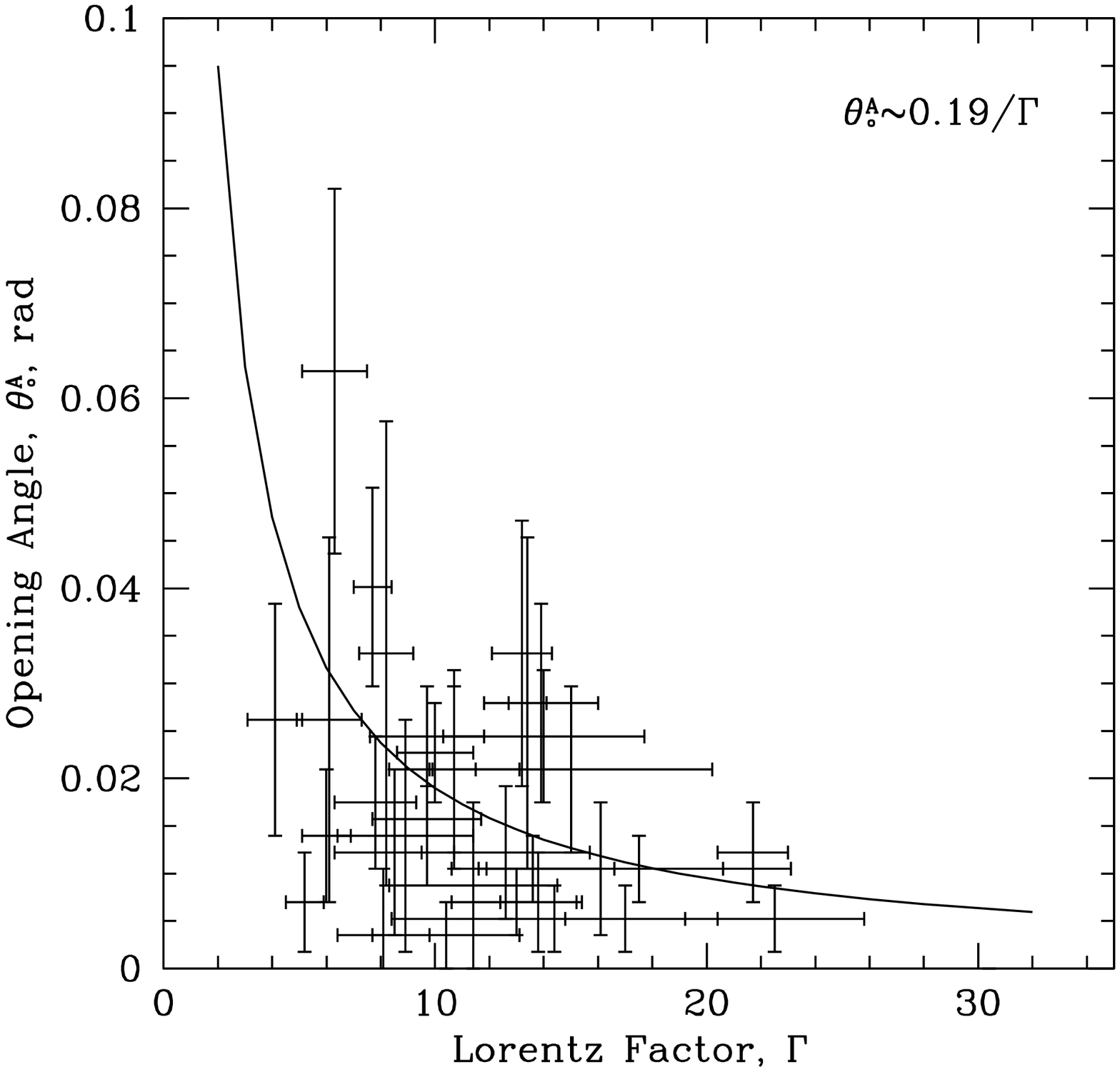}{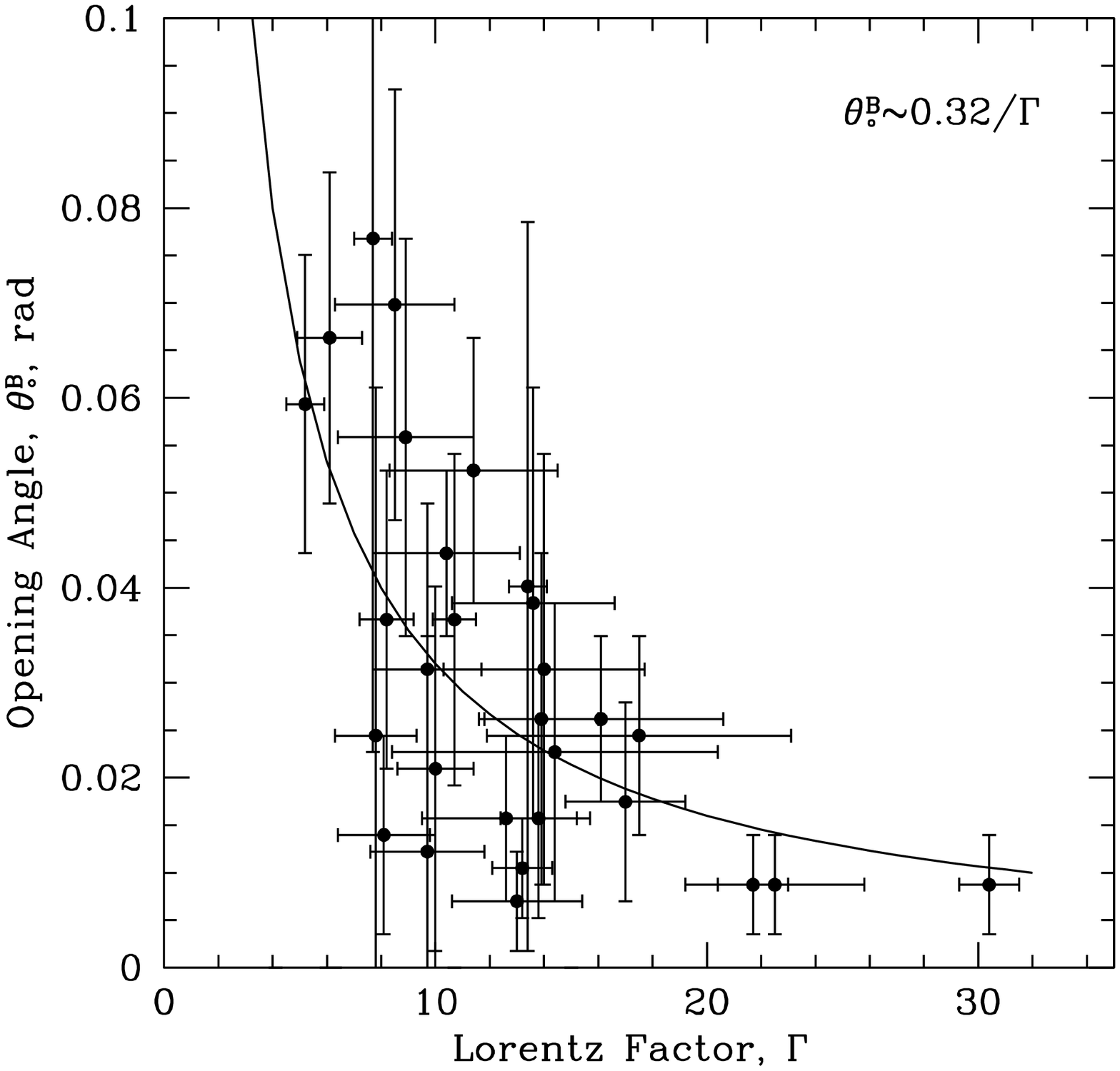}
\caption{Intrinsic opening semi-angle vs.\ Lorentz factor for $\theta_\circ^{\rm A}$ ({\it left}) and $\theta_\circ^{\rm B}$ ({\it right}). The solid curves represent the best fit to the relation
$\theta_\circ= const/\Gamma$.}
\label{OpenA}
\end{figure*}

Several studies \citep[J05,][]{PUSH09,CB13}, which use different samples of AGN with relativistic jets and different methods to estimate jet parameters, show that the product $\Gamma\theta_\circ$ can be approximated by a constant, $\rho$, implying a physical connection between the two jet parameters.
Such a connection is in fact expected in standard theories of the formation, acceleration, and 
collimation of relativistic jets \citep [e.g.,][]{BK79,Sasha09}. Figure~\ref{OpenA} plots $\theta_\circ$ versus $\Gamma$, with $\theta_\circ$ computed by different methods, A (left) and B (right).
Analysis of the correlation between $\theta_\circ^{\rm A}$ and $\theta_\circ^{\rm B}$ with 1/$\Gamma$
gives correlation coefficients, $\varrho$, equal to 0.40 and 0.60, respectively. These values reject
the null hypothesis of zero correlation at a significance level $\zeta$=0.05, with $f_{\rm A}$=31
 and $f_{\rm B}$=27 degrees of freedom, respectively,
according to the {\it t}-test. We use a gradient-expansion algorithm to compute a non-linear least squares fit to the data presented in Table~\ref{ParAve} and Figure~\ref{OpenA} for the function
$\theta_\circ$=$\rho/\Gamma$. This yields best-fit results of $\rho_{\rm A}$=0.19$\pm$0.07 and $\rho_{\rm B}$=0.32$\pm$0.13. \cite{CB13} have performed a more sophisticated analysis of the $\Gamma$-$\theta_\circ$ relation based on the MOJAVE sample, including the effects of relativistic velocity shear and Doppler beaming. They find a best-fit value of $\rho\sim$0.2, which appears to be robust independent of the method of deriving the jet parameters. In addition, \cite{CB13} conclude that the assumption that $\Gamma\theta_\circ$ is generally constant in the relativistic jets of AGN gives a new way to calculate the physical parameters of jets. However, as mentioned by these authors,  inability 
to determine $\Gamma$ and $\theta_\circ$ to high accuracy could obscure any difference in $\rho$
for different sub-classes of AGN. 

\subsection{Physical Jet Parameters of Sub-Classes of AGN}\label{subclass}

We calculate the weighted averages of the physical parameters for each sub-class in our sample using the results given in Table~\ref{ParAve}. 
\begin{equation}\label{Eave}
  \overline{\langle\delta\rangle} =\frac{\sum_{\rm i=1}^{\rm N_s}\frac{\langle\delta\rangle_{\rm i}}{\sigma^2_{\langle\delta\rangle_{\rm i}}}}{\sum_{\rm i=1}^{\rm N_s}\frac{1}{\sigma^2_{\langle\delta\rangle_{\rm i}}}};~~~~~
  \overline{\langle\Gamma\rangle} =\frac{\sum_{\rm i=1}^{\rm N_s}\frac{\langle\Gamma\rangle_{\rm i}}{\sigma^2_{\langle\Gamma\rangle_{\rm i}}}}{\sum_{i=1}^{\rm N_s}\frac{1}{\sigma^2_{\langle\Gamma\rangle_{\rm i}}}};~~~~~
  \overline{\langle\Theta_\circ\rangle} =\frac{\sum_{\rm i=1}^{\rm N_s}\frac{\langle\Theta_\circ\rangle_{\rm i}}{\sigma^2_{\langle\Theta_\circ\rangle_{\rm i}}}}{\sum_{i=1}^{\rm N_s}\frac{1}{\sigma^2_{\langle\Theta_\circ\rangle_{\rm i}}}};~~~~~
 \overline{\theta_\circ} =\frac{\sum_{\rm i=1}^{\rm N_s}\frac{\theta_\circ^{\rm i}}{\sigma^2_{\theta_\circ^{\rm i}}}}{\sum_{i=1}^{\rm N_s}\frac{1}{\sigma^2_{\theta_\circ^{\rm i}}}},
\end{equation}
where $N_s$ is the number of sources in the sub-class. For the opening semi-angle we obtain two
values, $\overline{\langle\theta_\circ\rangle}^{\rm A}$ and $\overline{\langle\theta_\circ\rangle}^{\rm B}$, corresponding to the two different
methods for estimating $\theta_\circ$. The uncertainties of the parameters are derived as weighted standard deviations with respect to the means given in eq.~\ref{Eave}, with weights inversely proportional to the squares of the uncertainties of the parameters used to compute the means. Table~\ref{ClassParm} gives the results of the calculations: 1 - name of AGN sub-class;
2 - number of sources in the sub-class, $N_s$;
3 - Doppler factor, $\overline{\langle\delta\rangle}$, and its uncertainty;
4 - Lorentz factor, $\overline{\langle\Gamma\rangle}$, and its uncertainty;
5 - viewing angle, $\overline{\langle\Theta_\circ\rangle}$, and its uncertainty;
6 - opening semi-angle of the jet, $\overline{\langle\theta_\circ\rangle}^{\rm A}$ according to 
method A and its standard deviation; 7 - opening semi-angle of a jet, $\overline{\langle\theta_\circ\rangle}^{\rm B}$ according to method B and its standard deviation. 

\begin{deluxetable}{rcrrrrr}
\singlespace
\tablecolumns{7}
\tablecaption{\small\bf Average Physical Parameters of Jets for Different Sub-Classes
\label{ClassParm}}
\tabletypesize{\footnotesize}        
\tablehead{
\colhead{Sub-Class}&\colhead{$N_s$}&\colhead{$\overline{\langle\delta\rangle}$}&\colhead{$\overline{\langle\Gamma\rangle}$}&\colhead{$\overline{\langle\Theta_\circ\rangle}$}&\colhead{$\overline{\theta_\circ^{\rm A}}$}&\colhead{$\overline{\theta_\circ^{\rm B}}$}\\
\colhead{}&\colhead{}&\colhead{}&\colhead{}&\colhead{deg}&\colhead{deg}&\colhead{deg}\\
\colhead{(1)}&\colhead{(2)}&\colhead{(3)}&\colhead{(4)}&\colhead{(5)}&\colhead{(6)}&\colhead{(7)}
}
\startdata 
FSRQs&21&13.1$\pm$6.3&11.6$\pm$3.1&1.4$\pm$0.3&0.6$\pm$0.3&1.0$\pm$0.6 \\
BLLacs&11&10.6$\pm$2.9&7.4$\pm$2.1&1.8$\pm$1.2&0.9$\pm$0.6&1.4$\pm$0.9 \\
RGs&2&2.5$\pm$0.4&8.0$\pm$0.4&14.2$\pm$2.4&1.7$\pm$0.5&\nodata \\
\enddata
\end{deluxetable}

Table~\ref{ClassParm} does not reveal any statistically significant differences between parameters of the FSRQs and BLLacs in our sample. However, the parameters follow the same tendency as found previously by J05, that FSRQs possess higher Doppler and Lorentz factors and smaller viewing and intrinsic opening angles than BLLacs. A higher value of $\delta$ for the FSRQs knots is
supported by significantly higher brightness temperatures of all knots in the FSRQs in comparison to $T_{\rm b,obs}$ of the BLLac knots. A wider opening angle in BLLacs with respect to FSRQs was reported by \cite{PUSH09} in the MOJAVE sample. This could explain the significantly greater
difference between the observed brightness temperatures of the cores and jet components in the BLLacs (see \S~\ref{STB}). While the cores in BLLacs have the same level of, or even stronger, Doppler boosting 
than the cores in FSRQs according to $T_{\rm b,obs}$ (Figure ~\ref{figTB3}, {\it right}), the level of Doppler boosting decreases for jet components, which have lower $T_{\rm b,obs}$ values than jet knots in the FSRQs (Figure ~\ref{figTB3}, {\it left}). This could be caused by more rapid expansion of the jet in the vicinity of the core in the BLLacs. A broader jet might be more subject to deceleration (by interaction with the external medium) close to the core, as we have tentatively found for the BLLacs (\S~\ref{SAccel}). The parameters of the RGs appear to be consistent with the expectations of the unified AGN scheme \citep{UP95}. Their Lorentz factors are comparable to those of the FSRQs and BLLacs, but their Doppler factors are significantly smaller owing to the wider viewing angle. A moderate Doppler factor in the RGs is consistent with the brightness temperatures of the RGs cores, which are close to $T_{\rm eq}^{\rm max}$. However, we stress that these sources are among the few RGs that are bright $\gamma$-ray sources, and hence may not be representative of the overall RG population. 
 
We have not detected superluminal motion in two sources in our sample, 3C~84 and Mkn421,
although both objects are bright $\gamma$-ray sources \citep{3FLG}. The latter can imply that relativistic 
motion in their jets is somehow hidden from us, given the results of several studies finding that
$\gamma$-ray sources have higher Doppler factors on average
than the general population of AGN \citep[e.g.,][]{J01,LV03,SAV10,LIST16}. We have estimated the jet parameters of 
3C~84 and Mkn421 based on the two assumptions that (1) the intrinsic opening semi-angle of the jet is equal 
to $\overline{\theta_\circ^{\rm A}}$ of the sub-class to which the source belongs, and (2) the relation 
$\Gamma\theta_\circ\sim$0.2 is universal for $\gamma$-ray bright AGN. Parameters of 3C~84 and Mkn~421 thus derived are given in Table \ref{ParAve}. The above assumptions, along with the values of $\theta_{\rm p}$, lead to rather small viewing angles of both 3C~84 and Mkn~421 with respect to $\overline{\langle\Theta_\circ\rangle}$ of the corresponding sub-class. This results in high Doppler factors of the sources. Although the Doppler factor of Mkn~421 given in Table \ref{ParAve} is similar to that used by \cite{Mrk12} to fit its SED during a multi-waveband campaign 2009 with a leptonic emission model, the parameters of the 3C~84 and Mkn~421 jets presented in Table \ref{ParAve} imply that proper motions of $\sim$0.35 and $\sim$0.25~mas per month, respectively, should be observed in these objects. Such speeds have not been detected in existing observations despite intense monitoring (e.g., this paper; \citealt{LIST16,BLASI13}). This creates discrepancy between low Doppler factors corresponding to apparent speeds of TeV BLLacs seen on parsec scales and high Doppler factors needed to explain their high energy behavior, known as ``Doppler crisis''. However, the
significant deceleration of BLLac jets that we have found in the vicinity of the core
(see Figure~\ref{AccelL}) might be a factor toward resolving the ``Doppler crisis'' in VHE-emitting BLLacs such as
Mkn421. It is also possible that the broad emission structure in 3C~84, and possibly Mkn~421, corresponds to a broad 
``sheath'' where the flow is only mildly relativistic. In this case, our estimates of the Doppler factor
would not be valid.

\section{Summary}

We have reported the results of an analysis of jet kinematics in 33 blazars and 3 radio galaxies observed
with the VLBA at 43~GHz based on 1929 total intensity images obtained from 2007 June to 2013 January within the VLBA-BU-BLAZAR program. The primary conclusions are summarized as follows:
\begin{enumerate}
\item We have determined apparent velocities of 252 jet features with respect to the VLBI core, which is assumed to be a stationary feature in the jet. The apparent velocities range from $0.02c$ to $78c$. The highest reliable apparent speeds $\sim 30c$, are observed in several quasars (CTA~26, 1406-076, 1510-089, and CTA~102) and two BLLac objects (3C~66A and 0235+164). Out of these 252 components, 54 are classified as quasi-stationary features. Among 198 knots with statistically significant motion, 31\% exhibit non-ballistic motions. We have derived the epoch of passage through the VLBI core for 86\% of the moving knots. 
\item The distributions of observed brightness temperatures in the host galaxy frame for all features in the jets at all epochs are statistically different for the three sub-classes, while the distributions of $T_{\rm b,obs}^{\rm s}$ of the cores for 
the FSRQs and BLLacs could be drawn from the same population with a probability of 73\% according to
the K-S test, although the distribution for the BLLacs peaks at a higher temperature than that of the FSRQs and RGs. 
There is a clear trend for the BLLacs to have cores
with higher values of $T_{\rm b,obs}^{\rm s}$ and knots with lower values than those in the FSRQs and RGs. 
\item We do not find a statistically significant difference in the distributions of apparent velocities of the FSRQs and BLLacs when each source is represented by the maximum velocity among its knots. The radio galaxies and quasars have the largest percentage ($>50\%$) of knots with unidirectional motion, while the BLLacs have the largest fraction ($>50\%$) of knots with direction of motion
between 30$^\circ$ and 60$^\circ$ with respect to the line connecting the core and the avarage position of the knot.
\item The distribution of normalized maximum flux densities of moving knots in the FSRQs is different from that of the BLLacs
according to the K-S test, with the FSRQs containing more knots with maximum flux densities exceeding the average flux density of the core. The distributions of relative sizes of moving knots at the maximum flux densities are similar for the BLLacs and FSRQs, with twice the core size a common value. These results, along with the finding of higher brightness temperatures of the FSRQ jet components relative to those in the 
BL~Lacs, imply that the FSRQ jets have higher average Doppler factors than those of the BLLac jets. 
\item Features with non-ballistic motion in the quasars exhibit significant acceleration parallel to the jet axis within 5~pc (projected) of the core, with maximum acceleration at 2-3~pc that exceeds the acceleration perpendicular to jet by a factor of $>3$. This supports the conclusion that the acceleration is intrinsic,
corresponding to an increase of the Lorentz factor of the jet flow in the FSRQs on deprojected scales of tens of parsecs from the core. Knots with non-ballistic motion in the BLLac jets tend to decelerate within 4~pc (projected) from the core, which could be caused by a significant curvature of their trajectories on these scales. Although the statistics for the BLLacs is based on a rather small sample, we speculate that 
deceleration of the BLLac jets close to the core could explain the lower measured
brightness temperatures of jet features with respect to those of the cores. 
\item The distribution of the average projected linear distances of knots classified as quasi-stationary features downstream of the core peaks at distances $<1$~pc. The positions of these quasi-stationary knots appear to fluctuate back-and-forth either along or perpendicular to jet. According to the K-S test, those that shift in position along a direction transverse to the jet tend to be brighter than those that ``slosh'' along the jet.
\item We have constructed a reliable sample of jet features with well-determined  timescales of variability and 
superluminal apparent speeds. These parameters were employed to calculate Doppler, $\delta_{\rm var}$, and Lorentz factors, $\Gamma$, and viewing angles, $\Theta_\circ$, of individual jet features, as well as their mean values for each object and each sub-class. We have determined the opening semi-angle, $\theta_\circ$, of each jet based on the projected opening angle and scatter of $\Theta_\circ$ of individual features, in objects with several knots observed in the jet. We find that the derived parameters of jets in 11 of 13 sources common to both our sample and the sample of J05, agree within uncertainties with $\delta_{\rm var}$, $\Gamma$, $\Theta_\circ$, and $\theta_\circ$ determined previously by J05 in a somewhat similar manner.
Two sources, 0528+134 and OJ~287 are exceptions, appearing to have changed their levels of activity significantly since the period analyzed by J05. 
\item The distribution of Doppler factors based on values of $\delta_{\rm var}$ derived for the reliable superluminal knots in the jets, ranges from 2 to 60 for the FSRQs, with a weighted average of $\sim$13 and the
highest value found for 1510$-$089. The distribution for the BLLacs has the same range with the highest value of 
$\delta_{\rm var}\sim60$ derived for 0235+164, and an average of $\sim$11. Similar behavior is observed for the distributions of Lorentz factors of the jets, with a
range of 4 to 38 and average of $\sim$12 for the FSRQs and a range of 2 to 32 and average $\sim$7 for the BLLacs. The distributions of minimum viewing angles for both the FSRQs and BLLacs jets range from nearly zero to
$5^\circ$, with an average of about $1.5^\circ$.
\item We find that the relation $\Gamma\theta_\circ\sim$0.2 holds independently of the method for
deriving the opening angle of the jet. This has the important implication that {\emph relativistic jets
with higher Lorentz factors have narrower opening angles}, as expected in most hydrodynamical jet models.
\item The highest intrinsic brightness temperatures in the cores exceed the equipartition values, indicating that the energy density in radiating particles dominates that of the magnetic field. Such conditions are infrequent, occurring during periods of
of enhanced nonthermal activity.
\item Comparison of the physical parameters of the jets for different sub-classes suggests that among
bright $\gamma$-ray sources, the FSRQs and BLLacs have similar ranges of the the physical parameters.  
However, in average, the FSRQs possess the highest Doppler and Lorentz factors, and smallest 
viewing and opening angles, although differences between the FSRQs and BLLacs jets fall within the 
uncertainties.
\end{enumerate}
\acknowledgements
We thank the referee for constructive comments, which helped to improve the paper. We thank J.\ Romney and R.\ C.\ Walker for valuable advice regarding the receiver setup for our observations.
The research at Boston University was supported by NASA through a number of Fermi Guest Investigator program
grants, most recently NNX14AQ58G. The St.Petersburg University group acknowledges support from Russian Science Foundation grant 17-12-01029. The research at the IAA--CSIC was supported in part by the MINECO through grant AYA2016--80889--P, and several previous ones. Iv\'an Agudo acknowledges support by a Ram\'on y Cajal grant of the Ministerio de Econom\'ia y Competitividad (MINECO) of Spain. The VLBA is an instrument of the Long Baseline Observatory.
The Long Baseline Observatory is a facility of the National Science Foundation operated by Associated Universities, Inc. 
\software{Difmap \citep{Difmap}, AIPS \citep{AIPS96}, DATAN \citep{DATAN}}
\appendix
 Here we discuss the parsec-scale jet behavior at 43~GHz of individual sources in our sample from 2007 June to 2013 January. Prior to this for the interested  reader we offer a short description of existing VLBA observations at 43~GHz of the sources between the completion of the J05 study in 2001 April and the initiation of the current project in 2007 June. As mentioned above, 13 sources (3C~66A, 3C~111, 0420$-$014, 3C~120, 0528+134, OJ~287, 3C~273, 3C~279, 1510$-$089, 3C~345, BL~Lac, CTA~102,
and 3C~454.3) are common to both samples. Out of these 13 sources, three blazars, 3C~273, 3C~279, and 1510$-$089, were observed roughly monthly with the VLBA at 43~GHz between the periods of the studies, along with monitoring 2-3 times per week
with the {\it Rossi X-ray Timing Explorer (RXTE)} at 2.4--10~keV. Analysis of these data for 3C~279 is presented in \cite{Ch08}.
The radio galaxies 3C~111 and 3C~120  were monitored in a similar manner from 2001 May to 2006 June, accompanied by the blazars 0420$-$014 and OJ~287 used as calibrators for the VLBA observations. The results of these observations for 3C~120 and 3C~111 can be found in \cite{Ch09,Ch11}, while the long-term kinematics of OJ~287 is discussed in \cite{IVAN12}. Monthly monitoring with the VLBA of blazars BL~Lac, CTA~102, and 3C~454.3 resumed in 2005 June within a project designed to study polarization
behavior simultaneously in the parsec-scale jet and optical bands. Results for BL~Lac and 3C~454.3 are presented in \cite{MAR08} and \cite{J10}, respectively. In 2006 July this ``polarization'' sample was augmented with additional sources, CTA~26, OJ~287, 1156+295, and 3C~446, three of which were not observed by J05 but all are members of the current project.  

\section {Notes on Individual Sources}\label{Notes}
{\it 3C 66A:} The parsec scale jet at 43~GHz of this BL~Lac object is extended to the south up to 3~mas from the core
(Fig.~\ref{maps}). The jet is modelled by four components, $A1, A2, A3$, and $A4$, in addition to the core, $A0$
(Fig.~4\_1, {\it left}). These might be associated with knots $C4, C2, C1$, and $A2$, respectively, found in J05, based on 
distance and position angle with respect to the core. Although the knots gradually move, as did $C4, C2$, and $C1$, 
values of $<R>$ and $<\Theta>$ of these knots in 1998-2001 correspond to within 1$\sigma$ uncertainty to those 
listed in Table~\ref{Parm}. This agrees with their classifications as quasi-stationary in position over $\sim$15~yr. 
We have detected two moving knots, $B1$ and $B2$. The existence of three stationary knots 
within 0.6~mas from the core makes it difficult to resolve such moving features when they are near the core. 
In addition, their low brightness at 43~GHz (Fig.~4.1, {\it right}) does not allow us to follow them 
beyond $\sim$1.5~mas of the core. This results in a limited region of the jet (between 0.7 and 1.5~mas) over which moving knots can be detected reliably. The variability of the flux at 37~GHz (from the whole source) and 
of the core at 43~GHz (Fig.~4.1, {\it right}) does not show a clear connection with the appearance of the knots. There is a flux increase of $A0$ and $A1$ near the end of 2010 and in the second half of 2011; however, our data do not allow us to identify possible disturbances connected with these events. Figure~2.1 shows 
a sequence of images displaying the most prominent features detected during the analysis. 

{\it 0235+164:} The parsec-scale jet of this BL~Lac object is strongly core-dominated (Fig.~\ref{maps}). However,
weak moving knots with high apparent speeds were detected previously at 43~GHz by \cite{J01} and \cite{Piner06}.
\cite{J01} found a very high apparent speed, up to $30c$, while \cite{Piner06} have measured a range of apparent 
speeds from 8 to $26c$. In both studies, a wide projected opening angle of the jet was reported, with the position 
angle of the jet ranging between +5$^\circ$ and $-$75$^\circ$. We have detected 3 moving knots in the jet, 
$B1$, $B2$, and $B3$ (Fig.~4.2, {\it left}). Knot $B2$ is the feature $Qs$ analyzed previously by \cite{IVAN11b}. 
The knots show a very wide range of apparent speeds (Table~\ref{beta}). $B1$ and $B2$ differ significantly in 
position angle with respect to the core (Fig.~2.2). In fact, $B2$ moves in the opposite direction 
to that of $B1$. As was noted by \cite{IVAN11b}, $B2$ is the brightest moving feature ever detected in 0235+164,
with a position angle, $\Theta\sim$160$^\circ$, never seen in this source previously. The latter supports the idea that the jet axis 
of 0235+164 points directly along the line of sight. After fading of $B2$, a relatively strong feature, $B3$, 
appeared in the jet at a position angle similar to that of $B2$ (Fig.~\ref{maps}). Knot $B3$ has 
complex motion and significantly lower apparent speed than $B1$ and $B2$. Epochs of ejection of $B1, B2$, and $B3$ 
coincide with strong flares in the 37~GHz light curve, which correlates very well with the core light curve according to visual inspection (Fig.~4.2, {\it right}).  

{\it 3C 84:} The parsec-scale jet of the radio galaxy 3C~84 has a very complex structure at 43~GHz 
(Fig.~\ref{maps}). The jet is dominated by two features, the core $A0$ and knot $C2$, which are located on 
the northern and southern edges, respectively, of the main emission of the jet. $C2$ moves with a subluminal speed of $\sim0.2c$
with respect to the core. There are several other knots, $C5, C6$, and $C7$, that move at similar subluminal speeds 
(Fig.~4.3, {\it left}), with $C6$ having the highest proper motion of $0.36c$. Knot $C4$ has 
a significantly different position angle and the most stable distance with respect to the core, while knot $C3$ appears to
move upstream.  Knot $C4$, along with the main part of the jet and weak diffuse emission on 
the north west side, forms a ring-like structure seen at many epochs (e.g., Fig.~\ref{maps}). 
Figure~4.3, {\it right} presents the 37~GHz light curve and light curves of the brightest knots. 
The 37~GHz light curve shows a significant increase in flux, by $\sim$50\%, in the middle of 2011. 
This brightening coincides with an increase in the flux density of $C2$, evolution of which is shown in 
Figure~2.3, although this knot alone cannot explain such a dramatic brightening 
in the mm-wave light curve. Knot $C8$ was ejected in 2011, but was too weak and the core variability too moderate 
during the event to account for the observed rise of the mm-wave flux. Rather, the increase in flux must have 
occurred in multiple locations in the jet.    

{\it 0336-019:} The parsec-scale jet of this quasar contains two stationary features, $A1$ and $A2$ (Fig.~\ref{maps}),  
in addition to the core, $A0$. We have detected three superluminal knots, $B1$, $B2$, and $B3$. Knot $B1$ has the highest apparent 
speed, $\sim$30~c, and could be associated with knot \#14 of \cite{LIST13}, especially if it decelerates as it moves downstream,  
according to these authors. The apparent speed of $B2$ is higher by a factor $>$2 than that of $B3$ (Fig.~4.4, {\it left}). 
Knots $B2$ and $B3$ have apparent speeds similar to those of knots $C3$ and $C4$ reported by \cite{MATTOX01}. Both $B2$ and $B3$ 
undergo a strong acceleration after passage through stationary feature $A1$ (Fig.~4.4, {\it left}). The ejection of $B1$ and $B2$ 
precedes flares in the core and in the 37~GHz light curve (Fig.~4.4, {\it right}). The structure of the most prominent mm-wave outburst in 2010-2011 is complex. The core exhibits two strong flares, while only one knot, $B3$, has appeared in the jet during this 
time, according to our images (Fig.~2.4). $B3$ is most likely associated with the core flare in 2010. However, 
it is possible that another knot, responsible for the core flare in 2011, was ejected in 2011 Summer. This knot is designated 
in Figure~4.4, {\it left} and Table~\ref{beta} as $B3^*$. The ejection of $B3^*$ also might explain the brightening of $B3$ in early 2012 (Fig.~4.4, {\it left}). In the case of $B2$, which shows non-ballistic motion, as does $B3$, an ejection of knot $B2^*$ shown in Figure~4.4, {\it left} is possible as well. However, the 37~GHz light curve is less supportive of such an interpretation than
for $B3$ and $B3^*$. Stationary feature $A2$ appears to be located at a bend, where the jet direction turns from east to northeast (Fig.~2.4). \cite{MATTOX01} have reported a stationary feature, $C2$, located at the bend. The difference in 
the position of knots $A2$ and $C2$ is too large to be explained by the difference in frequencies of observation (the position of 
$C2$ was determined by averaging measurements at 15, 22, and 43 GHz, with an assumption of the same position of the stationary core at all three frequencies), since $A2$ is located $\sim$0.7~mas farther downstream than $C2$. Most likely, the position of the bend varies on a long 
timescale, perhaps reflecting a change in physical conditions either inside or outside the jet. 

{\it 3C~111:} This radio galaxy possesses a prominent radio jet directed to the northeast, 
with a number of bright moving and quasi-stationary features (Fig.~\ref{maps}). We have identified two knots, $A1$ and $A2$, within 0.5~mas of the core, 
that have subluminal speeds. In the case of $A2$, the apparent motion is toward the core. We classify these knots as stationary features. Most likely, 
$A1$ and $A2$ are related to $A1$ and $A2$ reported in J05, since they have similar values of $<R>$, $<\Theta>$, and $<S>$, 
although in the case of $A2$ the  average distance from the core listed in Table~\ref{Parm} exceeds that given in J05 by slightly more than the 
1$\sigma$ uncertainty. However, the current upstream motion of $A2$ suggests that the positions of $A1$ and $A2$ fluctuate about the 
average with an amplitude of $\sim$2$\sigma$ on a timescale of years. We detect 13 superluminal knots, 
$K3-K15$ (Fig.~4.5, {\it left}). We use the designation of moving knots adopted by \cite{Ch11} and continued by \cite{Tom12}. 
Knots $K5-K7$ are the brightest and longest-lived superluminal features in the jet during the observations presented here. $K5$ and $K6$ 
are connected with the brightest outburst seen in the 37~GHz light curve in late 2007 to early 2008, while knots $K7$-$K9$ 
are related to the outburst in the beginning of 2009 (Fig.~4.5, {\it right}). The motion of knots $K3-K9$ within 1.5~mas of the core 
was discussed in \cite{Ch11}. Additional VLBA data, which track the motion of $K3$ and $K5-K7$ beyond 2~mas from the core, do not affect the time 
of ejection of the knots, although acceleration/deceleration of the motion is observed farther downstream. It is possible that 
knots $K5$ and $K6$ underwent an interaction at $\sim$1~mas from the core that resulted in the appearance of features $A3$ and $A4$ with 
fairly stable parameters over $\sim3$~yr (Table~\ref{Parm}). These features disappeared as knot $K10$ approached them. Knots 
$K10$ and $K11$ discussed in \cite{Tom12} continue to move ballistically at later epochs. Their ejection times occurred during the 
rising branch of a moderate radio outburst in the beginning of 2011 (Fig.~4.5, {\it right}). Knots $K12$-$K15$ move ballistically 
within 2~mas of the core, with a speed of $\sim5c$, similar to $K10$ and $K11$, although the ejections of $K12$-$K14$ are not  
associated with significant variability at 37~GHz. The evolution of knots $K12$-$K15$ is shown in Figure~2.5. An extremely high
intrinsic brightness temperature of the core of 3C~111, $T_{\rm b,int}>1.26\times$10$^{12}$~K, was observed on 2012 May 26,
near the time of the ejection of $K15$, although a very moderate mm-wave outburst coincides with this event (Fig.~4.5, {\it right}).
 
{\it 0420$-$014:} During our monitoring, the inner jet of this quasar lies along a direction $\sim$100$^\circ$ (Fig.~\ref{maps}), 
which is very different from that found in previous observations at 43~GHz, $\sim-140^\circ$ (J05), and at
longer wavelengths \citep{SILKE00,LIST13}. The core of 0420$-$014 has a size comparable with the resolution at 43~GHz, which suggests 
that it is not a point source, and may possess structure more complex than a circular Gaussian that we cannot resolve. 
We have detected 6 moving knots with apparent speeds ranging from 10 to $20c$ and one quasi-stationary feature, $A1$,  
in addition to the core (Fig.~4.6, {\it left}). Unfortunately, the motion of each knot in our 43 GHz images can be followed 
along the section of the jet with $\Theta\sim$~100$^\circ$  only up to 0.5~mas from the core; beyond this distance the knot disappears, 
although it is possible that it reappears later at PA$\sim -140^\circ$ to $-160^\circ$ as a weak and diffuse feature. In addition, 
knots are seen on images only after they pass stationary feature $A1$, which is located $\sim0.15$~mas from the core. 
Figure~2.6 shows a sequence of images, which highlights the evolution of the brightest knot, $B3$. A visual comparison between 
variations at 37~GHz and in the core at 43 GHz indicates a strong correlation between the light curves (Fig.~4.6, {\it right}), which implies a tight connection of the 37~GHz light curve with appearance and fading of superluminal knots.

{\it 3C~120:} We began monitoring 3C~120 under the VLBA-BU-BLAZAR program in 2012 January. This limits the number of epochs analyzed here to 10. More recent results of the 
jet kinematics of 3C~120 can be found in \cite{Carolina15a}. The inner jet of the radio galaxy includes three stationary features, 
$A1, A2$, and $A3$, within 0.5~mas of the core (Fig.~4.7, {\it left}). Although one year of monitoring is insufficient to confirm 
stationarity, J05 have reported a stationary knot, $A1$, with similar parameters as those of $A2$ listed in Table~\ref{Parm}, 
while $A3$ could be associated with stationary feature $D$ located $0.72\pm0.25$~mas from the core on 15~GHz images of 3C~120 \citep{Leon10}. 
However, one would expect the distance between the core and the jet feature to decrease with wavelength due to core opacity effects. 
Knot $A1$ is, perhaps, a quasi-stationary feature; it continues to be seen on 43~GHz images in 2013-2014 \citep{Carolina15a}. 
We have detected two superluminal knots, $C1$ and $C2$, with particularly high apparent speed up to $\sim9c$, while the maximum speed of 
jet components in 3C~120 found by the MOJAVE survey is $6.5\pm0.2c$ \citep{LIST13}. The fastest knot, $C2$, has a complex structure 
in 2012 January (Figure~\ref{maps}) and fades quickly (Fig.~2.7), which make its parameters less reliable. However, the times 
of ejection of the knots fall on the rising branches of outbursts in the 37~GHz light curve, with $C1$ connected with the more prominent event 
(Fig.~4.7, {\it right}). Very rapid motion of $C1$ and $C2$ agrees with the finding by \cite{Carolina15a} that knots ejected in 2007-2011 
have velocities corresponding to the extreme end of the apparent speed distribution in the jet of 3C~120. We do not see superluminal 
knots ejected in 2012, but in mid-2012 a dramatic outburst began at millimeter wavelengths, which correlates with an increase in the flux of the core and stationary features.  
 
{\it 0528+134:} The parsec-scale jet of this quasar has a prominent bend at $\sim$0.9~mas from the core (Fig.~\ref{maps}). 
Seven moving components, $C1, C2, C3, c4, B1, B2$, and $B3$, are detected from 2007 June to 2013 January (Fig.~4.8, {\it left}). 
Knots move ballistically, with $B1$ having extremely high 
apparent speed, $\sim$80c. Although $B1$ possesses consistent flux and PA across epochs, the size of the knot 
is rather large (0.21$\pm$0.07~mas) and it is detected  only within 0.5~mas of the core, which adds to the proper motion an unknown, large systematic uncertainty to the formal statistical uncertainty given in Table~\ref{beta}.   
Knot $c4$ has formed most likely as the result of interaction of $C2$ and $B2$ when the knots aproach the bend. The light curve at 37~GHz shows that the prominent outburst near the end of 2007 
is closely connected with the appearance of knots $C3$ and $B1$-$B3$ (Fig.~4.8, {\it right}). The latter
have ejection times that are the same within their uncertainties. The mm-wave outburst starting in late 2011 might be associated with knot $B4$, first detected on the 2012 October image. A sequence of images 
(Fig.~2.8) shows the evolution of knots $B2$ and $B3$ and blending of faster moving $B2$ with $C2$ near the bend.

{\it 0716+714:} The parsec-scale jet of this BL~Lac object is strongly core dominated (Fig.~\ref{maps}), with frequent ejections 
(at least every half year) of a superluminal knot, the apparent speed of which can reach $25c$ (Fig.~4.9, {\it left}). This has already been noted by \cite{LIZA08}, who analyzed the jet kinematics of 0716+714 during an active state in 2004. We have detected 8 moving knots in the jet. 
Knots $B5$ and $B6$ correspond to $K1$ and $K3$, respectively, discussed by \cite{VLAR13}. In addition to the core, the jet possesses two quasi-stationary features, $A1$ and $A2$, which can be associated with knots $b_{43}(a_{22})$ and $d_{43}(b_{22})$, respectively, described 
by \cite{SILKE09}. Although features $A1$ and $A2$ are stationary with respect to distance from the core, their average position 
angles have very large standard deviations, implying motion of the features transverse to the jet direction. This agrees with 
the result reported by \cite{SILKE09} that jet components in 0716+714 have steady radial separations from the core but change significantly in position angle. However, detection of bright fast-moving knots $B5$ and $B6$, which are observed from 0.2 to 1~mas from the core, contradicts 
these authors' conclusion that jet components do not show long-term outward motion. In addition, we calculate a high brighness temperature, $T_{\rm b,obs}\sim$10$^{13}$~K, of the core during the peak of the core outburst
in 2011 preceding the ejection of $B6$. The average value of $\Theta$ of $A1$ is different from that of $A2$ 
(see Table~\ref{Parm}), indicating curvature of the jet from eastward near the core to northward farther downstream.
Figure~Fig.~4.9, {\it right} shows that all flares in the radio light curve of 0716+714 at 37~GHz are associated with ejection of 
superluminal knots, except for the flare in 2008, although the latter coincides with an increase of the core flux. Therefore, most likely, 
we have missed a component in the jet due to a 3.5 month gap in monitoring in the first half of 2008, while at epoch 2008 July 6 the 
jet possesses a $\sim50$~mJy knot downstream of $A2$, which could be related to this flare.  Moving knots have different position angles 
within a wide range from $-2^\circ$ to $60^\circ$. In 2012 October a bright knot, $B8$, appeared in the jet, most likely
connected with the 37~GHz and core outburst in the second half of 2012. Figure 2.9 plots a sequence of VLBA images that shows 
the evolution of knots $B4-B7$ ejected along different position angles.

{\it 0735+178:} The parsec-scale jet of this BL~Lac object possesses a stationary feature, $A1$, in addition to the core (Fig.~\ref{maps}) 
and slowly moving knot $C1$, which are most likely knots $ca$ and $cb$, respectively, reported by \cite{SILKE10} and described by these authors 
as a ``staircase'' structure. The knots appear to be associated with bends in the jet. The average position angle of $A1$ has a large standard 
deviation (Table~\ref{Parm}), which supports the idea that stationary features in BLLacs in general, and 0735+178 in particular, exhibit motion 
perpendicular to the jet ridge line. We have detected two moving knots, $B1$ and $B2$, with high apparent speeds (Fig.~4.10, {\it left}). 
$B1$ has non-ballistic motion and decelerates as it approaches $C1$ (Fig.~2.10). Its average apparent speed is similar to 
that of knots 6 and 8 in \cite{LIST13}. Ejection of $B2$ coincides with the start of a strong outburst in the 37~GHz light curve at the beginning of 2012 (Fig.~4.10, {\it right}), which can be associated as well with bright knot $B3$ that appeared near the end of 2012. 

{\it 0827+243:} A high apparent speed of $\sim25c$ has been reported previously in the parsec-scale jet of this quasar \citep{J01,Piner06}.
Analysis of extended X-ray emission led \cite{J04} to conclude that the jet remains highly relativistic out to $>800$~kpc from the BH.
We have detected 4 moving knots, $B1$-$B4$, within 1~mas of the core, with apparent speeds ranging 
from 10 to $18c$ (Fig.~4.11, {\it left}). A similar range of apparent speeds is obtained at 15~GHz farther downstream \citep{LIST13}. At many epochs, a brightness excess is detected $\sim0.15$~mas from the core that we identify as a stationary 
feature, $A1$ (Fig.~\ref{maps}). Knots $B2$ and $B3$ have ballistic motion to the southeast. $B1$ accelerates beyond $A1$, 
while $B4$ decelerates beyond $A1$ to a subluminal speed during the first half of 2012 at a distance of $\sim0.27$~mas, and 
then accelerates to the previous proper motion. Visual inspection of flux variations at 37~GHz and in the VLBI core suggests a 
tight connection between mm-wave flares and the appearance of superluminal knots in the jet (Fig.~4.11, {\it right}), although 
the measured maximum brightness of knots does not correlate with the amplitudes of the flares. The latter implies that an 
interaction between the core and a knot might play a role in the strength of a flare, supporting the idea that the core is a 
physical structure, e.g., a conical shock \citep[e.g.,][]{dm88,TIM13}. Figure~2.11 shows the evolution of knots $B3$ and $B4$ associated with the strongest mm-wave outburst near the end of 2010. A strong outburst in the core and at 37 GHz is observed also near the end of 2012.

{\it 0829+046:} In this BL Lac object, a stationary feature, $A1$, is detected very close to the core 
at the majority of epochs (Fig.~4.12, {\it left}). Although $A1$ has a stable distance with respect to $A0$, its PA, $\Theta$, varies
significantly. This represents another case among BL~Lac objects with possible transverse motion in the jet. We classify the 
weak knot $A2$ as a stationary feature, although its position fluctuates near $\sim0.4$~mas from the core, and the knot  
disappears after the passage of $B3$ through this location. Knot $A3$ seems to be located at a bend in the jet. It brightens 
significantly as a superluminal knot approaches (Fig.~\ref{maps}). $B3$ decelerates to a subluminal speed between $A2$ and 
$A3$. In addition to $B3$, we have detected three superluminal knots, $B1$, $B2$, and $B4$. $B1$ and $B2$ have non-ballistic 
motion, while $B1$ accelerates beyond $A3$ and $B2$ decelerates as it approaches $A3$. The times of ejection of $B2$-$B4$ are 
coincident with outbursts in the 37~GHz light curve (Fig.~4.12, {\it right}). Figure~2.12 plots a sequence of VLBA images showing the evolution of $B3$.

{\it 0836+710:} This quasar has the highest redshift in our sample. During the time interval analyzed here, we have detected 
three superluminal features, $B1-B3$, in the innermost jet, while between 0.6 and 1.5~mas from the core only weak emission can 
be seen at a few epochs (Fig.~\ref{maps}). Beyond 1.5~mas, a prominent feature, $C1$, moves at much slower apparent speed 
than the innermost knots, although $B2$ decelerates beyond 0.3~mas from the core (Fig.~4.13, {\it left}). Knot $B3$ is 
of special interest as a superluminal feature that is brighter than the core at several epochs (Fig.~4.13, {\it right}). 
$B3$ is knot $K11$, associated by \cite{SGJGranada13} with a bright $\gamma$-ray outburst observed in 2011. There is a close 
visual correspondence between the 37~GHz and core light curves. The light curves peak simultaneously about one year earlier 
than the ejection time of $B3$. This suggests that the high flux density of $B3$ may be the result of an interaction between 
the knot and a slower moving feature ejected before $B3$ that we cannot resolve from $B3$ on our images. During the mm-wave 
outburst, a stationary feature, $A1$, appeared on the opposite side of the core and persisted for more than 2~yr. Note that 
modelling of the core by components with an elliptical Gaussian distribution does not remove the feature. The sequence of VLBA 
images in Figure~2.13 shows the location of $A1$ at different epochs, as well as the evolution of $B3$ and $C1$ with time. 

{\it 0851+202:} The jet of this BL~Lac object, OJ~287, possesses a stationary feature, $a$, $\sim~0.2$~mas from the core, which is 
the average distance of stationary features $A1$ and $A2$ found by J05 at $\sim$0.1 and $\sim$0.3~mas, respectively, 
during 1998-2001. Properties of $a$ are discussed in \cite{IVAN11a,IVAN12} and \cite{JEFF17}, where the feature is designated 
as $C1$, $a$, and $S$, respectively. There is some similarity between the properties of $a$ and $A0$. Knot $a$ is brighter 
than the core over almost the entire period analyzed here (Fig.~4.14, {\it right}).
According to \cite{IVAN12}, the appearance of $a$ in 2004 heralded a change in the 
projected direction of the innermost jet by $>$90$^\circ$. The majority of the data used
here have been analyzed in \cite{IVAN12} and \cite{JEFF17}, and we use the results of modelling of images for epochs before 2011 
August by \cite{IVAN12} and after 2011 August by \cite{JEFF17}. A few epochs have new models, and we include several 
additional epochs. We have detected 7 superluminal knots, $B1$-$B7$ that are visible on images only beyond $a$ 
(Fig.~4.14, {\it left}). Knots $B1$ and $B2$ correspond to $h$ and $j$, respectively, of \cite{IVAN12}. $B2$ is detectable 
up to 2~mas from the core (Fig.~\ref{maps}) and decelerates slightly at later epochs. Knot $B3$ is the sum of 
knots $l$ and $m$, as can be inferred from Figure 1 in \cite{IVAN12}. This results in a significantly lower speed of $B3$ 
compared to $\beta_{\rm app}$ of $l$ and $m$, although $l$ and $m$ were not considered as reliable features. $B4$ ---knot $n$ in \cite{IVAN12} --- slightly decelerates as it approaches quasi-stationary feature $A3$, which possesses 
similar parameters as feature $A3$ in J05. The strongest mm-wave outburst seen in the 37~GHz light curve is 
coincident with the flaring of $a$ (Fig.~4.14, {\it right}), most likely due to passage of $B3$ and $B4$ through the knot. 
Knot $a$ reaches an extremely high brightness temperature, $T_{\rm b,obs}\sim$1.6$\times$10$^{13}$~K, during this mm-wave event. 
\cite{IVAN11a} associate the passage of $B3$ and $B4$ through $a$
with a $\gamma$-ray flare \citep{IVAN11a} of OJ~287. $B4$ is the fastest knot among $B1$-$B7$, 
while $\beta_{\rm app}$ values of $B5$-$B7$ fall in the low end of the apparent speed distribution of the parsec-scale jet. 
The sequence of images plotted in Figure~2.14 shows the evolution of knots $B4$-$B6$.

{\it 0954+658:} The parsec-scale jet of this BL~Lac object exhibits curvature at $\sim$0.5~mas from the core (Fig.~\ref{maps}), where moving knots seem to decelerate (Fig.~4.15, {\it left}). The multi-frequency 
variability of 0954+658, along with the parsec-scale jet behavior from 2008 to 2011, were studied by \cite{DASHA14}. 
Although there are several new epochs added and new models created for a few epochs within the period mentioned above, 
the kinematics of knots $B1$-$B4$ is almost the same as for $K1$-$K4$ in \cite{DASHA14}, while $B5$ corresponds to prominent 
feature $K8$ associated with the brightest $\gamma$-ray outburst of the blazar. We do not register features corresponding 
to $K5$-$K7$ of \cite{DASHA14}, since they are weak and detected at fewer than 5 epochs. In addition, at some epochs these knots 
might be related to a quasi-stationary feature, $A1$, located $\sim0.2$~mas from the core. The light curve at 37~GHz has erratic behavior, with the most notable flares corresponding to an increase of the flux in the core (Fig.~4.15, {\it right}). Figure~2.15 shows a sequence of images featuring the motions of knots $B5-B7$, while sequences of images at earlier epochs can be found in \cite{DASHA14}.
 
{\it 1055+018:} \citet{SDSS08} have classified this blazar as a BL~Lac type object based on its optical properties, while \citet{LIST13} identify it as a highly polarized quasar (HPQ) based on radio properties. The source is strongly 
core-dominated on parsec scales and possesses a stationary feature, $A1$, that is close to the core (Fig.~\ref{maps})
and the second-brightest component after the core (Fig.~4.16, {\it right}). We have detected 3 moving features 
in the jet. The fastest knot, $B1$, appeared in the jet in the beginning of the sharp mm-wave outburst in the second 
half of 2009. Knot $b1$, with the slowest apparent speed, seems to have formed behind $B1$, and hence represents a trailing 
feature (Fig.~4.16, {\it left}). Knot $C1$, observed beyond 1~mas from the core (Fig.~4.16, {\it left}), 
moves with an apparent speed of 7c, similar to that found for superluminal features of the blazar in 
the MOJAVE survey \citep{LIST13}. A strong increase of the core flux in the middle of 2011, as well as the subsequent 
brightening of $A1$ coinciding with a long-lasting outburst in the 37~GHz light curve (see Fig.~4.16, {\it right}), might be associated with the ejection of a superluminal knot in early 2011. However, there is no feature detected beyond $A1$ during the epochs analyzed here (Fig.~2.16). 

{\it 1101+384:} This BL~Lac object, Mkn~421 is a well-known bright TeV/$\gamma$-ray source. As for many BLLacs, its parsec-scale structure is dominated by quasi-stationary features, the 
most prominent of which (other than the core) is $A1$,
located $\sim$0.4~mas from the core. Knots $A1$ and $A2$ (Fig.~\ref{maps}) can be associated with features $D$ and $C3$, 
respectively, found by \cite{BLASI13} in well-sampled VLBA observations at 43~GHz in 2011 that included some epochs from our 
program. They are also likely to be related to features \#8 and \#11, respectively, in 15~GHz MOJAVE images \citep{LIST13}. There is no superluminal motion 
detected in the parsec scale jet of Mkn~421 (e.g., \citealt{Piner10}); the maximum apparent speed found in the MOJAVE survey 
is $\sim0.3~c$. In fact, $A2$ shifts in position at a similar apparent speed, but toward the core (Fig.~4.17, {\it left}). 
The appearance of another subluminal knot, $B2$, in 2011 coincides with a flare in the core and mm-wave flare at 37~GHz
(Fig.~4.17, {\it right}). The time of ejection of $B2$ in early 2010 coincides with high activity and correlated 
variability at X-ray and TeV energies \citep{Aleksic15}. It is possible that another knot, $B1$, with the 
highest apparent speed, $\sim$1c, was ejected near the beginning of 2010. However, its reality is doubtful, 
since $B1$ is not detected beyond $A1$. Figure~2.17 shows a sequence of VLBA images with positions of knots 
$B1$ and $B2$ marked. In early 2012, the flux of the VLBI core started to rise in unison with the total 37~GHz 
flux (Fig.~4.17, {\it right}). The mm-wave light curve contains a sharp peak in the middle of 2012 when strong $\gamma$-ray 
activity of Mkn421 was detected by the {\it Fermi} LAT \citep{DAO12}. A new knot, $B3$, seen in the 2013 January image 
(Fig.~\ref{maps}) can be associated with this outburst, although whether this association is reliable is not yet clear. 

{\it 1127$-$145:} This quasar is classified as a young radio source with a gigahertz-peaked spectrum (GPS),  
\citep[e.g., ][]{ANETA08}. It has a prominant radio jet that curves on parsec scales from east to 
northeast \citep{LIST13}. The parsec-scale jet aligns with the large-scale X-ray/radio jet extending over 
300~kpc \citep{ANETA07}. The core, $A0$, dominates the parsec-scale jet at 43~GHz (Fig.~\ref{maps}). There are two 
slowly moving components, $A1$ with a subluminal speed, located close to the core, and $A2$, which is probably located 
close to the position of the bend in the jet. The location of $A2$ is similar to that of knots $C2$ and \#3 observed 
in 1996-1997 at 22~GHz \citep{J01} and in 1995-2011 at 15~GHz \citep{LIST13}, respectively, although $A2$
moves upstream while \#3 moves at a slow downstream velocity. Most likely, the apparent motion of $A2$ is 
related to the propagation of knots moving rapidly through the bend. We have detected five fast moving features, $D1$, $C1, C2$, 
$B1$, and $B2$, with apparent speeds ranging from 8 to $24c$ (Fig.~4.18, {\it left}). $C2$, the brightest of these, decelerates 
beyond 0.5~mas from the core for $\sim1.5$~yr and then accelerates to the initial speed. Although 
the faint knot $B1$ is not detected at all epochs at which it should be seen according to its trajectory, $B1$ has similar behavior as $C2$. Figure~2.18 presents a sequence of images showing the evolution of $C2$ and $B1$. Knot $B2$ 
appears to be connected with a strong outburst in the VLBI core in late 2009 to early 2010 (Fig.~4.18, {\it right}) and fades very quickly, in contrast to $C2$. Unfortunately, the 37~GHz light curve of 1127$-$145 is not available due to its 
southern declination, which limits monitoring of the source at the Mets\"ahove Radio Observatory.
     
{\it 1156+295:} We trace the parsec-scale jet of this quasar up to 1~mas (Fig.~\ref{maps}). Four moving components, 
$B1$-$B4$, are detected from 2007 June to 2013 January (Fig.~4.19, {\it left}). We classify feature $b1$ as a trailing knot behind bright component $B1$, although the motion of $b1$ is not statistically significant. 
Figure~4.19, {\it right} shows at least three outbursts in the 37~GHz light curve during this period, which are most 
likely connected with flares in the core and ejection of superluminal knots. It appears that disturbances $B2$ and $B3$ 
originated during the 2010 outburst. Knots $B1$-$B3$ move non-ballistically, with an acceleration beyond 
$\sim0.15$~mas from the core. Figure~2.19 identifies knots $B2$  and $B3$ on a sequence of VLBA images.  
Knots $B1$, $B2$, and $B3$ correspond to components  $C2$, $C4$, and $C3$, respectively, 
discussed in \cite{VENKATESH14}, although the identification of knots $B2$ and $B3$ near the core is not unique and slightly different from that of $C4$ and $C3$. In addition, trailing knot $b1$ is identified in \cite{VENKATESH14} as a stationary feature, $S1$. 

{\it 1219+285:} The jet of this BL Lac object is directed to the east, with a slight curvature toward the southeast beyond 
1~mas from the core (Fig.~\ref{maps}). We have found several quasi-stationary features within 3~mas of the core that 
move subluminally  either toward or upstream from the core (Fig.~4.20, {\it left}). The emission of the jet at 43~GHz is dominated by the core and knot $A1$ (Fig.~4.20, {\it right}), which is the closest stationary feature to the core. Although the existence of several stationary features makes it 
difficult to detect moving knots, bright knot $B1$, seen well beyond $A2$, moves with an apparent speed of $\sim5c$. Its evolution is presented in a sequence of images in Figure~2.20.  

{\it 1222+216:} During quiescent states, the parsec-scale jet of this quasar can be modelled by the core $A0$ and a presumably stationary feature $A1$, 
located $\sim0.15$~mas from the core (Figure~\ref{maps}). During the prominent mm-wave outburst in 2010 (Fig.~4.21, {\it right}),
bright emission at 43~GHz started to strengthen in the jet beyond $A1$ (Fig.~2.21). We have detected four superluminal knots, 
$B1$-$B3$, and $b3$ (Fig.~4.21, {\it left}). The latter is most likely a trailing feature behind a strong and fast disturbance 
seen as knot $B3$, which has a different trajectory from $B1$ and $B2$. Knots $B2$ and $B3$  have non-ballistic motion, 
with acceleration after passing $A1$ (Fig.~4.21, {\it left}). A strong outburst of very high-energy emission ($E>100$~GeV) 
with a short timescale of variability was observed by MAGIC during this period of jet activity \citep{Aleksic11} 
when $B1$-$B3$ were ejected. Note that knots $B2$ and $B3$ reported here 
correspond to $B1$ and $B2$, respectively, discussed in \cite{J15}. 

{\it 3C 273:} The quasar 3C~273 possesses a prominent jet, which can be traced up to 10~mas from the core in our 43~GHz 
images (Fig.~\ref{maps}). At the beginning of the period analyzed here, two bright features, $B1$ and $B2$, can be 
identified in the jet, with $B2$ having a faster apparent 
speed  by a factor of $\sim1.5$ (see Table~\ref{beta}). A significant velocity gradient across the jet has been 
reported previously by J05 and \cite{SAV06}, with a faster speed for knots moving along the southern edge. Beyond $\sim2$~mas 
from the core, $B1$, $B2$, and $B3$ --- 
the latter ejected later with a slightly higher apparent speed than $\beta_{\rm app}$ of $B2$, --- form a complex structure, 
which can be modelled by a single diffuse feature (Fig.~\ref{maps}). We have detected 10 moving, $D1$, $B1-B9$, and 2 stationary, 
$A1$ and $A2$, knots in addition to the core. Although $A1$ and $A2$  are not identified at all available epochs, they have 
similar positions to those of $A1$ and $A2$, respectively, found by J05 in 1999-2001. Most likely, these features represent 
long-lived structures in the jet. The moving knots show a range of apparent 
speeds from 5 to $12c$, with some knots moving non-ballistically (Fig.~4.22, {\it left}). Especially interesting is the motion of 
$B8$, with strong deceleration near $A2$ and acceleration beyond $\sim1$~mas from the core. The jet was extremely active 
in 2009 when at least 5 knots, $B4$-$B8$, were ejected closely in time. These events coincided with a series of strong 
$\gamma$-ray outbursts detected by the {\it Fermi} LAT \citep{ABDO10}. Note that the brightness temperature of $B5$ is extremly
high, $T_{\rm b,obs}\sim1.3\times$10$^{13}$~K, in early 2010 at a distance of
$\sim0.5$~mas from the core, which coincides with the location of stationary feature $A2$. In addition, the maximum intrinsic brightness temperature of the core of 3C~273, $T_{\rm b,int}^{\rm max}$=3.25$\times$10$^{12}$~K, the highest value derived for the sample (Table~\ref{ParAve}), was observed on 2009 November 28, which coincides with the highest mm-wave activity 
and $\gamma$-ray outbursts (see above). Figure~2.22 displays a 
sequence of images during this period, with identifications of knots across the epochs. The brightness of two of them, 
$B4$ and $B7$, exceeds the brightness of the core by a factor of $\sim3$ at 
the first epochs of detection (Fig.~4.22, {\it right}). The 37~GHz light curve confirms strong activity at millimeter wavelengths. 
The light curve contains two sharp outbursts near the beginning of 2009 and 2010, which are clearly connected with 
the appearance of $B4$ and $B7$ in the jet. The outbursts are superimposed on a high level of total flux, which suggests 
that the jet was already in an excited state before the exceptional culmination in 2009-2010. 

{\it 3C 279:} The parsec-scale jet of this quasar underwent a remarkable change of direction in late 2010 when 
extremely bright superluminal knot $C31$ was ejected (Fig.~\ref{maps}). We follow the historical designation of moving 
components at 43~GHz started by \cite{UNWIN89} and \cite{ANN01}, and continued by \cite{J04}, J05, \cite{Ch08}, and \cite{VLAR08}. Figure~2.23 reveals the 
appearance of the knot at PA$\sim -210^\circ$, nearly transverse to the usual direction of the jet, $\sim -130^\circ$ 
\citep{J05,LIST13}. The aberrant PA of the innermost jet in 2011 was reported by \cite{LU13} based on a VLBI observation at 
230~GHz. Figure~2.23 shows the evolution of $\Theta$ of $C31$ as it curved toward the normal direction of the jet 
with increased separation from the core. Note that $C31$ might be associated with knot $A1$ discussed by \cite{3C279A}, 
although additional epochs suggest a later time of ejection of $C31$ than that of $A1$. On the other hand, the uncertainty 
in $T_\circ$ is significant because the distance of the knot in 2011 with respect to the core changed only from 0.1 to 0.2~mas, 
while its PA rotated from $-210^\circ$ to $-170^\circ$. During the 2007 June -- 2013 January period, 
we detect 9 moving features, $C24$-$C32$ (Fig.~4.23, {\it left}). All knots observed at $\ge10$ epochs move non-ballistically, 
with different acceleration/deceleration properties. The ejection of each component appears to be associated with a 
flare in the VLBI core and 37~GHz light curve (Fig.~4.23, {\it right}). However, the largest outburst at 37~GHz in 2012 Summer
coincides with a strong flare in $C31$ when the knot became brighter than the core at a distance of $\sim0.25$~mas. 
The knot also possesses a high brightness temperature, $T_{\rm b,obs}>10^{13}$~K. The core underwent a flare at 
this time as well, most likely connected with the ejection of a new knot that we see at epochs later than 
those considered here. We propose two possible explanations of the strong flare in $C31$ several parsecs from the core: 
(1) $C31$ crossed the line of sight, maximizing its Doppler beaming factor, in 2012 Summer; this is supported by a very slow speed of the knot in the first 
half of 2012 and rapid acceleration afterward (Fig.~4.23, {\it left}); and (2) in 2012 Summer $C31$ interacts with the second 
brightest moving feature in the jet, $C32$, which was ejected earlier but moves more slowly than $C31$. Knot $C32$ 
can be associated with feature $A2$ discussed in \cite{3C279A}.   

{\it 1308+326:} This quasar (formerly classified as a BL Lac object) contains a bright radio jet to the northwest, 
which bends to the north beyond 0.5~mas from the core (Fig.~\ref{maps}). We have detected four superluminal knots, 
which move with apparent speeds ranging from 5 to $14c$ (Fig.~4.24, {\it left}), and a stationary feature, $A1$, located 
close to core. The MOJAVE survey gives significantly higher apparent speeds ($>20~c$) for knots detected beyond 1~mas 
from the core \citep{LIST13}, while knots tracked closer to the core have similar apparent speeds as we find. 
This supports the idea of acceleration of the jet several parsecs from the core \citep{Homan15}. We directly measure an 
increase in the apparent velocity of $B2$, while its projected PA remains constant. A moderate mm-wave outburst visible in the 37~GHz light curve in 2009--2010 is coincident 
with an increase in the flux of the core (Fig.~4.24, {\it right}) and ejection of knots $B2$-$B4$, with $B4$ ejected earlier than $B3$ according to our analysis. A small flare in the core in 2011, followed by a brightening of $A1$ in the beginning of 2012, can be associated with a new 
knot, $B5$, detected in the 2013 January map (Figure~\ref{maps}). Figure~2.24 shows the evolution of knots $B2$-$B4$. 

{\it 1406$-$076:} The inner parsec-scale jet of this quasar is directed to the west with a slight curvature to
the southwest (Figure~\ref{maps}). Its 43~GHz emission is dominated by the core (Fig.~4.25, {\it right}). We have detected four superluminal 
knots, $C1$, and $B1$-$B3$, and a stationary feature, $A1$, in addition to the core (Fig.~4.25, {\it left}). Knots $C1$ and $B1$-$B3$ move ballistically and show a significant difference in apparent speeds, from 5 to $30c$. A possible interaction of $B2$ and $B3$ with $A1$ in the second half of 2009 and 2010, respectively, 
might have occurred, as supported by an increase of the flux of $A1$ near those epochs (Fig.~4.25, {\it right}). Figure~2.25 presents a sequence of images, which delineates the paths of $B1$-$B3$. 

{\it 1510$-$089:} The 43~GHz core of this quasar dominates the parsec-scale jet at many epochs (Fig.~4.26, {\it right}). We identify two quasi-stationary features, $A1$ and $A2$, in addition to the core, $A0$ (Figure~\ref{maps}). The parameters of these features are similar to those of knots $A1$ and $A2$ in J05 within the 1$\sigma$ uncertainties. Five moving knots, $B1$-$B5$, ejected between 2007.5 and 2013.1, possess high apparent speeds, up to $30c$ (Fig.~4.26, {\it left}). Propagation of knot $B2$ down the jet 
was analyzed in \cite{MAR10} in connection with the strong $\gamma$-ray outburst in 2009. The appearance of $B4$ in the jet coincides with another high $\gamma$-ray state of the quasar in 2011 Autumn \citep{Orienti13}, while ejection of $B5$ occurred simultaneously with a TeV flare detected by MAGIC \citep{Aleksic14}. Knots $B4$ and $B5$ correspond to $K11$ and $K12$ discussed in \cite{Aleksic14}, respectively. Although $\beta_{\rm app}$ of $B4$ is slightly different from that of $K11$, the latter is caused by  
deceleration of $B4$ at later epochs. Epochs of ejections of knots $B2$, $B4$, and $B5$ correspond to the most prominent peaks in the light curve at 37~GHz (Fig.~4.26, {\it right}). It is interesting to note that, before the ejection of the brightest knot, $B4$, both features $A1$ and $A2$ moved away from the core at 5 successive epochs (Fig.~4.26, {\it left}) at a similar apparent speed of $7.0\pm0.3c$. 

{\it 1611+343:} Three superluminal features, $C1$-$C3$, are detected in this quasar, with the fastest knot, $C1$, being most distant from the core (Fig.~4.27, {\it left}). There are three quasi-stationary knots, $A1$-$A3$, in addition to the core, although $A1$ is a peculiar feature detected upstream of the core before the brightest moving knot, $C3$, appeared in the jet. A similar knot, which we designate as $A1$ as well, is observed in 2012 during a strong mm-wave outburst (Fig.~4.27, {\it right}), which correlates with a brightening of the core.  The jet executes a bend from south to southeast near the presumably stationary feature $A2$ (Figure~\ref{maps}). Relatively slow motion of knots $C2$ and $C3$ is visible in the sequence of images presented in Figure~2.27.

{\it 1622$-$297:} Figure~\ref{maps} shows a typical parsec-scale image of this quasar at 43~GHz. 
Five superluminal knots detected in the jet have apparent speeds within 6-$11c$, with the knot having the most reliably measured speed, $B4$, moving at the slowest rate (Fig.~4.28, {\it left}). This knot is also the brightest feature in the jet after the core, and its ejection is coincident with a strong core-region flare in 2010 (Fig.~4.28, {\it right}). Evolution of the knot is shown in the sequence of images presented in Figure~2.28.

{\it 1633+382:} The parsec-scale jet of this quasar can be traced up to 3-4~mas from the core at 43~GHz, with a strong dominance by the emission within 1~mas (Figure~\ref{maps}). We have detected 7 moving knots with similar apparent speeds of 5-$7c$. In the 2007 maps, feature $A1$ is found $\sim0.27$~mas from the core (Table~\ref{Parm}), while bright knot $B1$ is not detected at these epochs, although an extrapolation of its kinematics suggests that $B1$ should be visible $\sim0.1$~mas closer to the core than $A1$ (Fig.~4.29, {\it left}). Perhaps $A1$ was blended with $B1$; such a possibility is 
supported by the angular size of $B1$, $>0.2$~mas. The evolution of $B1$ is presented in the sequence of images shown in Figure~2.29. Modelling of images in 2010-2011 yields two superluminal knots, $B2$ and $B3$,  
near the core, which are not detected beyond 0.2~mas from the core. However, 
the VLBI core light curve exhibits an outburst around the time of ejection of the knots, which corresponds to an outburst in the 37~GHz light curve (Fig.~4.29, {\it right}). In contrast, an even stronger outburst seen in the core and at 37~GHz in 2011 Spring did not produce disturbances detectable in our images. This implies that, if every core outburst is connected with a disturbance propagating through the core \citep[e.g.,][]{SAV02}, not all disturbances ``survive'' the passage through the core, at least in terms of their 43~GHz emission. The next bright mm-wave outburst started in the second half of 2012 and 
became the brightest mm-wave event during 2007-2012. This can be associated with knot $B4$, which appeared in the end of 2012.  

{\it 3C~345:} This quasar has a prominent parsec-scale jet (Figure~\ref{maps}) with a puzzling core region, which consists of two 
features: the eastern knot, which we associate with the core, $A0$, and western knot $A1$, located $\sim0.15$~mas from the core. 
The knots have similar brightness and amplitude of variability (Fig.~4.30, {\it right}). Although the quasar is a candidate binary SMBH system \citep[e.g.,][]{LR05}, the dual core region is located far from the putative binary SMBH system according to an estimated distance of $\sim30~$pc from the central engine to the radio core at 15~GHz \citep{PUSH12}. We have detected 13 moving features 
in the jet. The separation of knots from the core within 2~mas from $A0$ is presented in Figure~4.30, {\it left}. Although the majority of 
knots execute complex motion with acceleration and deceleration, the average apparent speed is quite stable, $\sim10c$. 
The brightest moving knots, $B2, B3$, and $B5$, are connected with the strongest 
mm-wave outburst observed in 2009-2010 (Fig.~4.30, {\it right}). Knots $B2$ and $B3$ can be identified with knots $Q9$ and $Q10$ in \cite{SCHIN12}, which these authors associate with $\gamma$-ray activity of the quasar in 2009. Knots $B4$ and $b4$ appeared in the jet at distances farther then 0.5 and 1~mas from the core, respectively. We classify $b4$, as well as $c1$ behind knot $C1$, 
as trailing features \citep{IVAN01}. The evolution of knots $B1$-$B3$, $C1, C2$, and $D1$-$D3$ is presented in the sequence of images in Figure~2.30. Diffuse knots $D1$-$D3$ form a single feature at later epochs, as shown in Figure~\ref{maps}. 

{\it 1730$-$130:}  The parsec-scale jet of this quasar is dominated by the core region at 43~GHz, with a broad jet structure extending to the north (Figure~\ref{maps}). We have detected four superluminal knots, $B1$-$B3$ and $D$, with a wide range of apparent speeds, from 7 to $24c$ (Fig.~4.31, {\it left}). In addition to the core, $A0$, there is a quasi-stationary feature, $A1$, seen at several epochs near $\sim0.3$~mas from the core. $A1$ could be responsible for deceleration of knot $B1$ around 0.3~mas,
while there is a noticeable acceleration beyond $A1$. During the last several epochs presented here, a new knot, $b2$, is seen beyond 0.5~mas from the core that could correspond to a structure formed behind the fastest superluminal knot, $B2$. The sequence of images in Figure~2.31 shows the evolution of $B1$-$B3$ across epochs, while light curves of the core and knots are presented in Figure~4.31, {\it right}. Visual inspection of the light curves indicates that the ejection of $B2$ and $B3$ is connected with a strong mm-wave outburst at 37~GHz in late 2010 to early 2011.    

{\it 1749+096:} The parsec-scale jet of this BL~Lac object, directed to the north, has a wide apparent opening angle, $\sim50^\circ$, as can be inferred from Table~\ref{Parm}. Four superluminal knots, $B1$-$B4$, are detected in the jet with apparent speeds ranging from 6 to $18c$ (Fig.~4.32, {\it left}). The slowest knot, $B4$, is the brightest and most compact moving feature in the jet, and has been associated with $\gamma$-ray activity of the blazar \citep{DASHA13}. We classify knot $A1$ (Fig.~\ref{maps}), detected close to the core at many epochs, as a quasi-stationary feature, although its position angle with respect to the core varies across the entire projected opening angle of the jet. Knot $B3$ accelerates after it passes $A1$. The time of ejection of knots coincides with outbursts in the 37~GHz light curve (Fig.~4.32, {\it right}). The sequence of images in Figure~2.32 highlights the evolution of $B3$ and $B4$. Figure~4.32, {\it left} shows the appearance of a new superluminal knot, $B5$, which was ejected in 2012 Autumn, according to later images that are not presented here. 

{\it BL Lac:} The innermost jet of BL~Lac consists of three bright compact features, $A1$-$A3$, which have properties of stationary knots, in addition to the core, $A0$ (Figure~\ref{maps}). Knots $A1$ and $A2$ are imaged as a single feature, $C7$, at 15~GHz in the MOJAVE survey, which has been interpreted as a quasi-stationary recollimation shock \citep{COHEN14}. These stationary features make very difficult the detection of moving knots within 0.4~mas of the core. We identify some knots, e.g., $B5$ and $B6$, according to a succession of significant changes of brightness of $A0$, $A1$, and $A2$ (Fig.~4.33, {\it right}), see also \citealt{ANN16}, where $B5$ and $B6$ are designated as $K11a$ and $K11b$, respectively). Another problem in the identification of knots is a gap in the jet emission at 43~GHz between 0.5 and 1~mas from the core, noted previously by \cite{BACH06}. As is apparent in Figure~4.33, {\it left}, out of 8 moving knots, $B1$-$B8$, which should be detected in this region according to their kinematics, only 50\% are seen at some epochs. Beyond 1~mas from the core knots become diffuse, which further increases the difficulty of identification (Fig.~2.33). Nevertheless, we have determined apparent speeds of more than 10 moving knots, with velocities ranging from  0.5 to $12c$. Knots $B7$ and $B8$, ejected in 2012, have the highest apparent speeds. These knots correspond to features $K12a$ and $K12b$, respectively, discussed in \cite{ANN16}, and were ejected during dramatic $\gamma$-ray, X-ray, and mm-wave activity in BL~Lac. During this period, the brightness temperature of the core of BL~Lac twice reached the value of 2$\times$10$^{13}$~K, with the maximum intrinsic brightness temperature $>$9.81$\times$10$^{11}$~K (Table~\ref{ParAve}). 
 
{\it 3C~446:} This quasar possesses a prominent parsec-scale jet extending eastward up to 3~mas from the core at 43~GHz. The jet undergoes two notable bends, at $\sim0.4$~mas toward the south and then to the northeast $\sim1.5$~mas from the core (Figure~\ref{maps}). The kinematics of the quasar have not been studied well, especially at high frequencies. The MOJAVE survey \citep{LIST13} has revealed a range in apparent speeds from 8 to $18c$. We have detected 7 moving knots, $B1$-$B5$, $C1$, and $C2$, with similar apparent speeds (Fig.~4.34, {\it left}); the most common is $\beta_{\rm app}\sim14c$ and the maximum is $22c$, 
characterizing the jet as highly relativistic. The ejections of knots $B1$-$B5$ appear to be connected
with a strong mm-wave outburst starting in 2007 and lasting for $\sim4$~yr (Fig.~4.34, {\it right}). As can be seen in Figure~4.34, {\it left}, the bends affect the motions of knots. $C1$ and $C2$ decelerate near the 0.4~mas bend, speed up later, and decelerate again near the 1.5~mas bend. Although we distinguish knot $B3$ from $B2$ based on separation from the core versus time, these could be the same feature if $B2$ experiences a significant jump in position near 0.4~mas when new knot $B4$ appears in the jet (Fig.~2.34). 

{\it CTA~102:} The jet kinematics of this quasar are well studied, e.g., by J05, \cite{FROMM13}, and \cite{Carolina15b}. We find a significant range of apparent speeds in the jet, 3 to $27c$, and a stationary feature, $A1$, in addition to the core (Fig.~\ref{maps}), that can be associated with stationary feature $A1$ noted by J05 $\sim0.12$~mas from the core. \cite{FROMM13} have reported a stationary feature at 0.1~mas from the core as well, and interpreted it as a recollimation shock. Knot $D1$, with the slowest apparent speed in our data (Fig.~4.35, {\it left}), is most likely knot $C$ in J05, and the feature at $\sim2$~mas in \cite{FROMM13}. Neither the latter nor $C$ show motion exceeding the uncertainties. All fast moving knots, $B1$-$B4$, are ejected during the rising branches of mm-wave outbursts (Fig.~4.35, {\it right}). The data presented here are almost the same as those discussed in \cite{Carolina15b} for the 2007-2012 period, although a few new epochs have been added and some new models obtained. Therefore, knots $B2$-$B4$ can be associated with $N1$-$N3$, respectively, and $A1$ with $C1$ from \cite{Carolina15b}. We have found a moving feature $B1$, which is not reported in \cite{Carolina15b}. Its evolution is shown in the sequence of images in Figure~2.35 along with the ejection and separation from the core of the brightest knot, $B2$. The mm-wave outburst starting near the end of 2012 appears to be connected with dramatic $\gamma$-ray activity of the quasar, which is accompanied by the ejection of a new superluminal knot, $N4$ \citep{Carolina15b}, based on later data than those presented here. 

{\it 3C~454.3:} This quasar is one of the most active blazars over the past 15~years.
A dramatic X-ray/UV/optical/mm-wave outburst was observed in 2005 
\citep[e.g.,][]{MASSIMO07}, followed by multifrequency outbursts in 2007 \citep{RAI08}
and 2008 \citep{MASSIMO09}. The latter two were monitored at $\gamma$-ray energies by the {\it AGILE} satellite \citep{DONNA09, VER09}.  In 2009 December, 3C~454.3 became the brightest source in the $\gamma$-ray sky, \citep[e.g.,][]{PACCI10}, and an exceptional $\gamma$-ray outburst in 2010 November \citep{ABDO11} was accompanied by dramatic activity in the quasar across the electromagnetic spectrum \citep[e.g.,][]{VER11,ANN12}. All of these high-energy outbursts from 2005 to 2010 were associated with ejections of superluminal knots into the 43~GHz
parsec-scale jet, $KI$-$KIII$ \citep{J10}, and $K09$, $K10$ \citep{J13}. Analysis of parsec-scale jet behavior at longer wavelengths during the same period of time can be found in \cite{SILKE13, LIST13, LIST16}. In addition to superluminal knots moving downstream the jet, \cite{SILKE13} have found an arc-like structure around the VLBI core
of 3C~454.3, which expands with a superluminal speed as well. Here we present the evolution of the knots, including some new epochs and later data than those discussed in \cite{J10,J13}. To be consistent with the previous studies of the jet kinematics at 43~GHz \citep{J01,J05}, which give the history of superluminal ejections in 3C~454.3 from 1995 to 2013, we number moving knots following the designation in J05. During 2007-2013, we detect five superluminal knots, $B7$-$B11$ (Fig.~4.36, {\it left}). Knots $KI$-$KIII$ in \cite{J10} can be associated with $B7$-$B9$, respectively, and $K09$ and $K10$ in \cite{J13} with $B10$ and $B11$, respectively. In general, lower jet velocities are measured relative to 1995-2004. No new jet events were detected in 2011-2012, when the multifrequency emission became quiescent. The bright feature $C$ located $\sim0.6$~mas from the core has an especially long history, first discussed by \cite{PT87}. Although in the current data $C$ moves at $\sim3c$ toward the core, the knot is, most likely, a quasi-stationary feature of the jet, fluctuating from 0.5 to 0.7~mas with respect to the core as strong knots pass through the core, as observed in 1995 \citep{J01} and 2010 (Fig.~4.36, {\it left}). This suggests that the motion of the core itself might be responsible for such fluctuations. The sequence of images in Figure~2.36 shows merging of $B8$ and $B10$ with $C$, while Figure~4.36, {\it right} reveals that it is common for superluminal knots in the jet of 3C~454.3 to be of similar brightness, or even  brighter than, the core over several months. A very high brightness temperature of the core was observed during the multi-wavelength outbursts in 2010 Spring and Autumn, with the maximum intrinsic brightness temperature of the core, $T_{\rm b,int}$=1.65$\times$10$^{12}$, derived on 2010 October 24, which coincides with the start of the most prominent $\gamma$-ray outburst in 3C~454.3 \citep{ABDO11}.

{\it Facilities:} VLBA, Mets\"ahovi Radio Obs.

\end{document}